\documentstyle[epsfig,12pt]{article}
\newlength{\dinwidth}
\newlength{\dinmargin}
\catcode`@=11

\def\@citex[#1]#2{\if@filesw\immediate\write\@auxout{\string\citation{#2}}\fi
  \def\@citea{}\@cite{\@for\@citeb:=#2\do
    {\@citea\def\@citea{,\penalty\@m}\@ifundefined
      {b@\@citeb}{{\bf ?}\@warning
       {Citation `\@citeb' on page \thepage \space undefined}}%
\hbox{\csname b@\@citeb\endcsname}}}{#1}}

\def\citer{\@ifnextchar [{\@tempswatrue\@citexr}{\@tempswafalse\@citexr[]}}

\def\@citexr[#1]#2{\if@filesw\immediate\write\@auxout{\string\citation{#2}}\fi
  \def\@citea{}\@cite{\@for\@citeb:=#2\do
    {\@citea\def\@citea{--\penalty\@m}\@ifundefined
       {b@\@citeb}{{\bf ?}\@warning
       {Citation `\@citeb' on page \thepage \space undefined}}%
\hbox{\csname b@\@citeb\endcsname}}}{#1}}
\catcode`@=12
\def\B{{\cal B}}
\begin{document}
\begin{flushright}  
DESY 98-041\\
April 1998
\end{flushright}

\begin{center}
{\large\bf Experimental Tests of Factorization in  Charmless 
Non-Leptonic Two-Body $B$ Decays}
\end{center}
\vspace*{1.0cm}
\centerline{\large\bf A.~Ali}
\centerline{\bf Deutsches Elektronen-Synchrotron DESY, 22607 Hamburg, 
Germany}
\vspace*{0.3cm}
\centerline{\large\bf G.~Kramer\footnote{Supported
by Bundesministerium f\"ur Bildung und
Forschung, Bonn, under Contract 057HH92P(0) and EEC Program ``Human Capital
 and Mobility'' through Network ``Physics at High Energy Colliders'' under
Contract CHRX-CT93-0357 (DG12COMA).} and 
 Cai-Dian L\"u\footnote{Alexander von Humboldt Foundation Fellow.}}
\centerline{\bf II. Institut f\"ur Theoretische Physik, Universit\"at 
Hamburg, 22761 Hamburg, Germany}
\date{}

\vspace*{1.0cm}
\centerline{\Large\bf Abstract}
\vspace*{1cm}
Using a theoretical framework based on the next-to-leading order QCD-improved
effective Hamiltonian and a factorization Ansatz for the  hadronic
matrix elements of the four-quark operators, we reassess branching fractions 
in two-body non-leptonic decays $B \to PP, PV, VV$, involving the lowest
lying light pseudoscalar $(P)$ and vector $(V)$ mesons in the standard model.
 We work out the parametric
dependence of the decay rates making use of the currently
available information on the weak mixing matrix elements, form factors, 
decay constants and quark masses. Using the sensitivity of the decay 
rates on the effective number of colors, $N_c$, as a criterion of
theoretical predictivity, we classify all the current-current (tree) 
and penguin transitions in five different classes. The recently measured
charmless two-body $B \to PP$ decays $(B^+ \to K^+ \eta^\prime, B^0 \to K^0
\eta^\prime, B^0 \to K^+\pi^-, B^+ \to \pi^+ K^0$ and charge conjugates) are 
dominated by the $N_c$-stable QCD 
penguins (class-IV transitions) and their estimates are consistent with 
data. The measured charmless $B \to PV$  $(B^+ \to \omega K^+, ~B^+ \to 
\omega h^+)$ and $B\to VV$ transition $(B \to \phi K^*)$, on the other 
hand, belong to the penguin (class-V) and tree (class-III) transitions.
The class-V penguin transitions are 
$N_c$-sensitive and/or involve large 
cancellations among competing amplitudes making their decay rates in general 
more difficult to predict. Some of these transitions 
may also receive significant contributions from annihilation
and/or final state interactions. We propose a number of tests of the 
factorization framework in terms of the ratios of branching ratios for 
some selected $B \to h_1 h_2$ decays involving light hadrons $h_1$ and 
$h_2$, which depend only moderately on the form 
factors. We also propose a set of measurements to determine the effective 
coefficients of the current-current and QCD penguin operators. The potential
impact of $B \to h_1 h_2$ decays on the CKM phenomenology is emphasized
by analyzing a number of decay rates in the factorization framework.
\newpage

\section{Introduction}

Recent measurements by the CLEO Collaboration \cite{cleo,cleobok} of a 
number 
of decays of the type $B\to h_1h_2$, where $h_1$ and $h_2$ are light hadrons 
such as $h_1h_2= \pi\pi, \pi K, \eta' K, \omega K$, 
have triggered considerable theoretical 
interest in understanding two-body non-leptonic B decays.
These decays involve the so-called tree
(current-current) $b \to (u,c)$ and/or $b \to s$ (or $b \to d$) penguin 
amplitudes 
with, in general, both the QCD and electroweak penguins participating.
The appropriate theoretical framework to study these decays is that
of an effective theory based on the Wilson operator product expansion
\cite{Wilson} 
obtained by integrating out the heavy degrees of
freedom, which in the standard model (SM) are the top quark and $W^\pm$ 
bosons. This effective theory allows to separate the short- and long-distance
physics and one can implement  
the perturbative QCD improvements systematically in this
approach. Leading order corrections have been known for quite some 
time \cite{AM74} and in many cases this program has been completed 
up to and including the next-to-leading order corrections \cite{BBL96}.  
Present QCD technology, however, does not allow to 
undertake a complete calculation of the exclusive non-leptonic decay rates  
from first principles, such as  provided by the lattice-QCD 
approach, as this requires the knowledge of the hadronic matrix elements 
$<h_1h_2 | {\cal H}_{eff} |B>$, where $ {\cal H}_{eff}$ is an effective 
Hamiltonian consisting of the four-quark and magnetic moment operators.
These are too complicated objects to be calculated with the 
current lattice-QCD methods. Hence, a certain amount of model building 
involving these hadronic matrix elements is at present unavoidable.
 
The approach which has often been employed in non-leptonic heavy hadron 
decays is based on factorization \citer{Feyn65,BSW87}.
With the factorization Ansatz, the matrix elements 
$<h_1h_2 | {\cal H}_{eff} |B>$ can be expressed as a product of two factors 
$<h_1 | J_1 |B><h_2 | J_2 |0>$. 
The resulting matrix elements of the current operators $J_i$  
are theoretically  more tractable and have 
been mostly calculated in well-defined theoretical frameworks, such as 
Lattice-QCD \citer{APE96,UKQCD97}, QCD sum rules 
\citer{ABS94,Ball98} and potential models 
\cite{BSW87},\citer{ISGW,NS97}; some are also available from 
data on semileptonic and leptonic decays \cite{PDG96}. One can then make 
quantitative predictions
in this framework taking into account the theoretical and experimental 
dispersion in the input parameters in the decay rates.
 Factorization holds in the 
limit that one ignores soft non-perturbative effects. 
The rationale of this lies in the phenomenon of color-transparency 
\cite{bjorken}, in which one expects intuitively that a pair of fast moving 
(energetic) 
quarks in a color-singlet state effectively decouples from long-wavelength 
gluons. In the decays $B\to h_1h_2$, with typically $E_{h_{1,2}}\sim 
O(m_B/2)$, the energy of the quarks leaving the interaction is large and 
soft final state interactions should be small and hence factorization 
should be a good approximation. 
Final state interactions generated by hard gluon exchanges are, however,
perturbatively calculable and can be included. 
 Phenomenology of the factorization
hypothesis in the decays $B \to D^{(*)} \pi (\rho)$, $B \to J/\psi K^{(*)}$ 
and related ones, involving the so-called current-current amplitudes, 
has been worked out and compared with the existing data with the tentative 
conclusion that data in these decays can be described in terms
of two phenomenological parameters, $a_1$ and $a_2$ \cite{BSW87}, whose
values seem to be universal \cite{NS97,BHP96}.   

 The decays $B \to h_1 h_2$ have been studied repeatedly in the factorization
framework \citer{cpbook,Deandrea93}.  However, with the
measurements of some of the $B \to h_1 h_2$ decays \cite{cleo,cleobok},
theoretical interest in this field has resurged.
In particular, NLO-improved perturbative framework with updated 
phenomenological input has been used in a number of recent papers 
\citer{ag,CT98} to study the CLEO data.
We would like to take a closer look at the non-leptonic two-body decays $B 
\to h_1 h_2$, in which QCD and/or electroweak penguins are expected to
play a significant role.
  
There are several theoretical issues involved in $B \to h_1 h_2$ decays,
which one does not encounter in the 
transitions $B \to H_1 h_2$, where $H_1$ is an
open $(D^{(*)},D_s^{(*)})$ or bound $(J/\psi,\eta_c,\chi_c)$ charmed 
hadron, or in decays such as $B \to D_s^{(*)}D^{(*)}$,
which are governed by the current-current (tree) amplitudes.
 In the case of the induced $b \to s$ and $b \to d$ transitions, penguins
play an important role. Of these 
penguin transitions, the ones involving the top-quark 
can be reliably calculated in perturbation theory as they represent 
genuine short-distance contributions.
 The rest of the penguins, which involve both the  
charm- and light-quarks, also have genuine short-distance
contributions which can be calculated using perturbation theory.
Their importance in the context of direct CP asymmetries has been
emphasized repeatedly in the literature \cite{kps,kp}. However, in 
principle, such penguin 
amplitudes may also involve significant non-perturbative (long-distance) 
contributions. Arguments for an enhanced role of non-perturbative penguin
effects have been advanced in the literature \cite{Martinellifudge}.
In simpler cases, such as the electromagnetic  
decays $B \to X_s + \gamma$ and $B \to K^* + \gamma$, charm-penguins
are likewise present and they   
introduce $1/m_c^2$ (and higher order) power corrections akin to the 
long-distance effects being discussed in non-leptonic decays. In these 
cases, one finds that the $1/m_c^2$ power corrections are  negligible 
\citer{Voloshinbsg,BIR97}. The same holds for the non-resonant
$B \to X_s \ell^+ \ell^-$ decays \cite{BIR97}.
The pattern of the $1/m_c^2$-corrections  
remains to be investigated systematically for non-leptonic $b \to (s,d) q 
\bar{q}$ decays. However, it is suggestive that the next-to-leading
order QCD-improved framework based on factorization can explain most of the 
recent CLEO data without invoking a significant non-perturbative penguin
contribution \cite{ag,acgk}. With improved measurements, this aspect will
surely be scrutinized much more quantitatively.

A related issue is that of the current-current $b \to c \bar{c} s$ and 
$b \to c \bar{c} d$ transitions feeding into the $b \to s q\bar{q}$
and $b \to d q \bar{q}$ transitions, respectively, by (soft) final state 
interactions (FSI) \citer{falketal98,delepine98}.
While in the oft-studied case of $B \to K \pi$ decays, these effects
are not found to be overwhelming for decay rates, yet, in general, it is
not difficult to imagine situations where FSI may yield the dominant 
contribution to a decay width. There are three ways in which the amplitude
for a decay in the factorization approach can become small: (i)  the
effective coefficients of the various operators entering into specific 
decays are small reflecting either their intrinsic (perturbative) values,
implying they are small for $N_c=3$, or their $N_c$-sensitivity meaning that
they are small for some phenomenologically relevant value of $N_c$, 
(ii)  due to CKM-suppression,  
(iii) due to delicate cancellations among various competing Feynman 
diagrams, resulting into an amplitude which is effectively small. Using 
$N_c$, the effective 
number of colors, as a variable parameter, it becomes immediately clear
that some linear combinations of the effective coefficients 
entering in specific decays are particularly
sensitive to $N_c$ and they indeed become very small for certain values of
$\xi=1/N_c$. 
This then implies that other contributions such as the ones coming
from FSI and/or annihilation may become important. 
A good case to illustrate this is the decay $B^\pm \to K^\pm K$, whose 
decay rate may be enhanced
by an order of magnitude due to FSI \cite{fleischer98} and/or annihilation
\cite{annihilationold} contributions.

 In this paper, we undertake a comprehensive study, within the factorization 
framework, of all the two-body decay modes of 
the type $B\to PP$, $B\to PV$ and $B\to VV$ where $P(V)$ is a light 
pseudoscalar (vector) meson in the flavor $U(3)$ nonet. Concentrating on 
the lowest lying $0^-$ and
$1^-$ mesons, there are some seventy-six (76) such decays (and an equal 
number involving the charge conjugate states). The branching ratios of
these decays are found to vary over four orders of magnitude.
We calculate their decay rates (branching ratios) and work out the most
sensitive parametric dependence of these quantities. In many cases the
factorized amplitudes are small due to the reasons mentioned in the
preceding paragraph. While this by itself does not imply an intrinsic
inability to calculate, it becomes difficult to be confident if the rate is 
additionally unstable,  requiring a good deal of
theoretical fine tuning in the factorization approach.  We 
list all such two-body decay modes here and caution about drawing too
quantitative conclusions on their widths based on the factorized amplitudes 
alone. We think that the
sensitivity of some of the effective coefficients $a_i$ on $N_c$ and the
fine tuning required in some amplitudes can be used as a criterion of
predictivity of $B \to h_1 h_2$ decay rates in the factorization approach. 
The pattern of
color-suppression in current-current amplitudes has been previously used
to classify the $N_c$-sensitivity of these decays into three classes 
\cite{BSW87}. We extend this to
also include the penguin-dominated decays, which belong either to 
$N_c$-stable
(class-IV) or $N_c$-sensitive (class-V) decays. In addition, 
penguin-dominated decay 
amplitudes involving large cancellations are also included in class-V.
All penguin-dominated $B \to PP$ decays belong to class-IV. This class 
includes in particular
the decays $B^0 \to K^+ \pi^-$, $B^+ \to K^+\eta^\prime$, $B^0 \to K^0 
\eta^\prime$ and $B^+ \to \pi^+ K^0$, measured recently by the CLEO 
collaboration
\cite{cleo} (here and in what follows, charge conjugate decays are
implied). On the other hand, the recently measured  
$B \to PV$ and $B \to VV$ decay modes by CLEO \cite{cleobok} are in  
class-V ($B^+ \to \omega K^+$ and $B \to K^* \phi)$ or tree-dominated 
class-III $(B^+ \to \omega \pi^+)$. Possibly some of these, and 
many more examples of class-V decays worked out by us here, indicate that 
the factorization-based approach is rather uncertain in these decays and one 
may have to develop more powerful methods to make theoretically stable 
predictions in this class.
Factorization approach is expected to do a better job in accounting for 
class-IV decays - a claim which is  pursued here 
and which is supported by present data.

We propose tests of factorization in $B\to h_1 h_2$
decays through measuring a number of ratios of the branching ratios 
which   
depend only on the form factors but are otherwise insensitive to other  
parameters, such as the effective coefficients $a_i$
 and hence $N_c$, quark masses, QCD-scale parameter
and CKM matrix elements. The residual model dependence of these ratios
on the form factors is worked out in two
representative cases: (i) the Bauer-Stech-Wirbel (BSW) model
\cite{BSW87}
and (ii) a hybrid approach, based on Lattice-QCD/Light-cone QCD sum rules,
specifically making use of the results obtained in the frameworks of
lattice-QCD \cite{FS97,UKQCD97} and the Light-Cone QCD
sum rules \cite{ABS94,Ball98}. The proposed ratios will
test factorization and determine the form factors.

A quantitative test of the factorization approach lies in a consistent
determination of the effective coefficients $a_i$ of this 
framework. The QCD perturbative contributions to $a_i$ can be calculated
in terms of the renormalized Wilson coefficients 
in the effective Hamiltonian governing the decays $B \to h_1 h_2$.
Then, there are non-perturbative contributions which have to be 
determined phenomenologically.
Of these $a_1$ and $a_2$ govern the current-current amplitudes
and they should be determined in $B \to h_1 h_2$ decays without any {\it 
prior} prejudice.
Four of the $a_i$'s ($a_3,...,a_6$) govern the QCD-penguin amplitudes 
and four more ($a_7,...,a_{10}$) govern the electroweak-penguin
amplitudes. We propose measurements 
of selected branching ratios (and their ratios) to determine  
the effective coefficients $a_1,a_2,a_4$ and
$a_6$ from the first six from data on $B \to h_1 h_2$ decays in 
the future. Since the Wilson 
coefficients of the electroweak penguin operators in the SM are rather 
small in magnitude (except for $C_9$), which in turn yield very small
branching ratios for these decays, a determination of $a_7,...,a_{10}$ is 
a formidable proposition. The coefficient $a_9$ can be determined and we
propose several decays to measure this. We also list decay modes in which
electroweak penguins (hence $a_7,...,
a_{10}$) do play a noticeable role, and work out their corresponding 
branching ratios.

Finally, we explore the potential impact of the  $B\to h_1 h_2$ decays  on 
the 
phenomenology of the Cabibbo-Kobayashi-Maskawa (CKM) matrix \cite{CKM}.
Here, we discuss relations of the type put forward by 
Fleischer and Mannel \cite{fm} (see, also \cite{fleischer})
involving the decay rates of $B^0 \to K^+ \pi^-$
and $B^+ \to K^0 \pi^+$,
which can be used to determine $\cos \gamma$, where $\gamma$
is one of the angles of the CKM unitarity triangle, in terms of the
ratio of the tree-to-penguin amplitudes $z\equiv T/P$ and $\delta$, the
strong phase shift difference involving these amplitudes. A bound 
on $\sin^2 \gamma$ can be obtained, assuming that there are just the tree and
QCD-penguin amplitudes:
\begin{equation}
 R \equiv \frac{\Gamma(B^0 \to \pi^\mp K^\pm)}{\Gamma(B^\pm \to
\pi^\pm K^0)} = 1-2\, z \cos \gamma \cos \delta + z^2 \geq \sin^2 
\gamma~. 
\end{equation}
 From this, constraints on $\gamma$ of the form
\begin{equation}
 0^\circ \leq \gamma \leq \gamma_0 ~~\vee ~~ 180^\circ - \gamma_0 \leq \gamma
\leq 180^\circ ~
\end{equation}
follow, where $\gamma_0$ is the maximum value of
$\gamma$, which are complementary to the ones from the CKM unitarity fits 
\cite{aliapctp97,Parodi98}. There are similar
relations involving the decays $B \to PV$ and $B \to VV$, where 
$P=\pi,K$ and $V=\rho,K^*$. A determination of the angle $\gamma$, however,
requires the knowledge of $z_i$ and $\delta_i$ in these processes. Also,
the effect of the electroweak penguins has to be included.
Having a definite model, whose consistency can be checked
in a number of decays, one could 
determine (within a certain range) the values of  $z_i$ and $\delta_i$. 
Given data, this would allow us in turn to determine 
$\gamma$ in a number of two-body non-leptonic $B$ decays.
 We draw 
inferences on the angle $\gamma$ based on existing data on $R$, and in line
with \cite{ag}, we show that the allowed values of $\gamma$
(or the CKM-Wolfenstein \cite{Wolfenstein83} parameters $\rho$ and $\eta$) 
from this analysis are
consistent with the ones following from the CKM unitarity fits.
Similar analysis can be carried out for the decays $B \to PP, PV,VV$, 
where now $P=\pi^0,\pi^\pm$ and $V=\rho^0,\rho^\pm$. Measurements of
these decays and their ratios would allow to draw 
inferences on the angle $\alpha$. We illustrate this
in the context of our model. 
The other kind of relations discussed by us involve ratios of the decay 
rates dominated by the $b\to s$ and $b\to d$ penguins, respectively.
As pointed out in ref.~\cite{gr}, these ratios 
can be used to determine the ratio of the CKM matrix 
elements $|V_{td}/V_{ts}|$. Since this
CKM-ratio will, in principle, be measured also in $B^0$ - $\overline{B^0}$ 
mixings and
radiative and semileptonic rare $B$ decays \cite{aliapctp97,aag98}, one 
could check the consistency of such determinations to reach quantitative 
conclusions about the QCD dynamics at work in non-leptonic decays.
However, it is conceivable that some of the non-leptonic decays may already
provide interesting information on $V_{td}$ before the other mentioned 
processes are actually measured. While not competitive in terms of 
eventual theoretical precision, non-leptonic decays are nevertheless quite 
instructive in this respect for the current CKM phenomenology. 

This paper is organized as follows:
In section 2, we discuss the effective Hamiltonian together with the quark 
level matrix elements and the numerical values of the Wilson coefficients 
$C_i^{eff}$ in the effective Hamiltonian approach. 
In section 3, we introduce the factorization Ansatz, define the relevant 
matrix elements and discuss their evaluation in the BSW model and in the 
hybrid lattice QCD/QCD sum rule approach.
The matrix elements for the three classes $B\to PP$, $B\to PV$ 
and $B\to VV$, obtained in the factorization approach,  are relegated to 
Appendix A, B and C, respectively.
Section 4 contains a discussion of the various input parameters (CKM matrix
elements, quark masses, hadronic form factors and mesonic constants). The
numerical input we use in the estimates of branching ratios 
are collected in various tables. In section 5, we  tabulate the
values of the phenomenological parameters $a_i$ for three values of
the effective number of colors $(N_c=2,3, \infty)$ for the four cases
of interest $b \to s$, $\bar{b} \to \bar{s}$, $b \to d$ and $\bar{b} \to 
\bar{d}$. This serves to show both the relative magnitude of the 
effective coefficients of the various
operators in $B \to h_1 h_2$ decays in the factorization approach and 
also the stability of these coefficients against $N_c$. The 
classification of the $B \to h_1 h_2$ decays is also discussed here.
We also discuss the contribution of the annihilation amplitudes and list
some decays of potential interest. 
Section 6 contains the numerical results for the branching ratios which
we tabulate for three specific values of the effective number of colors
$N_c=2,3,\infty$.
The parametric dependence on $\xi=1/N_c$ is shown for some representative 
cases in various figures and compared with data, whenever available.
In section 7, we list a number of ratios of branching ratios to test the 
hypothesis of 
factorization and give their values for the two sets of form factors 
(in the BSW and the hybrid Lattice-QCD/QCD sum rule approaches). 
We also discuss the determination of the effective coefficients 
$a_1,...,a_6$ here through a number of relations.
We estimate these ratios and make comparisons with data, whenever available.
Potential impact of the $B \to h_1 h_2$ decay rates on CKM phenomenology
are also discussed here.
Finally, we conclude in section 8 with a summary and outlook.

\section{The Effective Hamiltonian}

\subsection{Short-distance QCD corrections}
We write the effective Hamiltonian $H_{eff}$
for the $\Delta B=1$ transitions as
\begin{equation}
\label{heff}
{\cal H}_{eff}
= \frac{G_{F}}{\sqrt{2}} \, \left[ V_{ub} V_{uq}^*
\left (C_1 O_1^u + C_2 O_2^u \right)
+ V_{cb} V_{cq}^* 
\left (C_1 O_1^c + C_2 O_2^c \right) -
V_{tb} V_{tq}^* \,
\left(\sum_{i=3}^{10}
C_{i} \, O_i + C_g O_g \right) \right] \quad ,
\end{equation}
where
$q=d,s$ and $C_i$ are the Wilson coefficients evaluated at the
renormalization scale $\mu$. We
specify below the operators in  ${\cal H}_{eff}$ for $b \to s$ 
transitions (for $b \to d$ transitions, one has to make the replacement
$s \to d$):
\begin{equation}\begin{array}{llllll}
O_1^{u} & = &\bar s_\alpha\gamma^\mu L u_\alpha\cdot \bar 
u_\beta\gamma_\mu L b_\beta\ , &
O_2^{u} & = &  \bar s_\alpha\gamma^\mu L u_\beta\cdot \bar 
u_\beta\gamma_\mu L b_\alpha\ , \\ 
O_1^{c} & = &\bar s_\alpha\gamma^\mu L c_\alpha\cdot \bar
c_\beta\gamma_\mu L b_\beta\ , &
O_2^{c} & = &  \bar s_\alpha\gamma^\mu L c_\beta\cdot \bar
c_\beta\gamma_\mu L b_\alpha\ , \\ 
O_3 & = & \bar s_\alpha\gamma^\mu L b_\alpha\cdot \sum_{q'}\bar
 q_\beta'\gamma_\mu L q_\beta'\ ,   &
O_4 & = & \bar s_\alpha\gamma^\mu L b_\beta\cdot \sum_{q'}\bar 
q_\beta'\gamma_\mu L q_\alpha'\ , \\
O_5 & = & \bar s_\alpha\gamma^\mu L b_\alpha\cdot \sum_{q'}\bar 
q_\beta'\gamma_\mu R q_\beta'\ ,   &
O_6 & = & \bar s_\alpha\gamma^\mu L b_\beta\cdot \sum_{q'}\bar 
q_\beta'\gamma_\mu R q_\alpha'\ , \\
O_7 & = & \frac{3}{2}\bar s_\alpha\gamma^\mu L b_\alpha\cdot 
\sum_{q'}e_{q'}\bar q_\beta'\gamma_\mu R q_\beta'\ ,   &
O_8 & = & \frac{3}{2}\bar s_\alpha\gamma^\mu L b_\beta\cdot 
\sum_{q'}e_{q'}\bar q_\beta'\gamma_\mu R q_\alpha'\ , \\
O_9 & = & \frac{3}{2}\bar s_\alpha\gamma^\mu L b_\alpha\cdot 
\sum_{q'}e_{q'}\bar q_\beta'\gamma_\mu L q_\beta'\ ,   &
O_{10} & = & \frac{3}{2}\bar s_\alpha\gamma^\mu L b_\beta\cdot 
\sum_{q'}e_{q'}\bar q_\beta'\gamma_\mu L q_\alpha'\ ,\\
O_g &=& (g_s/8\pi^{2}) \, m_b \, \bar{s}_{\alpha} \, \sigma^{\mu \nu}
      \, R  \, (\lambda^A_{\alpha \beta}/2) \,b_{\beta}
      \ G^A_{\mu \nu}~. 
\label{operators}
\end{array}
\end{equation}
Here $\alpha$ and $\beta$ are the $SU(3)$ color indices and 
$\lambda^A_{\alpha \beta}$, $A=1,...,8$ are the Gell-Mann matrices;
$L$ and $R$ are the left- and right-handed projection operators with
$L(R)=1 - \gamma_5 ~(1 + \gamma_5)$,
and $G^A_{\mu \nu}$ denotes the gluonic field strength tensor.
The sum over $q'$ runs over the quark fields that are active at the scale 
$\mu=O(m_b)$, i.e., $(q'\epsilon\{u,d,s,c,b\})$. The usual tree-level 
$W$-exchange contribution in the effective theory corresponds to $O_1$
(with $C_1(M_W) = 1+O(\alpha_s)$) and $O_2$ emerges due to the QCD 
corrections. The operators $O_3,\ldots,O_6$ arise from 
the QCD-penguin diagrams which contribute in order $\alpha_s$
through the initial 
values of the Wilson coefficients at $\mu \approx M_W$ \cite{k2} and 
operator mixing due to the QCD corrections \cite{QCD_SD}.
 Similarly, the operators $O_7,\ldots,O_{10}$ arise from
the electroweak-penguin diagrams. Note that we neglect the effects of 
the electromagnetic penguin operator which we did not list explicitly.
The effect of the weak annihilation and exchange diagrams will be 
discussed later.

The renormalization group evolution from $\mu \approx M_W$ to $\mu
\approx m_b$ has been evaluated
in leading order in the electromagnetic coupling and in the 
NLL precision in the strong coupling $\alpha_s$
\cite{Burasmartinelli}. Working consistently to the NLL precision, the
coefficients $C_1,...,C_{10}$ are needed in NLL precision, while it is
sufficient to use the LL value for $C_g$. These
coefficients depend on the renormalization scheme used.  
To obtain numerical values for the $ C_i$ we must specify the input
parameters. We
fix $\alpha_s(M_z)=0.118$, $\alpha_{ew}(M_z)=1/128 $ and
$ \mu=2.5~GeV $. Then in the naive dimensional regularization (NDR)
scheme, we have:
\begin{equation}\begin{array}{llllll}
 C_1 & = & 1.117\ ,  &  C_2 & = & - 0.257\ , \\
 C_3 & = & 0.017\ ,  &  C_4 & = & -0.044\ , \\
 C_5 & = & 0.011\ ,  &  C_6 & = & -0.056\ , \\
 C_7 & = & -1\times 10^{-5}\ , &  C_8 & = & 5\times 10^{-4}\ , \\
 C_9 & = & -0.010\ ,  & C_{10} & = & 0.002\ , \\
 C_g^{eff} & = & -0.158~.
\end{array}\end{equation}
Here, $C_g^{eff}= C_g + C_5$. 
 From the electroweak coefficients $C_7,...,C_{10}$, only $ C_9$ has a 
sizable value compared to 
the coefficients of the QCD penguins; its major contribution arises
from the $Z$ penguin. Note that the scale $(\mu)$ and scheme-dependence
of the Wilson coefficients will cancel against the corresponding dependences
in the matrix elements of the operators in ${\cal H}_{eff}$, as shown
explicitly in \cite{Burasmartinelli}. Since, the matrix elements given below
are obtained in the NDR-scheme, we have listed the values of the Wilson
coefficients $C_i$ also in this scheme.

\subsection{Quark-level Matrix Elements}

In the NLL precision, the matrix elements of 
${\cal H}_{eff}$ are
to be treated at the one-loop level. The one-loop matrix
elements can be rewritten in terms of the tree-level matrix elements of
the effective operators  
\begin{equation}
\langle sq'\bar q'\vert {\cal H}_{eff}\vert b\rangle =
\sum_{i,j} C_i^{eff}(\mu)
\langle sq'\bar q'\vert O_j \vert b\rangle ^{\rm tree}  .
\label{me1}
\end{equation}

In the NDR renormalization scheme and for $SU(3)_C$, the effective 
coefficients multiplying
the matrix elements $< sq'\bar q'\vert O_j^{(q)}\vert b>^{\rm tree}$
become ( $r_V^T$ and $\gamma_V^T$ are the transpose of the matrices 
given below) 
 \begin{eqnarray}
C_1^{eff} &=& C_1 + 
\frac{\alpha_s}{4\pi} \, \left( r_V^T +
 \gamma_{V}^T \log \frac{m_b}{\mu}\right)_{1j} \, C_j  +\cdots ,
\nonumber \\
C_2^{eff} &=& C_2 +  
\frac{\alpha_s}{4\pi} \, \left( r_V^T +
 \gamma_V^T \log \frac{m_b}{\mu} \right)_{2j} \, C_j   +\cdots ,
\nonumber \\
C_3^{eff} &=& C_3 -\frac{1}{6} \frac{\alpha_s}{4\pi} \, (C_t + C_p + C_g) +
\frac{\alpha_s}{4\pi} \, \left(  r_V^T +
\gamma_V^T \log \frac{m_b}{\mu}\right)_{3j} \, C_j +\cdots ,  
\nonumber \\
C_4^{eff} &=& C_4 +\frac{1}{2} \frac{\alpha_s}{4\pi} \, (C_t + C_p + C_g) +
\frac{\alpha_s}{4\pi} \, \left( r_V^T +
\gamma_V^T \log \frac{m_b}{\mu}\right)_{4j} \, C_j   +\cdots ,
\nonumber \\
C_5^{eff} &=& C_5 -\frac{1}{6} \frac{\alpha_s}{4\pi} \, (C_t + C_p + C_g) +
\frac{\alpha_s}{4\pi} \, \left(  r_V^T +
\gamma_V^T \log \frac{m_b}{\mu}\right)_{5j} \, C_j   +\cdots ,
\nonumber \\
C_6^{eff} &=& C_6 +\frac{1}{2} \frac{\alpha_s}{4\pi} \, (C_t + C_p + C_g) +
\frac{\alpha_s}{4\pi} \, \left(  r_V^T +
\gamma_V^T \log \frac{m_b}{\mu} \right)_{6j} \, C_j   +\cdots ,\nonumber \\
C_7^{\rm eff} & = &  C_7 +  \frac{\alpha_{ew}}{8\pi} C_e ~, \nonumber\\
C_8^{\rm eff} & = &  C_8~, \nonumber\\
C_9^{\rm eff} & = &  C_9 +  \frac{\alpha_{ew}}{8\pi} C_e ~, \nonumber\\
C_{10}^{\rm eff} & = &  C_{10}~.
\label{ceff}
\end{eqnarray}

We have separated the contributions $C_t$, $C_p$, and $C_g$
arising from the penguin-type diagrams of 
the current-current operators $O_{1,2}$, the penguin-type diagrams of
the operators
$O_3$-$O_6$, and the tree-level diagram of the 
dipole operator $O_g$, respectively. Note also that we follow the 
procedure of ref.~\cite{ag} of including the tree-level diagrams $b \to s 
g \to s q^\prime \bar{q^\prime}$ associated with the operator $O_g$ into the
contribution $C_g$ appearing in the expressions for $C_i^{eff}$. So, we
have the hadronic matrix elements of four-quark operators only.
 The process-independent contributions 
from the vertex-type diagrams are contained in the matrices
$r_V$ and $\gamma_V$. Here 
$\gamma_V$ is that part of the anomalous matrix which is due
to the vertex  (and self-energy) corrections. This part can
be easily extracted from $\hat\gamma^{(0)}$ in ref. \cite{Burasmartinelli}:
\begin{equation}
\label{gammamatrix}
\gamma_V = \left(
\begin{array}{cccccc}
-2 & 6 & 0 & 0 & 0 & 0 \\
6 & -2 & 0 & 0 & 0 & 0 \\
0 & 0  &-2 & 6 & 0 & 0 \\
0 & 0  & 6            & -2           & 0 & 0 \\
0 & 0  & 0            & 0            & 2 & -6 \\
0 & 0  & 0            & 0            & 0 & -16 \\
\end{array}
\right) \quad .
\end{equation}
The matrix $r_V$ contains constant, i.e.,  momentum-independent, parts
associated with the vertex diagrams. This matrix
can be extracted from the matrix $\hat{r}$ defined
in eqn.~(2.12) (and given explicitly in eqn.~(4.6)) 
by Buras et al. in ref. \cite{Burasmartinelli}:
\begin{equation}
\label{rmatrix}
r_V = \left(
\begin{array}{cccccc}
\frac{7}{3} & -7 & 0 & 0 & 0 & 0 \\
-7 & \frac{7}{3} & 0 & 0 & 0 & 0 \\
0 & 0  & \frac{63}{27} & -7 & 0 & 0 \\
0 & 0  & -7            & \frac{7}{3}          & 0 & 0 \\
0 & 0  & 0            & 0            & -\frac{1}{3} & 1 \\
0 & 0  & 0            & 0            & -3 & \frac{35}{3} \\
\end{array}
\right) \quad .
\end{equation}
Note that the $\mu$ dependence and the scheme dependence
of the vertex correction diagrams
are fully taken into account in eqn.~(\ref{ceff}) by the terms
involving the matrices $\gamma_V$ and $r_V$, respectively.
There are, however, still scheme-independent, process-specific terms
omitted as indicated by the ellipses, and we refer to \cite{ag} for 
a discussion of these omitted terms in exclusive two-body $B$ decays.

 The quantities $C_t$, $C_p$, and $C_g$ are given in  the NDR scheme
(after $\overline{MS}$ renormalization) by   
\begin{eqnarray}
C_t &=& -\sum_{q' =u,c}\frac{V_{q'b}V_{q'q}^*}{V_{tb}V_{tq}^*}\left[
\frac{2}{3} + \frac{2}{3} \log \frac{m_{q'}^2}{\mu^2}
- \Delta F_1 \left( \frac{k^2}{m_{q'}^2} \right) \right]C_1\quad ,\label{ct}\\
C_p &=& C_3 \, \left[ \frac{4}{3} 
+ \frac{2}{3} \log \frac{m_s^2}{\mu^2}
+ \frac{2}{3} \log \frac{m_b^2}{\mu^2}
- \Delta F_1 \left( \frac{k^2}{m_s^2} \right) 
- \Delta F_1 \left( \frac{k^2}{m_b^2} \right) 
\right] \quad  \nonumber \\
&& + (C_4 + C_6) \, \sum_{i=u,d,s,c,b} \, \left[ \frac{2}{3} \, \log
\frac{m_i^2}{\mu^2} - \Delta F_1 \left( \frac{k^2}{m_i^2} \right) \,
\right] \quad .\label{cp}\\
\label{cg}
C_g &=& - \frac{2 m_b}{\sqrt{< k^2 >}} \, C_g^{eff} \quad ,
\end{eqnarray}
with $C_g^{eff}=C_g+C_5$.
 The function $\Delta F_1(z)$ is defined as
\begin{equation}
\label{deltaf1}
\Delta F_1(z) = -4 \, \int_0^1 dx \, x(1-x) \, \log \left[
1-z \, x(1-x) - i \epsilon \right] \quad .\label{12}
\end{equation}

 The corresponding electroweak coefficient $C_e$ is given by
\begin{equation}
\label{celweak}
C_e  = - \frac{8}{9}(3 C_2+ C_1)
\sum_{q' =u,c}\frac{V_{q'b}V_{q'q}^*}{V_{tb}V_{tq}^*}
\left(\frac{2}{3} + \frac{2}{3} \ln
        \frac{m_{q'}^2}{\mu^2}-\Delta F_1\left(\frac{k^2}{m_{q'}^2}
\right)\right) .
\end{equation}

Note that the quantities $C_t$ and $C_e$ depend on the CKM matrix elements. 
In addition, the  coefficients $C_i^{\rm eff}$ depend on $k^2$, where $k$ is 
the momentum transferred by the gluon, photon or $Z$ to the 
quark-antiquark pair $q^\prime \overline{q^\prime}$ in 
$b \to q q^\prime \overline{q^\prime}$. In two-body decays 
any information on $k^2$ is
lost in the factorization assumption. However, given a specific model
for the momentum distributions of the quark-antiquark pair inside the
hadron, the partonic distributions calculated here can be folded with this 
distribution, as, for example, has been done in \cite{SW}.
 Since, we are interested 
here in the decays $B \to h_1 h_2$, where $h_1,h_2$ are light mesons, it
is not unreasonable to assume that this smearing will be very similar
in all the decays being considered. In particular,  
$\langle k^2 \rangle$ is expected to be comparable in these decays. However,
the actual value of $\langle k^2 \rangle$ is model dependent.
 From simple two-body kinematics \cite{Deshpande}
or from the investigations in ref.~\cite{SW} one expects $k^2$ to be
typically in the range
\begin{equation}
\frac{m_b^2}{4}\stackrel{<}{\sim} k^2 \stackrel{<}{\sim}\frac{m_b^2}{2}\ .
\label{kk_range}
\end{equation}
As we shall see later, branching ratios considered here are not sensitive to 
the value of $k^2$ if it is varied in a reasonable range.

\section{Factorization Ansatz for the hadronic matrix 
elements of the four-quark operators}
We have now to calculate the hadronic matrix elements of the type
$\langle h_1 h_2 | O_i | B \rangle$, where $O_i$ are the  four-quark 
operators
listed in the preceding section. These will be calculated in the 
factorization assumption, which in the present context has been explained
in a number of papers (see, for example, ref.~\cite{ag}). To recapitulate 
briefly, the hadronic  
matrix elements involving four-quark operators are split into a product of 
two matrix elements of the
generic type $\langle h_1 | \bar{q}b | B \rangle$ and $\langle h_2|
\bar{q'}q' |0\rangle$, where Fierz transformation  is used so that
the flavor quantum numbers of the quark currents match those of the
hadrons. Since fierzing yields operators which are in the color 
singlet-singlet and 
octet-octet forms, this procedure results in general in matrix elements which
have the right flavor quantum numbers but involve both the 
singlet-singlet and octet-octet operators.
No direct experimental information is available on the latter. In the 
factorization approximation, 
one discards the color octet-octet piece and compensates this by introducing
a phenomenological parameter which determines the strength of the 
singlet-singlet contribution, renormalizing it from its perturbative
value. The hadronic matrix 
elements resulting from the factorization are  calculated in a model or 
determined from data, if available.

To set our notation and introduce some auxiliary quantities which we
shall need for numerical calculations, we illustrate the salient features
of our framework below.
When a pseudoscalar meson is a decay product, such as in the decay $B \to 
PP$, there are additional contributions from the $(V+A)$
penguin operator $O_6$ and $O_8$. After Fierz reordering and factorization
they contribute terms which involve a matrix element of the
quark-density operators between a pseudoscalar meson and the
vacuum. For $O_6$ involving $b\to s$ transition (in $b\to d$ transition 
$s$ is replaced by $d$), for example, this is given by
\begin{equation}
< P_1 P_2\vert O_6 \vert B > = - 2 \sum_q \Bigl(
< P_1 \vert \bar s R q \vert 0> < P_2 \vert\bar q L b \vert B > +
[ P_1 \leftrightarrow P_2]
\Bigr) \ . \end{equation}

Using the Dirac equation, the matrix elements entering here can be
rewritten in terms of those involving the usual $(V\!-\!A)$ currents,
\begin{equation}
   < P_1 P_2 \vert O_6 \vert B > =
   R[P_1,P_2] < P_1 P_2 \vert O_4 \vert B > +
[ P_1 \leftrightarrow P_2] \ ,
\label{O6_vs_O4} \end{equation}
with
\begin{equation}
R[P_1,P_2] \equiv
\frac{2M^2_{P_1}}{(m_{q}+m_{ s})(m_b - m_{q})} \ .
\label{def_of_R} \end{equation}
Here, $m_{s}$  and $m_{q}$ are the current masses of the
quarks in the mesons $P_1$ and $P_2$.
The same relations work for $O_8$. Finally, one arrives at the form
\begin{eqnarray} < P_1\,P_2\vert {\cal H}_{eff}\vert B>
& = & Z_1 < P_1 \vert j^{\mu}\vert 0 >
          < P_2 \vert j_{\mu} \vert B> \nonumber \\
& + & Z_2 < P_2 \vert j'^{\mu}\vert 0 >
          < P_1 \vert j'_{\mu} \vert B> \ ,
\label{def_of_Z12} \end{eqnarray}
where $j_\mu$ and $j'_\mu$ are the corresponding (neutral or charged)
$V\!-\!A$ currents. The quantities $Z_{1}$ and $Z_{2}$
involve the effective coefficients, CKM factors and $G_F$.
The $0^- \to 0^-$ form factors are defined as follows:
\begin{equation}
\label{5}
\langle P_1(p_1) |\bar q  \gamma_\mu L b|  B(p_B)\rangle =\left 
[(p_B +p_1)_\mu - \frac{m_B^2-m_1^2}{q^2}q_\mu \right ] 
F_1 (q^2) 
+\frac{m_B^2-m_1^2}{q^2}q_\mu F_0 (q^2) ,
\end{equation}
where $q=p_B - p_1$. In order to cancel the poles at $q^2 =0$,
we must impose the condition
$$F_1(0)=F_0(0).$$
The pseudoscalar decay constants are defined as:
\begin{equation}
  \label{6}
  \langle P (p)|\bar q \gamma ^\mu L q'|0 \rangle = i f_P~ p^\mu.
\end{equation}
With this, we can write the required matrix element in its factorized form
\begin{equation}
  \label{7}
\langle P_1 P_2 | {\cal H}_{eff} |  B \rangle =  i
\frac{G_F}{\sqrt{2}} V_{qb}V_{q q'}^* \left( \frac{1}{N_c} C_i +C_j \right)
f_{P_2} (m_B^2-m_1^2) F_0^{B\to P_1} (m_2^2) +(1\leftrightarrow 2).
\end{equation}
The dynamical details are coded in the  quantities $a_i$, 
which we define as
\begin{equation}
a_i \equiv C_i^{eff} + \frac{1}{N_c} C_{i+1}^{eff} ~~(i=\mbox{odd});
~~a_i \equiv C_i^{eff} + \frac{1}{N_c} C_{i-1}^{eff} ~~(i=\mbox{even})
\label{def_of_ai} ,
\end{equation}
where $i$ runs from $i=1,...,10$. 
 Thus, we see that there are ten such quantities.
They depend on the SM-input parameters, including the CKM matrix elements. 
The non-factorizing contributions in the matrix elements
$\langle h_1 h_2 | O_i | B \rangle$ are modeled by 
treating $N_c$ as a phenomenological parameter.
 Note, that this is the only place where $N_c$
is treated as a phenomenological parameter. In particular, in the calculation
of $C_i^{eff}$, we have used the QCD value $N_c=3$. Insisting that there
are no non-factorization effects present amounts to setting $N_c=3$ in
calculating $a_i$. This is also referred to as ``naive factorization' and is
known not to work in decays such as $B \to (D,D^*)(\pi,\rho), J/\psi K^{(*)}$
\cite{NS97,BHP96}. In these decays only the coefficients $a_1$ and $a_2$
are determined. Note that QCD does not demand the equality of $a_1$ and 
$a_2$ from these decays and from the ones $B \to h_1 h_2$, though their
values may come out to be close to each other.
 Hence, all the ten
quantities $a_i$ should be treated as phenomenological parameters and fitted
from data on $B \to h_1 h_2$ decays. 

 Returning to the discussion of the hadronic matrix elements, we 
recall that when a vector meson is involved in a decay, such as in $B 
\to PV$ and $B \to VV$ decays, we need also the $B\to V$ form factors, which 
are defined as follows:
  \begin{eqnarray}
\langle V (p_V) | V_\mu-A_\mu  | \bar B^{0}(p_B)  \rangle &=&  
-\epsilon _{\mu\nu\alpha\beta}
\epsilon^{\nu*} p_B^\alpha p_V^\beta \frac{2 V(q^2)}{(m_B+m_V)}
\nonumber\\
&-&i\left( \epsilon_\mu^* -\frac{\epsilon^*\cdot q}{q^2}q_\mu\right) 
(m_B+ m_{V}) A_1 (q^2) \nonumber \\
&+&i\left( (p_B+p_V)_\mu -\frac{(m_B^2 -m_{V}^2)}{q^2}q_\mu\right)
(\epsilon^* \cdot q) \frac {A_2 (q^2) }{m_B+ m_{V}}\nonumber \\
&-&i\frac{2m_V (\epsilon^* \cdot q)}{ q^2} q_\mu A_0 (q^2) ,  
\label{s3}
\end{eqnarray}
where $q=p_B - p_V, $ and $\epsilon^*$  is the polarization vector 
of  $V$. 
To cancel the poles at $q^2=0$, we must have 
  \begin{equation}
  \label{s4}2m_V  A_0 (0)=
(m_B+ m_{V}) A_1 (0) 
-(m_B-m_{V}) A_2 (0) .
\end{equation}

The decay constants of the  vector mesons are defined as follows: 
\begin{equation}
  \label{9}
<V| \bar q \gamma_\mu q | 0 >=   f_{V}~ m_{V} \epsilon _\mu.
\end{equation}

  This completes the discussion of the factorization Ansatz. The various
input parameters needed to do numerical calculations, including the
form factors and meson decay constants, are discussed in the next section.

\section{Input parameters}

 The matrix elements for the decay $B \to h_1 h_2$ derived 
in the preceding section depend on the
effective coefficients $a_1,...,a_{10}$, quark masses, 
various form factors, decay constants, the CKM parameters,
the renormalization scale $\mu$ and
the QCD scale parameter $\Lambda_{\overline{\rm MS}}$.
We have
fixed $\Lambda_{\overline{\rm MS}}$ using 
the central value of the present world average
 $\alpha_s (M_Z)=0.118 \pm 0.003$ \cite{Schmelling96}. The 
scale $\mu$ is varied
between $\mu =m_b$ and $\mu =m_b/2$, but due to the inclusion of the NLL
expressions the dependence of the decay rates on $\mu$ is small and hence 
not pursued any further. 
To be specific, we use $\mu=2.5$ GeV in the following.
The dependence on the rest of the parameters is more pronounced and we
discuss them below giving the present status of these quantities.

\subsection{ CKM Matrix Elements}

The CKM matrix will be
expressed in terms of the Wolfenstein parameters \cite{Wolfenstein83}, 
$A$, $\lambda$, $\rho$ and  $\eta$. 
\begin{equation}
V_{CKM}\simeq \left( \begin{array}{ccc}
1-\frac{1}{2}\lambda^2   & \lambda  &  A \lambda^3 (\rho-i\eta) \\
    -  \lambda   &   1-\frac{1}{2}\lambda^2  & A \lambda^2 \\
 A \lambda^3 (1-\rho-i\eta)  & - A \lambda^2   &    1\\
\end{array}
\right)
\end{equation}
Since the first two are 
well-determined
with $A= 0.81 \pm 0.06, ~\lambda=\sin \theta_C=0.2205 \pm 0.0018$
\cite{PDG96}, we fix
them to their central values. The other two are correlated and are found 
to lie (at 95\% C.L.) in the range $0.25 \leq \eta \leq 0.52$ and
$-0.2 \leq \rho \leq 0.35$ from the CKM unitarity fits
\cite{aliapctp97}. We shall show the dependence of the decay rates on the
parameters $\rho$ and $\eta$  in the allowed domain.
However, for illustrative purposes and if not stated otherwise, we shall use 
$\rho = 0.12, \eta = 0.34$, which are the ``best-fit" values
from the CKM unitarity fits \cite{aliapctp97}\footnote{The corresponding 
``best-fit" values obtained in \cite{Parodi98} $\rho\simeq 0.15$ and 
$\eta \simeq 0.34$ are very close to the ones being used here.}.

\subsection{Quark masses}
The quark masses enter our analysis in two different ways. First, they
arise from the contributions of the penguin loops in connection with the
function $\Delta F_1(k^2/m_i^2)$. We 
treat the internal
quark masses in these loops as constituent masses rather than current
masses. For them we use the following (renormalization scale-independent)
values:
\begin{equation}
\label{constmass}
m_b=4.88~\mbox{GeV}, ~~~m_{c}=1.5 ~\mbox{GeV}, ~~~m_s=0.5 ~\mbox{GeV}, 
~~~m_u=m_d=0.2 ~\mbox{GeV}~.
\end{equation}  
Variation in a reasonable range of these 
parameters does not change the numerical results of the branching ratios in
question, as also investigated  in \cite{ag}. The value of $m_b$ is fixed 
to be the current quark mass value
$\overline{m_b}(\mu=2.5~\mbox{GeV})= 4.88$ GeV, given below. Second, the 
quark masses  $m_b$, $m_s$, $m_d$ and $m_u$ appear through the equations of
motion when working out the (factorized) hadronic matrix elements.
In this case, the quark masses should be interpreted as current masses.
It is worthwhile to discuss the spread in the quark masses, as determined
from various calculational techniques and experiment. The top quark mass is 
now known rather precisely $\overline{m_t}(m_t)= 168 \pm 6 $ GeV.
Typical uncertainty on the $b$-quark 
mass $\delta (\overline{m_b}(\mu=2.5~\mbox{GeV}))= \pm 0.2$ GeV 
\cite{Gremmetal,Neubert97-24} is also small. Likewise,
the mass difference $m_b-m_c= (3.39 \pm 0.06)$ GeV \cite{Neubert97-24} is
well determined, which can be used to determine $m_c$ reasonably
accurately for the calculations being done here. Hence, to 
the accuracy of the present framework, 
the uncertainties in the decay rates related to $\delta m_t$, $\delta m_b$ 
and $\delta m_c$ are small and ignored.

 Light quark mass ratios have been
investigated in chiral perturbation theory
\cite{GaL85} and updated in \cite{Leutwyler96}, yielding:
 $m_u/m_d=0.553 \pm 0.043, ~m_s/m_d=18.9\pm 0.9, ~m_s/m_u=34.4 \pm 
3.7$. These ratios were converted into the quark masses by using the QCD sum 
rule
estimates of the $s$-quark mass of the  somewhat older vintage \cite{BPR95}: 
$\overline{m_s}(1 
~\mbox{GeV}) = 175 \pm 25 $ MeV, yielding $\overline{m_u}(1 ~\mbox{GeV}) 
= 5.1 \pm 0.9$ MeV, $\overline{m_d}(1 ~\mbox{GeV}) = 9.3 \pm 1.4$ MeV
\cite{Leutwyler96}.
Improved estimates based on QCD sum rules have been reported during the
last year, which include $O(\alpha_s^3)$-perturbative improvements
\cite{Silcher97}, improved estimates of 
$\Lambda_{\overline{{\rm MS}}}^{(3)}$ yielding
 $\Lambda_{\overline{{\rm MS}}}^{(3)} \simeq 
380$ MeV, and improvements in the estimates of the spectral functions
\cite{Colangelo97,Jamin97} lowering the $s$-quark mass. A contemporary 
representative values of the $s-$quark mass in the QCD sum rule approach is:
$\overline{m_s}(1 
~\mbox{GeV}) = 150 \pm 30 $ MeV \cite{Jamin97}.

\begin{table}
\label{table1}
\begin{center}
\caption{Input values in numerical calculations.}
\begin{tabular} {|c|c|}
\hline
names & Values\\
\hline
$\alpha_s (m_Z)$ & 0.118\\
$\mu$ & 2.5 GeV\\
A& 0.81\\
$\lambda $ & 0.2205\\
$\tau (B^+) $ & 1.62 ps\\
$\tau (B^0) $ & 1.56 ps\\
$m_t(m_t) $ & 168 GeV\\
$m_b(2.5GeV) $ & 4.88 GeV\\
$m_c(2.5GeV) $ & 1.5 GeV\\
$m_s(2.5GeV) $ & 122 MeV\\
$m_d(2.5GeV) $ & 7.6 MeV\\
$m_u(2.5GeV) $ & 4.2 MeV\\
\hline
\end{tabular}\end{center}\label{input}
\end{table}

 The corresponding estimates in the quenched 
lattice-QCD approach have been recently reported in a number of papers
\citer{CP-PACS97,Rajangupta98}. The lattice community likes to
quote the light quark masses at the scale $\mu=2 $ GeV, and in comparing 
them with the
QCD-sum rule results, quoted above for 1 GeV, one should multiply the 
lattice numbers by a factor $1.3$. Representative lattice-QCD values are:
$\overline{m_s}(2 ~\mbox{GeV}) = 100 \pm 12 $ MeV \cite{CP-PACS97}, 
 $\overline{m_s}(2 ~\mbox{GeV}) = 130 \pm 2 \pm 18$ MeV \cite{GGRT98}, and
$\overline{m_s}(2 ~\mbox{GeV}) = 110 \pm 20 \pm 11 $ MeV \cite{Rajangupta98}.
The error due to unquenching is largely unknown and for a discussion of the
given lattice-specific errors, we refer to the original literature. 
Taking the last of these values as fairly representative, one now has the
central value $\overline{m_s}(1 ~\mbox{GeV}) \simeq 140 $ MeV with a typical
error of $\pm 25$ MeV -- in reasonably good agreement with the QCD sum rule
estimates.
Using $\overline{m_b}(\mu=m_b)= 4.45$ GeV from the
central value in \cite{Gremmetal} and  
 \begin{equation}
\label{msbarmassleut}
\overline{m_s}(1 ~\mbox{GeV}) = 150 \ MeV \quad , \quad
\overline{m_d}(1 ~\mbox{GeV}) =9.3 \ MeV \quad , \quad
\overline{m_u}(1 ~\mbox{GeV}) = 5.1 \ MeV \quad ,
\end{equation}   
from the discussion above, the corresponding values at the
scale $\mu=2.5 $ GeV used in our calculations are given in 
Table~1.

 Varying the light quark masses by $\pm 20\%$ 
yields variation of up to $\pm 25\%$ in some selected decay rates (such as in
$B^\pm \to \eta^\prime K^\pm$ and $B^0 \to \eta^\prime K^0$, as also noted in 
\cite{acgk}). While this dependence should be kept in mind in fitting the 
quantities $a_i$ from precise data, this is clearly not warranted by present
data. Also, fitting the values of the quantities $a_i$ 
is not the aim of this paper. Hence, we shall fix all the current quark
masses to their values in Table~1.

\subsection{Form factors and hadronic coupling constants}
\begin{table}\begin{center}
\label{table2}
\caption{Form factors at zero momentum transfer in the BSW model
\cite{BSW87}. } \begin{tabular} {|c|ccccc|}
\hline
Decay & $F_1 =F_0$ & V & $A_1$ & $A_2$ & $A_0$\\
\hline
$B \to \pi$ & 0.33 &      &       & &\\
$B \to K$   & 0.38 &      &       & &\\
$B \to \eta$ & 0.145 &      &       & &\\
$B \to \eta^\prime$ & 0.135 &      &       & &\\
$B \to \rho$ &      & 0.33 & 0.28 &  0.28& 0.28\\
$B \to K^*$ &      & 0.37 & 0.33 &  0.33 & 0.32\\
$B \to \omega$  &      & 0.33 & 0.28 &  0.28& 0.28\\
\hline
\end{tabular}\end{center}
\end{table}
\begin{table}
\begin{center}
\label{table3}
\caption{Values of pole masses in GeV.}
\begin{tabular} {|c|c|c|c|c|}
\hline
Current & m($0^-$) & m($1^-$) &  m($1^+$)&  m($0^+$) \\
\hline
$\bar u b$ & 5.2789 & 5.3248 &5.37& 5.73\\
$\bar d b$ & 5.2792 & 5.3248 &5.37& 5.73\\
$\bar s b$ & 5.3693 & 5.41 &5.82 & 5.89\\ 
\hline
\end{tabular}\end{center}
\end{table}

 Finally, we discuss the numerical values of the form factors and coupling 
constants introduced 
in the previous section. Concerning the form factors, we shall use two
different theoretical approaches. The first is 
based on the quark model due to Bauer, Stech and Wirbel \cite{BSW87},
which has been found to be rather successful in accommodating data on
a number of exclusive decays.
In the BSW model, the meson-meson matrix elements of the currents are
evaluated from the overlap integrals of the corresponding wave functions.
The dependence of the form factors on the momentum transfer squared $Q^2$ 
(which in $B \to h_1 h_2$ decays equals  the 
mass squared of the light meson) is modeled by a single-pole Ansatz.
 The values of the
form factors in the transitions $B \to \pi$, $B \to K$, $B \to \eta$, 
$B \to \eta '$, $B \to \rho$, $B \to K^*$ and $B \to \omega$, evaluated
at $Q^2=0$ are given in Table~2. 
We assume ideal mixing for the ($\omega$, $\phi$) complex. This amounts 
to using in the quark language  $\phi =s\bar s$ and $\omega = 
\frac{1}{\sqrt{2}} (u\bar u+ d \bar d)$.
 Note, that to implement the
$\eta$-$\eta^\prime$ mixing, we shall use the two-mixing-angle formalism
proposed recently in \cite{Leutwyler97,HSLT97}, in which one has:
\begin{equation}
\label{etaetapmix}
|\eta \rangle = \cos \theta_8 |\eta_8\rangle - \sin \theta_0 |\eta_0 
\rangle ~,~~~
|\eta^\prime \rangle = \sin \theta_8 |\eta_8\rangle + \cos \theta_0 |\eta_0
\rangle ~.
\end{equation}
Here, $\eta_8$ and $\eta_0$ 
are, respectively, the flavor $SU(3)$-octet and -singlet components.
The relations for the pseudoscalar decay constants in this mixing
formalism involving the axial-vector currents $A_\mu^{8}$ and $A_\mu^{0}$
are:
\begin{eqnarray}
\label{etaetapconst}
\langle 0|A_{\mu}^{8} | \eta(p) \rangle = i f_{\eta}^{8} p_\mu,
~~~\langle 0|A_{\mu}^{8} | \eta^\prime (p) \rangle = i f_{\eta^\prime}^{8} 
p_\mu, \nonumber\\
\langle 0|A_{\mu}^{0} | \eta(p) \rangle = i f_{\eta}^{0} p_\mu, 
~~~\langle 0|A_{\mu}^{0} | \eta^\prime (p) \rangle = i f_{\eta^\prime}^{0}
p_\mu ~.
\end{eqnarray}
The best-fit values of the $(\eta$-$\eta^\prime)$ mixing parameters from
\cite{FK97} yields: $\theta_8=-22.2^\circ$, $\theta_0=-9.1^\circ$, $f_8=168$
MeV, and $f_0=157$ MeV, which are used to calculate the decay rates in
which $\eta$ and/or $\eta^\prime$ are involved.
In deriving the expressions for the decays involving $\eta$ and 
$\eta^\prime$, we include the anomaly term in $\partial_\mu  A^\mu$ and
the contributions of $b \to s gg \to s (\eta,\eta^\prime$) as calculated
 in \cite{acgk}.  
Definitions of the various matrix elements can be seen in the 
appendix and we refer to \cite{ag,acgk} for further discussions. The
values of the input pole masses used in calculating the form factors are
given in Table~3. However,  in the decays $B \to h_1 h_2$, only
small extrapolations from $Q^2=0$ are involved, hence the error due to the
assumed $Q^2$-dependence and/or the specific values for the pole masses is 
small.   

\begin{table}\begin{center}
\label{table4}
\caption{Form factors at zero momentum transfer from Lattice QCD and
Light-cone QCD sum rules. }
\begin{tabular} {|c|ccccc|}
\hline
Decay & $F_1 =F_0$ & V & $A_1$ & $A_2$ & $A_0$\\
\hline
$B \to \pi$ \protect\cite{Ball98} & $0.30 \pm 0.04$ &      &       & &\\
$B \to K \protect\cite{Ball98} $   & $0.35 \pm 0.05$ &      &       & &\\
$B \to \eta$ \mbox{(see text)} & $ 0.13 \pm 0.02$ &      &       & &\\
$B \to \eta^\prime $ \mbox{(see text)} & $0.12 \pm 0.02 $ &      &       
& &\\
$B \to \rho$ \protect\cite{UKQCD97} & & $0.35\pm 0.05$ & 
$0.27\pm 0.04$ & $0.26\pm 0.04$ & $0.30\pm 0.05$\\
$B \to K^*$ \protect\cite{ABS94} & & $0.48\pm 0.09$ & $0.35\pm 
0.07$ &  $0.34\pm 0.06$ & $0.39 \pm 0.10$\\
$B \to \omega$ (\protect\cite{UKQCD97} \& SU(3))  & & $0.35\pm 0.05$ & 
$0.27\pm 0.04$ & $0.26 \pm 0.04$ & $0.30\pm 0.05$\\ \hline
\end{tabular}\end{center}
\end{table}

The second and more modern approach to calculating decay form factors is a 
hybrid approach, in which often lattice-QCD 
estimates in the so-called $heavy \to light$ mesons, calculated at 
high-$Q^2$, are combined with the $Q^2$-dependence following from the 
light-cone QCD sum rule analysis \cite{ABS94,BB97}.
We refer to \cite{FS97} for detailed discussions, compilation of the
lattice-QCD analysis and references to the literature, and quote here the 
results from the UKQCD analysis \cite{UKQCD97}. For the  $B \to 
\pi$ form factor: $F_1(0)=F_0(0)=0.27 \pm 0.11$; for $B \to \rho$ form 
factors: 
$V(0)=0.35^{+0.06}_{-0.05}$, $A_1(0)= 0.27 ^{+0.05}_{-0.04}$, $A_2(0)=
0.26^{+0.05}_{-0.03}$, and $A_0(0)=0.30^{+0.06}_{-0.04}$.
The results from an improved light-cone QCD sum rule calculation 
\cite{Ball98} for $F_1(B \to \pi)=F_0(B \to \pi)$ and
$F_1(B \to K) = F_0(B \to K)$
are given in Table 4. The results for $F_1(B \to \eta)=F_0(B \to \eta)$ and
$F_1(B \to \eta^\prime)=F_0(B \to \eta^\prime)$ are calculated from the
$B \to \pi$ form factors from \cite{Ball98} taking into account 
additionally the $(\eta,\eta^\prime)$ mixing, as discussed 
earlier and further detailed in Appendix A.
 The results for
the $B \to K^*$ form factors have been obtained in the light-cone QCD sum 
rule in ref.~\cite{ABS94}, which yield:  
\begin{eqnarray}
\label{lcqcdsr}
\frac{A_1(0)^{B \to \rho}}{A_1(0)^{B \to K^*}} &=& 0.76 \pm 0.05~, \\ 
\nonumber
\frac{V(0)^{B \to \rho}}{V(0)^{B \to K^*}} &=& 0.73 \pm 0.05~,
\end{eqnarray}
which, in turn,  lead to the estimates $A_1(0)^{B \to K^*}=0.35\pm 0.07$ and
$V(0)^{B \to K^*}=0.48 \pm 0.09$. Assuming similar $SU(3)$-breaking in the
remaining  two form factors, and using the estimates for the 
corresponding
form factors in $B \to \rho$ quoted above, one gets: $A_2(0)^{B \to 
K^*}=0.34\pm 
0.06$ and $A_0(0)^{B \to K^*}=0.39\pm 0.10$. The values from this
hybrid approach are collected in Table~4. As for the form factors in the BSW 
model, we use a simple pole approximation for calculating the form
factors at $Q^2$ different from $Q^2=0$. However, for the decays of interest,
this extrapolation is small and one does not expect any 
significant error from this source.
For example, for the $B \to P$ form factors, using the parameterization of 
$F_{0,1}(Q^2)$  given
in eq.~(12) of ref.~\cite{Ball98}, the resulting difference
in the form factors is found to be less than 2\%.

\begin{table}
\begin{center}
\label{table5}
\caption{Values of decay constants in MeV. }
\begin{tabular} {|c|c|c|c|c|c||c|c|c|c|}
\hline
$f_\pi$ & $f_ K$ & $f_8$ & $f _0$  &  $f_{\eta }^c$  &  $f_{\eta '}^c$ &
 $f_\rho$   & $f_{K^*}$ & $f_ \omega$  & $f_\phi$  \\
\hline
   133  & 158    &  168  &  157   &      $-0.9$  &    $-2.3$  &
 210  &   214    &  195  & 233\\
\hline
\end{tabular}\end{center}
\end{table}

 The values for the pseudoscalar and vector decay constants are given
in Table~5.
 The values for $f_\omega, f_K, f_{K^\ast}$ and $ f_\pi$ coincide with the
central values quoted in \cite{NS97} extracted from data on the
electromagnetic decays of $\omega$ and
$\tau$ decays, respectively \cite{PDG96}.
The decay constants $f_{\eta'}^u$,  $f_{\eta'}^s$,
$f_{\eta}^u$ and $f_{\eta}^s$ defined in the appendix A
are obtained from the
values for $f_0$ and $f_8$, 
$\theta_8$ and $\theta_0$ for the $(\eta,\eta')$ mixing,
given earlier. 
The errors on the decay constants in Table~5 are small
(typically $(1 - 3)\%$), except on $f_{\eta'}^{(c)}$ and $f_{\eta}^{(c)}$
for which we use here the estimates from \cite{acgk} 
obtained using the QCD-anomaly method. These quantities have also been
determined from the $\eta_c-\eta^\prime-\eta$-mixing formalism and radiative
decays $J/\psi \to (\eta_c,\eta^\prime,\eta) \gamma$ and the two-photon
decay widths $(\eta_c,\eta^\prime,\eta) \to \gamma \gamma$ in ref.~\cite{ag}
with results similar to the
corresponding values obtained using the QCD-anomaly method \cite{acgk}.
For some recent determinations of these quantities, see also 
\cite{FKS98,AMT98}.

\section{Effective coefficients $a_i$ and a classification of $B \to h_1
h_2$ decays}
\subsection{Effective coefficients $a_i$}

  The effective coefficients $a_i$, which are specific to the factorization
approach, are the quantities of principal phenomenological interest.
Note that there are four types
of transitions that one encounters in the current-current and
penguin-induced decays $B \to h_1 h_2$: $b \to s$ $[\bar{b} \to \bar{s}]$,
and $b \to d$ $[\bar{b} \to \bar{d}]$. Numerical values
of $a_i$ $(i=1,...,10)$ for representative
values of the phenomenological parameter $N_c$ are displayed in
Table~6 and Table~7 for the $b \to s$ $[\bar{b} \to \bar{s}]$ and
 $b \to d$ $[\bar{b} \to \bar{d}]$ cases, respectively. A number of remarks
on the entries in these tables is helpful for a discussion of the  
branching ratios worked out later.

\begin{itemize}
\item The determination of $a_1$ and $a_2$ in the $b \to c$ current-current
transitions has received a lot of attention. It remains an open and
interesting question if $a_1$ and $a_2$ in the $b \to u$
transitions are close to their $b \to c$ counterparts, which have the
phenomenological values $a_1 \simeq 1$ and $a_2 \simeq 0.2$
\cite{NS97,BHP96}. These values correspond to the parameter $\xi \equiv 
1/N_c$ having a value around $0.4$. The decays $B \to \pi \pi$, $B \to 
\rho \pi$ and $B \to \omega \pi$ are well suited to determine these 
coefficients.
\item The coefficients $a_3$ and $a_5$ in the QCD-penguin sector
are smaller compared to $a_4$ and $a_6$. In particular, the combination
$a_3+a_5$ has a perturbative
value of $3 \times 10^{-4}$, i.e., for $N_c=3$, in all four cases resulting 
from large cancellations between $a_3$ and $a_5$. This
coefficient also shows extreme sensitivity to the parameter $N_c$, which
in the present model is a measure of non-factorizing effects. Hence,
for decays whose decay widths depend dominantly on these
coefficients, the factorization framework is not reliable.
The reason is simply that the neglected contributions, such as the
weak annihilation diagrams and/or feed down from final state interactions
to these channels, could easily overwhelm the perturbative factorizable 
contributions.

\item Concerning the effective coefficients of the electroweak operators,
we note that $a_7$, $a_8$
and $a_{10}$ are numerically very small. This again reflects their
perturbative magnitudes, i.e. the coefficients $C_i^{eff}$, as can be
seen in the columns for $N_c=3$. Varying $N_c$, one sees no noticeable
enhancement in these coefficients (except for $a_{10}$ but it remains  
phenomenologically small to have any measurable effect).
 Hence, electroweak penguins
enter dominantly through the operator $O_9$,
barring rather drastic enhancements (of $O(100)$)
in the matrix elements of the operators $O_7,O_8$ and $O_{10}$, which
we discount. No 
attempts will be made to determine these coefficients here. In fact,
in the context of the SM 
one could as well work with a much reduced basis in the effective theory
in which the coefficients $a_7$, $a_8$ and $a_{10}$ are set to zero.

\item The dominant coefficients are then $a_1$, $a_2$ (current-current
amplitudes), $a_4$, $a_6$ (QCD penguins) and
$a_9$ (electroweak penguin), which can be eventually determined from 
experiments and we discuss 
this programmatically later. Of these $a_1$, $a_2$ (and to a very high  
accuracy also $a_9$) do not depend on the CKM matrix elements. The 
dependence of $a_4$ and $a_6$ (likewise, the smaller parameters $a_3$ and
$a_5$) on the CKM factors enters through the
function $C_t$.
The numbers given in the tables for $a_i$ are obtained
for the CKM parameters having the values $\rho=0.12$ and $\eta=0.34$.
Note that $a_2$ depends strongly on $N_c$.
\end{itemize}

  This sets the stage for discussing the various branching ratios numerically
and comparison with the available data.

\begin{table}
\begin{center}
\label{table6}
\caption{Numerical values of effective coefficients $a_i$ for
 $b\to s$ [ $\bar b\to \bar s$] at $N_c=2,3,\infty$, where $N_c=\infty$
corresponds to $C_i^{eff}$. The penguin coefficients $C_3^{eff}
,...,C_7^{eff}$ and $C_9^{eff}$
are calculated for the Wolfenstein parameters $\rho=0.12$ and $\eta=0.34$.
 Note that the entries for $a_3$,...,$a_{10}$
have to be multiplied with $10^{-4}$.}
\begin{tabular}{|c|c|c|c|}
\hline
        & $N_c=2 $& $N_c=3 $   & $N_c=\infty $ \\
\hline
$a_1$  &0.99 [0.99]& 1.05 [1.05] &  1.16 [1.16] \\
$a_2$  &0.25 [0.25]&  0.053 [0.053] & $-0.33$ [$-0.33$] \\
$a_3$  &$-37-14i$ [$-36-14i$]  & 48 [48] & $218+  29i$ [$215+  29i$] \\
$a_4$  &$-402-72i$ [$-395-72i$]& $-439-77i$ [$-431-77i$]  & $-511-87i$
[$-503-87i$]\\
$a_5$  &$- 150-14i$ [$-149-14i$]& $-45$ [$-45$] & $165+ 29i$ [$162+
29i$]\\
$a_6$  &$- 547-72i$ [$-541-72i$]& $-575-77i$ [$-568-77i$] & $-630-87i$
[$-622-87i$]\\
$a_7$  &$ 1.3-1.3i$ [$1.4-1.3i$]& $0.5-1.3i$ [$0.5-1.3i$] & $-1.2-1.3i$
[$- 1.1-  1.3i$]\\
$a_8$  &$4.4 -0.7i$ [$4.4-0.7i$]& $4.6-0.4i$ [$4.6-0.4i$] & 5.0  [5.0] \\
$a_9$  &$-91- 1.3i$ [$-91-1.3i$]& $-94-1.3i$ [$-94-1.3i$] & $- 101-1.3i$
[$- 101- 1.3i$] \\
$a_{10}$&$-31-0.7i$ [$-31-0.7i$]& $-14-0.4i$ [$-14-0.4i$] &  20  [20] \\
\hline
\end{tabular}
\end{center}
\end{table} 
\begin{table}
\begin{center}
\label{table7}
\caption{Numerical values of effective coefficients $a_i$ for
 $b\to d$ [ $\bar b\to \bar d$] at $N_c=2,3,\infty$, where $N_c=\infty$
corresponds to $C_i^{eff}$. The penguin coefficients $C_3^{eff},...,
C_7^{eff}$ and $C_9^{eff}$
are calculated for the Wolfenstein parameters $\rho=0.12$ and $\eta=0.34$.
 Note that the entries for $a_3$,...,$a_{10}$ 
have to be multiplied with $10^{-4}$.}
\begin{tabular}{|c|c|c|c|}
\hline
        & $N_c=2 $& $N_c=3 $   & $N_c=\infty $ \\
\hline
$a_1$  &0.99 [0.99]& 1.05 [1.05] &  1.16 [1.16] \\
$a_2$  &0.25 [0.25]& 0.053 [0.053] & $-0.33$ [$-0.33$] \\
$a_3$  &$-  33-7i$ [$-  42-23i$]&         48 [48]    & $208+14i$
[$226+47i$] \\
$a_4$  &$-377-34i$ [$-423-116i$]& $-412-36i$ [$-461-124i$] & $-481-41i$
[$- 536- 140i$]\\
$a_5$  &$-145-14i$ [$-154-14i$] & $-45$ [$-45$] &  $155+  14i$ [$
173+47i$]\\
$a_6$  &$-523-34i$ [$-568-116i$]& $-548-36i$ [$-597-124i$] & $-600-41i$
[$-655- 140i$]\\
$a_7$  &$ 1.5-1.0i$ [$1.1-1.8i$]& $0.7-1.0i$ [$ 0.3-1.8i$] & $-1.0-1.0i$
[$- 1.4-  1.8i$]\\
$a_8$  &$4.5 -0.5i$ [$4.3-0.9i$]& $4.7-0.3i$ [$4.5-0.6i$]  & 5.0  [5.0] \\
$a_9$  &$-91- 1.0i$ [$-91-1.8i$]& $-94-1.0i$ [$-95-1.8i$]  & $- 101-1.0i$
[$- 101- 1.8i$] \\
$a_{10}$&$-30-0.5i$ [$-31-0.9i$]& $-14-0.3i$ [$-14-0.6i$]  &  20  [20]  
\\
\hline
\end{tabular}
\end{center}
\end{table} 
 Before discussing the numerical results and their detailed comparison 
with experiment and existing results in the literature, it is worthwhile to
organize the decays $B \to h_1 h_2$ in terms of their sensitivity on $N_c$ 
and anticipated
contributions due to the annihilation diagrams in some of these decays.

\subsection{Classification of factorized amplitudes}
In the context of the tree $(T)$ decays, a classification 
was introduced in \cite{BSW87}, which is used widely in the
literature in the analysis of $B$ decays involving charmed hadrons. 
These classes, concentrating  
now on the  $B \to h_1 h_2$ decays, are the following
\begin{itemize}
\item Class-I decays, involving those decays in which only a charged meson
can be generated directly from a singlet current, as in ${B}^0 \to 
\pi^+ \pi^-$, and the relevant coefficient for these decays is $a_1$.
This coefficient is stable against variation of $N_c$ (see Tables 6 and 7).
There are just five class-I decays: $B^0 \to \pi^-\pi^+$,  $B^0 \to 
\rho^-\pi^+$, $B^0 \to \rho^+\pi^-$, $B^0 \to \rho^-\rho^+$, and
exceptionally also $B^0 \to \rho^- K^+$.
\item Class-II decays, involving those transitions in which the meson
generated directly from the current is a neutral meson, like ${B}^0 \to
\pi^0 \pi^0$, and the relevant coefficient for these decays is $a_2$, which
shows a strong  $N_c$-dependence (see Tables 6 and 7). There are twelve
such decays $B^0 \to h_1^0 h_2^0$, where $h_1^0$ and $h_2^0$ are mesons from 
the set 
$\pi^0,\eta, \eta^\prime, \rho^0$ and $\omega$. The decays $B^0 \to \pi^0 
\eta^{(\prime)}$ exceptionally do not belong to this class, as their
decay amplitudes proportional to $a_2$ almost cancel due to the 
destructive interference in two tree diagrams having to do with the
configuration $\pi^0 \sim u\bar{u} - d\bar{d}$ and $\eta^{(\prime)} \sim
(u\bar{u} + d\bar{d}) +...$. Note that as $a_2$ has the smallest value
at $N_c=3$, all class-II decays have their lowest values at $N_c=3$.
 \item Class-III decays, involving the interference of class-I and 
Class-II decays, as in this case both a charged and a neutral meson is 
present both of which can be generated through the currents involved in
$H_{eff}$. An example of these decays is $B^+ \to \pi^+ \pi^0$,
and the relevant coefficient is $a_1 + r a_2$, where $r$ is 
process-dependent (but calculable in terms of the ratios of the form factors
and decay constants). For $r \leq 1$, the $N_c$-dependence of the 
class-III amplitudes is below $\pm 20\%$ w.r.t. the perturbative value.
 As we shall see, the quantity $r$ may
considerably enhance the $N_c$-dependence if $r$ is well in excess of 1.
This, in particular, is the case in $B^+ \to \rho^0\pi^+$
and $B^+ \to \omega \pi^+$ decays, where $r\simeq 2$; hence these 
Class-III decays show marked $N_c$-dependence. However, one should note
that the decay rates for this class do not have their minima at $N_c=3$,
but rather at $N_c=\infty$, reflecting the behavior of $a_1 + a_2$. There 
are eleven such decays involving
$B^+ \to (\pi^+,\rho^+)(\pi^0,\eta,\eta^\prime, \rho^0,\omega)$ and
exceptionally also the decay $B^+ \to K^{*+} \eta^\prime$, in which case
the penguin amplitudes interfere destructively. Its decay rate is,
however, rather stable w.r.t. the variation in $N_c$ but small due to
the CKM suppression. 
\end{itemize}

However, when QCD $(P)$ and electroweak penguins $(P_{\small \mbox{EW}})$ 
are also present, as is the
case in the decays $B \to h_1 h_2$ being considered, in general,
the above classification has to be extended. In this case,
the generic decay amplitude  
 depends on $T + P + P_{\small \mbox{EW}}$. If the amplitude is 
 still dominated by the tree amplitude, the 
BSW-classification given  above can be applied as before. For those 
decays which are dominated by penguin amplitudes, i.e.,
$T + P + P_{\mbox{EW}} \simeq P + P_{\mbox{EW}}$, the 
above classification used for the tree amplitude is no longer applicable.
 
For the penguin-dominated decays, we introduce two new  
classes:
\begin{itemize}
\item Class-IV decays, consisting of decays whose amplitudes involve 
one (or more) of the dominant penguin coefficients
$a_4$, $a_6$ and $a_{9}$, with  constructive interference 
among them. They  are stable against
variation in $N_c$ (see tables 6 and 7) and have the generic form:
\begin{eqnarray}
{\cal M}( B^0 \to h_1^\pm h_2^\mp) &\simeq&\alpha_1 a_1  +
\sum_{i=4,6,9}\alpha_i a_i~ +..., \label{classiv}
\\ \nonumber
{\cal M}( B^0 \to h_1^0 h_2^0) &\simeq&\alpha_2 a_2 + \sum_{i=4,6,9}\alpha_i
a_i~ +..., \\ \nonumber
{\cal M}( B^\pm \to h_1^\pm h_2^0) &\simeq&\alpha_1 (a_1 + r a_2) + 
\sum_{i=4,6,9}\alpha_i a_i~ +..., 
\end{eqnarray}
with the second ($P + P_{\mbox{EW}})$ term
dominant in each of the three amplitudes. The ellipses indicate possible 
contributions from 
the coefficients $a_3,a_5,a_7,a_8$ and $a_{10}$ which can be neglected
for this class of decays.
The coefficients $\alpha_j$ are process-dependent and  contain the CKM matrix
elements, form factors etc.
 The decays where $\alpha_1$ and $\alpha_2$ are zero are
pure penguin processes and are obviously included here.
The tree-dominated decays, discussed earlier,
also have a generic amplitude of the type shown above. However, in this
case the penguin-related coefficients $\alpha_j$ are numerically small
due to the CKM factors (specifically due to $V_{td} \ll V_{ts}$).

\end{itemize}
 Examples of Class-IV decays are quite abundant. In our classification,
all twelve $B \to PP$ decays dominated by penguin amplitudes are class-IV
decays. They  include decays
such as $B^+ \to K^+ \pi^0$, $B^+ \to K^+ \eta^{(\prime)}$, which 
involve
$a_1 + r a_2$ as the tree amplitude, and ${B}^0 \to {K}^0 \pi^0$,
and ${B}^0 \to {K}^0 \eta^{(\prime)}$, which involve $a_2$ from
the tree amplitude. Finally, the pure-penguin decays, such as $B^+ \to 
\pi^+ K^0$, $B^+ \to K^+ \bar{K}^0$ and ${B}^0 \to K^0 \bar{K}^0$
naturally belong here. There are altogether twenty nine such decays. The
decay $B^0 \to K^{*0} \eta^\prime$, in contrast to its $B^+$-counterpart,
is not a class-IV decay due to the destructive interference in the 
QCD-penguin amplitude. The variation in the decay rates belonging to
class-IV decays is less than $\pm 30\%$ compared to their perturbative
$(N_c=3)$ value.  
\begin{itemize}
\item Class-V decays, involve penguins with strong $N_c$-dependent  
coefficients $a_3, a_5,a_7$ and $a_{10}$, interfering   
significantly with one of the dominant penguin coefficients $a_4,a_6$ and
$a_9$ (analogous to the class-III decays $a_1 + r a_2$ dominated by tree 
amplitudes). Then, there are decays in which the dominant penguin 
coefficients $(a_4,a_6,a_9)$ interfere destructively. 
Their amplitudes can be written much like the ones in eq.~(\ref{classiv}),
except that the sum in the second term now goes over all eight penguin 
coefficients. Since these amplitudes involve large and delicate 
cancellations, they are generally not stable against $N_c$.
\end{itemize}
Examples of this class are present in $B \to PV$ and $B \to VV$ decays,
such as $B^\pm \to \pi^\pm \phi$, ${B}^0 \to \pi^0 \phi$,
${B}^0 \to \eta^{(\prime)} \phi$,  
$ {B}^0 \to \omega \phi$, $B^\pm \to \rho^\pm \phi$, ${B}^0 \to
\rho^0 \phi$, etc. In all these cases, the amplitudes are proportional to
the linear combination $[a_3 + a_5 -1/2 (a_7 + a_9)]$ (see Appendix B and C).
Examples of those where the amplitudes proportional to the dominant penguin 
coefficients interfere destructively are: $B^+ \to K^+ \phi$, $B^0 \to 
K^0 \phi$ etc. The above five classes exhaust all 
cases, though clearly there are some amplitudes where comparable $T$ and
penguin ($P + P_{\mbox{EW}})$ contributions are present.
They can be assigned to one of the classes depending on their tree
and/or penguin coefficients, the criterion being the $N_c$-dependence of
the decay rates.

Summarizing the classification, Class-I and Class-IV decays are relatively 
large, unless suppressed by the CKM factors, and 
stable against variation of 
$N_c$, which is a measure of non-factorizing effects in the present
model. Class-III decays are  mostly stable, except for the already 
mentioned $B \to PV$ decays.
Many Class-II and Class-V decays are rather unstable against variation of 
$N_c$ either due the dependence on the $N_c$-sensitive coefficients or
due to delicate cancellations. Many decays in Class-II and Class-V may 
receive significant contribution from the annihilation diagrams which we
discuss now. 
\subsection{Contribution of annihilation amplitudes}

 Annihilation (by which are meant here both $W^\pm$-exchange and 
$W^\pm$-annihilation) contributions are present in almost all decays of
the type $B \to h_1 h_2$ being considered here. However, their contribution
should be understood as power corrections in inverse powers of $m_b$ 
(equivalently in $1/m_B$) in $B$ decays.
In inclusive $B$ decays, their 
 contribution to the decay width relative to that of the 
parton model is determined by the factor 
\begin{equation}
\label{annicontr}
4 \pi^2 \frac{f_B^2m_B}{m_b^3} \simeq \left(\frac{2 \pi f_B}{m_b}\right)^2
 \simeq 5 \%,
\end{equation} 
where $f_B \simeq 200$ MeV is the $B$-meson decay constant.
The near equality of the lifetimes of $B^\pm$, $\bar{B}^0(B^0)$ and 
$\bar{B}_s^0(B_s^0)$ mesons shows that the above crude estimate is largely
correct, and that annihilation contributions are sufficiently 
power-suppressed in $B$-meson decays. For more sophisticated but in their
spirit essentially similar calculations, see, for example, \cite{NS96}.

 However, in exclusive two-body $B$-decays, 
the contribution to a particular channel depends on
the CKM factors and the dynamical quantities $a_i$, and in some cases the
non-annihilation contribution is enormously suppressed. In these channels, 
the annihilation diagrams, despite being power suppressed in $1/m_b^2$,
may yield the dominant contributions to the decay and must therefore be 
included in the rate estimates and CP-asymmetries.
Instead of working out the annihilation contribution in all the channels
discussed here, which necessarily introduces unknown hadronic quantities,
we do a classification of annihilation diagrams and list only those 
decays in which they are anticipated to be important.
 
 For the decays $B \to h_1 h_2$, we 
need to consider the following annihilation amplitudes:
\begin{itemize}
\item $W^\pm$-Exchange:  ${\cal M}(\bar{b} d \to \bar{u} u)
\Longrightarrow {\cal M}(B^0 \to (\bar{u}q) (\bar{q}u)) \propto 
a_2 \lambda^3$ ,
\item $W^\pm$-Annihilation:  ${\cal M}(\bar{b} u \to \bar{d} u)
\Longrightarrow {\cal M}( B^+ \to (\bar{d}q) (\bar{q}u)) \propto 
a_1 \lambda^3$ ,
\item $W^\pm$-Annihilation:  ${\cal M}(\bar{b} u \to \bar{s} u)
\Longrightarrow {\cal M}( B^+ \to (\bar{s}q) (\bar{q}u) )\propto
a_1 \lambda^4$,
\end{itemize}
where $\lambda = \sin \theta_C$.
Here, $q\bar{q}$ is a light quark-antiquark pair. These amplitudes can be 
termed as 
the tree-annihilation contributions. In addition, there are also the 
penguin-annihilation contributions which are important for certain decays.
For example, they feed dominantly to the decay $B^0 \to \phi \phi$.
 
There are yet more decays which can be reached via 
annihilation followed by rearrangement of the quark-antiquark pairs in 
the final state. Representative of these are the decays $B^\pm \to \phi 
\pi^\pm$, $B^\pm \to \phi \rho^\pm$ and
$B^0(\bar{B}^0) \to \phi \pi^0$, $B^0(\bar{B}^0) \to \phi \eta^{(\prime)}$,
$B^0(\bar{B}^0) \to \phi \omega$ and $B^0(\bar{B}^0) \to \phi \rho^0$.
 However, these rescattering effects (final state 
interactions) are expected to suffer from suppression due to the 
color-transparency argument used in defense of the factorization 
Ansatz. Since we have neglected these rescattering contributions in the
factorization amplitudes worked out in this paper, it is only consistent
that we also drop the annihilation contributions which feed into other 
channels through rescattering.

 We specify below those two-body $B$ decays which are accessible directly in 
annihilation processes and hence may have significant annihilation
contributions: 
\begin{itemize}
\item $B \to PP$ decays:       
$B^0 \to \pi^0 \eta^{(\prime)}$, $B^0 \to \eta \eta^{\prime}$.
\item $B \to PV$ decays: $B^0 \to \rho^0 \pi^0$, $B^0 \to 
\rho^0 \eta^{(\prime)}$, $B^0 \to \omega \pi^0$, $B^0 \to
\omega \eta^{(\prime)}$, $B^+ \to K^{*+} \bar{K}^0$, $B^+ \to K^{+} \phi$,
$B^0 \to K^{*+} {K}^-$, $B^0 \to K^{+} K^{*-}$.
\item $B \to VV$ decays: $B^0 \to \rho^0 \rho^0$, 
$B^0 \to \rho^0 \omega$, $B^0 \to \omega \omega$,
$B^0 \to \phi \phi$, $B^+ \to K^{*+} \bar{K}^{*0}$, 
$B^+ \to K^{*+} \phi$, $B^0 \to K^{*+} K^{*-}$.
\end{itemize}
Note, that in addition to the decay modes listed above, there are
quite a few others in the Class-I, Class-III and
Class-IV decays given in the tables, which  also have annihilation
contributions but in view of the large $T$ and/or $P+ 
P_{\small\mbox{EW}}$ contributions in these decays, the
annihilation contributions are 
 not expected to alter the decay rates in these
channels significantly and hence we have not listed them.

The annihilation amplitude can be written as 
\begin{equation}
<h_1h_2| {\cal H}_{eff} | B>_a = Z <h_1h_2| j^\mu | 0> <0|j_\mu |B>.
\end{equation}
If $h_1$ and $h_2$ are two pseudoscalars,
 the annihilation form factors are defined as 
\begin{equation}
 <P_1P_2| j^\mu | 0> =\left[ (p_1-p_2)^\mu - \frac{m_1^2-m_2^2}{Q^2}
Q^\mu\right] F_1^{P_1P_2} (Q^2) + \frac{m_1^2-m_2^2}{Q^2}
Q^\mu  F_0^{P_1P_2} (Q^2) ,
\end{equation}
where $Q=p_1+p_2$.
With this, we can write the required matrix element 
from the annihilation contribution (denoted here by a subscript) in its 
factorized form \begin{equation}\label{ann1}
<P_1P_2| {\cal H}_{eff} | B>_a = i\frac{G_F}{\sqrt{2}}V_{qb}V_{qq'}^* a_i
f_B (m_1^2-m_2^2) F_0^{P_1P_2} (m_B^2)~,
\end{equation}
where $a_i$, $i=1,2$.
Note that the annihilation amplitude in the decay $B \to P_1 P_2$ is 
proportional to the mass difference of the two mesons
in the final state. Hence, in the present
framework, there is no 
annihilation contribution to the decays such as ${B}^0 \to \pi^0 \pi^0$,
${B}^0 \to K^+ K^-$ etc.
Comparing this amplitude with the non-annihilation contributions given in 
eqn (\ref{7}), 
one finds that the annihilation amplitude in $B \to P_1 P_2$ decays is 
indeed suppressed by a hefty factor
\begin{equation}
  \label{factor}
 \frac{ (m_1^2-m_2^2)F_0^{P_1P_2} (m_B^2)}{(m_B^2-m_1^2)
F_0^{B\to P_1} (m_2^2) }. 
\end{equation}
The annihilation form factors are difficult to relate directly to 
experimental measurements but they can be modeled. We
expect $F_0^{P_1P_2}(0)$ to have a similar magnitude as the 
the corresponding form factors $F_0^{B\to P_1}(0)$, to which they are 
related by crossing, and which  we have listed  in Tables 2 and 4. 
Based on this, the annihilation form factors appearing in 
eqs.~(\ref{ann1}) and ~(\ref{factor}) are suppressed due to large momentum 
transfer at 
$q^2=m_B^2$, at which they have to be evaluated. The total suppression 
factor in $B\to PP$ decays is then ${\cal O}(m_{1,2}^4/m_B^4)$.
However, the effective coefficients $a_i$, $i=1,2$ entering in the 
annihilation amplitude are much
larger than $a_j$, $j=3,...,10$ governing the penguin-amplitudes. So, a part 
of the power suppression is offset by the favorable effective coefficients.

In the decays $B \to PV$ and $B \to VV$, we do not anticipate the
annihilation suppression as severe as in the decay $B \to PP$.
Concentrating on the decays $B \to PV$,  the annihilation form factors are
\begin{eqnarray}
 <P V| j^\mu | 0>& =&  \epsilon_{\mu\nu\alpha\beta}\epsilon^{*\nu}
p_P^\alpha p_V^\beta  \frac{2V (Q^2)}{m_P+m_V}\nonumber\\
&&-i\left[ \epsilon^*_\mu - \frac{(\epsilon^*\cdot Q)}{Q^2}
Q_\mu\right] (m_P+m_V) A_1 (Q^2)\nonumber\\
&&+ i\left[ (p_P-p_V)_\mu - \frac{m_P^2-m_V^2}{Q^2}
Q_\mu\right] (\epsilon^*\cdot Q) \frac{A_2 (Q^2)}{m_P+m_V}\nonumber\\
&&-i \frac{2m_V }{Q^2}
Q_\mu (\epsilon^*\cdot Q) A_0 (Q^2).
\end{eqnarray}
The annihilation matrix element in the factorization approximation 
can now be written as follows:
 \begin{equation}
<PV| {\cal H}_{eff} | B>_a = i \sqrt{2}G_F V_{qb}V_{qq'}^* a_i
f_B m_V (\epsilon^*\cdot p_B) A_0 (m_B^2).
\end{equation}
 From this, it is easy to see that for this class of decays the 
suppression 
factor is only due to the large momentum transfer involved in the form 
factors $A_0 (m_B^2)$. Hence, 
the annihilation diagrams can contribute more significantly  in 
the decay amplitude. For
 some of the channels for which the non-annihilation contributions are 
highly suppressed, the annihilation diagram can be easily dominant.
For example, the annihilation amplitude to the decay $B^+ \to K^{*+} \bar K^0$
is
\begin{equation}
<K^{*+}\bar K^0| {\cal H}_{eff} | B^+>_a= i \sqrt{2}G_F V_{ub}^*V_{ud} a_1
f_B m_{K^*} (\epsilon^*\cdot p_B) A_0 (m_B^2).
\end{equation}
If we take $A_0(0)=0.4$, $f_B=200$ MeV, the annihilation branching ratio 
is of the order $10^{-8}$ which is an order of 
magnitude higher than the branching ratio calculated with the penguin 
contribution alone. Other channels where the annihilation channel may 
play a significant role have been listed above.

For $B\to VV$ decays, the conclusion is quite similar to the one for 
the $B\to PV$ decays. However, as these decays involve 
yet more untested form factors, their numerical estimates
require a model for these form factors. The suspected channels in $B \to VV$ 
decays sensitive to annihilation contribution have been listed above.  We 
conclude that the decays most sensitive to the annihilation
channel are indeed the Class-II and Class-V decays, mostly involving
$\bar{B}^0(B^0)$ decays.


%
\section{Branching Ratios and Comparison with Data}
%

%
%
%
\begin{table}[htb]
\begin{center}
\label{table8}
\caption{$B\to PP$  Branching Ratios (in units of $10^{-6}$)
using the BSW [Lattice QCD/QCD sum rule] form factors, with
$k^2=m_b^2/2$, $\rho=0.12$, $\eta=0.34$, and $N_c=2,3,\infty$ in
the factorization approach. The last column contains  
measured branching ratios and
upper limits (90\% C.L.) \cite{cleo}.} 
\label{pp1}
\begin{tabular} {|l|c|c|c|c|c|}
\hline
Channel &  Class & $N_c=2$ &  $N_c=3$ & $N_c=\infty$ & Exp. \\
\hline
$B^0 \to \pi^+ \pi^-$ & I & $9.0 \;[11 ]$ &$10.0 \;[12 ]$ 
 & $12 \;[15] $ & $ < 15$ \\
$B^0 \to \pi^0 \pi^0$  & II &  $0.35 \;[0.42]$ &$0.12 \;[0.14]$ 
 &$0.63 \;[0.75]$ &$<9.3$ \\
$ B^0 \to \eta' \eta^\prime$  & II & $0.05 \;[0.07]$ &$0.02 \;[0.02]$
 &$0.09 \;[0.10] $ & $<47$\\
$ B^0 \to \eta \eta^\prime$  & II & $0.19 \;[0.22]$ &$ 0.08 \;[0.10] $ 
 &$0.29 \;[0.34]$ & $<27$\\
$ B^0 \to \eta \eta$  & II & $0.17 \;[0.20]$ &$0.10 \;[0.11] $ 
&$0.24 \;[0.29]$ & $<18$\\
$ B^+ \to \pi^+ \pi^0$ & III & $6.8 \;[8.1 ]$ &$5.4 \;[6.4 ]$
 &$3.0 \;[3.6 ]$ & $ < 20$\\
$ B^+ \to \pi^+ \eta^\prime$  & III & $2.7 \;[3.2 ]$ &$2.1 \;[2.5 ]$
 &$1.1 \;[1.4]$ & $<31$\\
$ B^+ \to \pi^+ \eta$  & III & $3.9 \;[4.7 ]$ &$3.1 \;[3.7 ]$
 &$1.9 \;[2.2 ]$ & $<15$\\

$ B^0 \to \pi^0 \eta^\prime$  & IV & $0.06 \;[0.07] $ &$0.07 \;[0.09] $ 
 &$0.11 \;[0.13] $ & $<11$\\
$ B^0 \to \pi^0 \eta$ & IV  & $0.20 \;[0.24] $ &$0.23 \;[0.27]$ 
 &$0.30 \;[0.36]$ & $<8 $\\

$ B^+ \to K^+ \pi^0$  &  IV & $9.4 \;[11 ]$  & $10 \;[12]$  &
  $12 \;[15]$  &$ <16 $ \\
$B^0 \to K^+ \pi^-$  & IV & $14 \;[16]$ &$15 \;[18]$ &
$18 \;[21]$ &$15^{+5}_{-4} \pm 1$\\
$B^0 \to K^0 \pi^0$  & IV & $5.0 \;[5.9 ]$  & $5.7 \;[6.8 ]$  &
  $7.4 \;[8.9 ]$  &$ <41$\\
$ B^+ \to  K^+ \eta^\prime$  & IV & $21 \;[25]$  & $25 \;[29]$  &
  $35 \;[41]$  &$ 65^{+15}_{-14}\pm 9 $ \\
$B^0 \to K^0 \eta'$  & IV &  $20 \;[24]$  & $25 \;[29]$  &
  $35 \;[41]$  &$ 47^{+27}_{-20}\pm 9 $\\
$ B^+ \to  K^+ \eta$  & IV & $2.0 \;[2.3 ]$  & $2.4 \;[2.7 ]$  &
  $3.4 \;[3.9 ]$  &$ <14$\\
$B^0 \to K^0 \eta$  & IV & $1.7 \;[1.9 ]$  & $2.0 \;[2.2 ]$  &
  $2.6 \;[3.0 ]$  &$ <33 $\\
$ B^+ \to \pi^+ K^0$  & IV & $14 \;[17]$ & $16 \;[20]$ 
& $22 \;[26]$ & $23^{+11}_{-10}\pm 4 $\\
$ B^+ \to K^+ \bar K^0$  & IV & $0.82 \;[0.95]$ & $0.96 \;[1.1 ]$
& $1.3 \;[1.5 ]$ & $ <21$\\
$ B^0 \to K^0\bar K^0$ & IV  &  $0.79 \;[0.92]$ &$0.92 \;[1.1]$
& $1.2 \;[1.4 ]$ & $ <17$\\
\hline
\end{tabular}\end{center}
\end{table}

The decay branching ratios are shown in Tables~8 - 11 for the
decays $B \to PP$, $B \to PV $ (involving $b \to d$ transitions),
$B \to PV $ (involving $b \to s$ transitions) and $B \to VV$,
respectively, for the two sets of form factors given in Tables~2 and 4.
The numbers shown for the hybrid Lattice-QCD/QCD sum rules correspond to
using $F_{1,0}^{B \to \pi} =0.36, ~F_{1,0}^{B \to K} =0.41,
~F_{1,0}^{B \to \eta} =0.16$ and $F_{1,0}^{B \to \eta^\prime} =0.145$.
The first two are slightly above the range determined in \cite{Ball98}
but within the (larger) range as determined from the lattice-QCD calculations
\cite{UKQCD97}. This choice is dictated by data, as discussed in detail 
below. The $k^2$-dependence of the branching ratios in the range
$k^2=m_b^2/2 \pm 2$ GeV$^2$
is small and hence the numbers in these tables are shown only for the
case $k^2=m_b^2/2$. The CKM parameters are fixed at their ``best-fit"  
values: $\rho=0.12, \eta=0.34$. All other parameters have their central
values, discussed in the preceding section. In these tables we give the 
averages of the branching fractions of $\bar{B}^0$ and $B^0$, and of 
$B^+$ and $B^-$, respectively. Hence, when we refer to branching 
fractions in the following sections we always mean the averages over the
$B$ and anti-$B$ decays.   
 The CP-asymmetries are, however,
in general quite sensitive to $k^2$ \cite{kp,kps}. We shall discuss this 
point in a forthcoming paper on CP asymmetries \cite{AKL98-2}.

  A number of observations are in order:
\begin{itemize}
\item There are so far five measured $B \to h_1 h_2$ decay modes in 
well-identified final states: 
$B^0 \to K^+ \pi^-$, $B^+ \to K^+ \eta^\prime$,
$B^0 \to K^0 \eta^\prime$, $B^+ \to \pi^+ K^0$, and $B^+ 
\to 
\omega K^+$, with their branching ratios (averaged over the charge 
conjugate modes) given in Tables 8 and 10.
In addition, the decay modes $ B^+ \to \pi^0 h^+ (h^+=\pi^+,K^+)$
with a branching ratio ${\cal B}(B^+ \to \pi^0 h^+)=(1.6 ^{+0.6}_{-0.5} 
\pm 0.4) \times 10^{-5}$ \cite{cleo}, the decay mode $B^+ \to \omega h^+
(h^+=\pi^+,K^+)$ with a branching ratio ${\cal B}(B^+ \to \omega h^+)= (2.5 
^{+0.8}_{-0.7} \pm 0.3) \times 10^{-5}$ \cite{cleobok}
and the decay modes $B \to K^* \phi$, averaged over $B^+$ and $B^0$
decays with a branching ratio  ${\cal B}(B \to K^* \phi)=(1.1 
^{+0.6}_{-0.5} \pm 0.2) \times 10^{-5}$ \cite{cleobok}, have also been 
measured.

\begin{table}[htb]
\begin{center}
\label{table9}
\caption{$B\to PV$ Branching Ratios (in units of $10^{-6}$) involving 
$b\to d$ (or $\Delta S =0$) transitions 
using the BSW  [Lattice QCD/QCD  sum rule] form factors, with
$k^2=m_b^2/2$, $\rho=0.12$, $\eta=0.34$, and $N_c=2,3,\infty$ in
the factorization approach. The last column contains
 upper limits (90\% C.L.) from \cite{cleo}.
 The upper limit on
the branching ratio for $B^+ \to \rho^+ \pi^0$ is taken from the
PDG tables \cite{PDG96}.} 
\label{pv1}
\begin{tabular} {|l|c|c|c|c|c|}
\hline
Channel &  Class & $N_c=2$ &  $N_c=3$ & $N_c=\infty$ & Exp. \\
\hline

\parbox[c]{3cm}{$B^0 \to \rho^-  \pi^+ $ \\$B^0 \to \rho^+  \pi^- $}
 &\parbox{1cm}{\centering I\\I} & 
\parbox[c]{3cm}{\centering$5.7 \;[6.6 ]$\\$21 \;[25]$ }& 
\parbox[c]{3cm}{\centering$6.4 \;[7.3 ]$ \\ $23 \;[28]$ }
 & \parbox[c]{3cm}{\centering $7.8 \;[9.0 ]$\\$28 \;[34]$} & $ \}<88$
\vspace{0.8mm}
\\
$B^0 \to \rho^0 \pi^0$ & II & $0.75 \;[0.88]$ & $0.07 \;[0.08] $
 &$1.4 \;[1.7 ]$ &$<18$\\
$B^0 \to \omega \pi^0$ & II & $0.28 \;[0.33]$ & $0.08 \;[0.10] $
 &$0.10 \;[0.12] $ &$<14$\\
$ B^0 \to \rho^0 \eta $  & II & $0.02 \;[0.03] $  &$ 0.02 \;[0.02] $
 &$0.06 \;[0.07]$ & $<13$ \\
$ B^0 \to \rho^0 \eta^\prime$  & II & $0.01 \;[0.01] $  &$0.001 \;[0.001] $
 &$0.03 \;[0.04] $ & $<23$ \\
$ B^0 \to \omega \eta $  & II & $0.46 \;[0.54]$  &$0.05 \;[0.06]$
 &$0.63 \;[0.74]$ & $<12 $ \\
$ B^0 \to \omega \eta^\prime$  & II & $0.29 \;[0.34]$  &$0.02 \;[0.02] $
 &$0.46 \;[0.54]$ & $<60$ \\
$ B^+ \to \rho^0 \pi^+ $  & III & $6.3 \;[7.3 ]$ &$3.9 \;[4.5 ]$
 &$0.89 \;[0.98] $ & $<58$\\  
$ B^+ \to \rho^+ \pi^0$   & III &$14 \;[16]$ &$13 \;[15]$
 &$11 \;[13] $ & $<77$ \\
$ B^+ \to \omega \pi^+ $  & III & $6.8 \;[7.9 ]$  &$4.2 \;[4.9 ]$
 &$1.0 \;[1.1]$ & $< 23$ \\
$ B^+ \to \rho^+ \eta $  & III & $6.3 \;[7.4 ]$  &$5.5 \;[6.5 ]$
 &$4.2 \;[5.0 ]$ & $<32$ \\
$ B^+ \to \rho^+ \eta ^\prime$  & III & $4.5 \;[5.3 ]$  &$4.0 \;[4.7 ]$
 &$3.0 \;[3.7 ]$ & $<47$ \\
$ B^0 \to \bar K^{*0} K^0$ & IV & $ 0.31 \;[0.36] $& $0.38 \;[0.44]$
& $0.55 \;[0.64]$ & $-$\\
$ B^+ \to  \bar K^{*0} K^+ $ & IV & $ 0.32 \;[0.37] $& $0.40 \;[0.46]$
& $0.57 \;[0.67]$ & $-$\\
$ B^+ \to K^{*+}  \bar K^0$ & V & $0.001 \;[0.002] $ & $0.0005 \;[0.0007 ]$
& $0.002 \;[0.002 ]$ & $-$\\
$ B^+ \to \phi \pi^+ $  & V & $0.040 \;[0.047] $& $0.005 \;[0.005 ]$
& $0.36 \;[0.43]$ &  $<5.0$\\
$ B^0 \to \phi \pi^0$  & V & $0.019 \;[0.023]$& $0.002 \;[0.003 ]$
& $0.17 \;[0.21]$ &  $<5.0$\\
$ B^0 \to \phi \eta $  & V & $0.008 \;[0.010 ]$& $0.0009 \;[0.001] $
& $0.073\;[0.087] $ &  $<9 $\\
$ B^0 \to  \phi \eta^\prime $  & V & $0.006 \;[0.007 ]$& $0.0007\;[0.0008 ]$
& $0.053 \;[0.064]$ &  $<3.1$\\
$ B^0 \to K^{*0} \bar K^0$ & V & $ 0.001 \;[0.002] $& $0.0004 \;[0.0006]$
& $0.002 \;[0.002]$ & $-$\\

\hline
\end{tabular}\end{center}
\end{table}
\item The branching ratios for $B^0 \to K^+\pi^-$ and $B^+ \to \pi^+ K^0$
are in good agreement with the CLEO data. Moreover, being class-IV decays,
they show only a small sensitivity on $\xi$. The estimated branching ratios
for $B^+ \to \pi^+ \pi^0 $ and $B^+ \to K^+ \pi^0$ are in agreement 
with the
respective upper bounds. The latter being a class-IV decay is again stable
w.r.t. the variation of $N_c$; the former (a class-III decay) varies by 
approximately a factor
2.3 as $N_c$ is varied. The branching ratio for the sum  $B^+ \to \pi^0 h^+$
is plotted as a function of $\xi=1/N_c$ in Fig.~\ref{fph} for the BSW model
form factors (dashed-dotted curve) and two different sets, corresponding to
the central values of the hybrid Lattice QCD/QCD-SR form factors (dashed 
curve) and for values which are closer to their theoretical range given
in Table 4 (dotted curve). We see that data for this mode is well explained.
\item We estimate the branching ratio for $B^0 \to \pi^+ \pi^-$ to be
around $1 \times 10^{-5}$ for the central values of the CKM parameters,
which could go down to about $5 \times 10^{-6}$ for $V_{ub}/V_{cb}=0.06$.
The present CLEO upper limit is in comfortable accord with our estimates
but we expect that this decay mode should be measured soon.
However, the decay $B^0 \to \pi^0 \pi^0$ is not expected to go above 
$10^{-6}$, which makes it at least a factor 10 below the present experimental
sensitivity.

\begin{table}[htb]
\begin{center}
\label{table10}
\caption{$B\to PV$ Branching Ratios (in units of $10^{-6}$) involving
$b\to s$ (or $|\Delta S| =1$) transitions
using the BSW  [Lattice QCD/QCD  sum rule] form factors, with
$k^2=m_b^2/2$, $\rho=0.12$, $\eta=0.34$, and $N_c=2,3,\infty$
in the factorization approach. The last column contains
the measured branching ratio and
upper limits (90\% C.L.) \cite{cleo}.}
\label{pv3}
\begin{tabular} {|l|c|c|c|c|c|}
\hline
Channel &  Class & $N_c=2$ &  $N_c=3$ & $N_c=\infty$ & Exp. \\
\hline
$B^0 \to  \rho^- K^{+} $   & I &$0.40 \;[0.46]$ &$0.45 
\;[0.52]$ & $0.56 \;[0.64]$ &$ <33$\\
$ B^+ \to K^{*+} \eta'$  & III &    $0.28 \;[0.39]$ &    $0.24 \;[0.29]$
  &    $0.33 \;[0.33]$  &$<130$\\

$B^0 \to K^{*+} \pi^-$   & IV & $6.0 \;[7.2 ]$& $6.6 \;[7.8 ]$
&$7.8 \;[9.3 ]$ &$ <67$\\
$B^0 \to K^{*0} \pi^0$  &  IV & $1.8 \;[2.0 ]$  &   
$2.2 \;[2.5 ]$  &  $3.2 \;[3.6 ]$  &$ <20$\\
$B^0 \to \rho^0 K^0$  &  IV &  $0.50 \;[0.58]$  & $0.49 \;[0.57]$  &
  $0.62 \;[0.73]$  &$ <30$\\
$ B^+ \to K^{*+ }\pi^0$  & IV &  $4.4 \;[5.4 ]$  &  $4.7 \;[5.9 ]$  &
  $5.6 \;[6.9 ]$  &$ <80$\\
$ B^+ \to \rho^0 K^+ $ &  IV &   $0.58 [0.67]$&    $0.50 [0.58]$
  &    $0.47 [0.55]$  &$ <14$\\
$ B^+ \to K^{*+} \eta$  & IV &    $2.2 \;[2.8 ]$ &    $2.2 \;[2.7 ]$
  &    $2.0 \;[2.4 ]$  &$<30$\\
$ B^0 \to K^{*0} \eta$  &  IV &  $2.0 \;[2.5 ]$ &    $2.1 \;[2.7 ]$
  &    $2.6 \;[3.1 ]$  &$<30$\\
$ B^+ \to  K^{*0} \pi^+ $  & IV & $5.6 \;[6.7 ]$& $6.9 \;[8.3 ]$
& $10 \;[12] $ &$<39$\\
$ B^+ \to \rho^+ K^0$  & IV & $0.03 \;[0.03] $& $0.01 \;[0.01] $
& $0.01 \;[0.02] $ &  $<64$\\
$ B^0 \to K^{*0} \eta'$  &  V &  $0.06 \;[0.12] $ &    $0.07 \;[0.07]$
  &    $0.41 \;[0.39]$  &$<39$\\
$ B^+ \to \phi K^+ $  & V & $16 \;[18]$& $8.3 \;[9.6 ]$
& $0.45 \;[0.53]$ &  $<5.0 $\\
$ B^0 \to \phi K^0 $  & V & $15 \;[18]$& $8.0 \;[9.3 ]$
& $0.44 \;[0.51]$ &  $<31$\\
$ B^0 \to \omega K^0 $ &  V & $2.8 \;[3.3 ]$ & $0.02 \;[0.02]$
  &    $8.9 \;[10 ]$  &$<57 $\\
$ B^+  \to \omega K^+ $  & V &   $3.2 \;[3.7 ]$ &    $0.25 
\;[0.28]$
  &    $11 \;[13]$  &$ 15^{+7}_{-6}\pm 2$\\
\hline
\end{tabular}\end{center}
\end{table}
 \item   We show the dependence of the branching ratios on
the input form factors and the parameter $\xi=1/N_c$ for
the decays $B^+ \to K^+ \eta^\prime$ and
$B^0 \to K^0 \eta^\prime$ in Figs.~\ref{fke} and \ref{fk0e}, respectively.
As can be seen in these figures, data tends to prefer somewhat larger values  
for the form factors $F_{1,0}$ than the central values given by the 
Lattice-QCD/QCD sum rules  in Table 4. 
However, the experimentally preferred values of the form factors all lie 
within the range allowed by the present theoretical estimates. Likewise,
the branching ratio increases as the $s$-quark mass decreases, as already
noted in \cite{ag,acgk}. Thus, for $m_s$ ($\mu=2.5$ GeV) $=100$ MeV, and
$F_{1,0}^{B \to \eta^\prime} =0.15$, there is no problem to accommodate the
CLEO data within the measured $\pm 1 \sigma$ range. As already discussed at
length in refs.~\cite{ag,acgk}, these decay modes are dominated by the
QCD penguin, and while the contributions of the anomaly terms are included
in the rate estimates, their role numerically is subleading.  
The decay modes $B^+ \to K^+ \eta^\prime$ and
$B^0 \to K^0 \eta^\prime$ show some    
preference for smaller values of $\xi$, though this is correlated with 
other input parameters and at this stage one can not draw completely
quantitative conclusions. 
Summarizing the $B \to PP$ decays, we stress that the 
factorization-based estimates described here are consistent with the
measured decay modes. All other estimated branching ratios are consistently
below their present experimental limits. However, we do expect the modes
$B^0 \to \pi^+ \pi^-$, $B^+ \to \pi^+ \pi^0$,
and $B^+ \to K^+ \pi^0$ to be measured soon.

\begin{table}[tbh]
\begin{center}
\label{table11}
\caption{$B\to VV$ Branching Ratios (in units of $10^{-6}$)
using the BSW  [Lattice QCD/QCD  sum rule] form factors, with
$k^2=m_b^2/2$, $\rho=0.12$, $\eta=0.34$, and $N_c=2,3,\infty$  in
the factorization approach. The last column contains
upper limits (90\% C.L.) mostly from \cite{cleo} except for 
the branching ratios for $B^0 \to \rho^+ \rho^-$,
$B^0 \to \rho^0 \rho^0$, $ B^+ \to \rho^+ \rho^0$, $B^0 \to K^{*0} 
\rho^0$ and $ B^+ \to K^{*+} \rho^0$, which  are taken from the PDG 
tables \cite{PDG96}.}
\label{vv1}
\begin{tabular} {|l|c|c|c|c|l|}
\hline
Channel & Class & $N_c=2$ &  $N_c=3$ &  $N_c=\infty$ & Expt.\\
\hline
$B^0 \to \rho^+  \rho^-$  & I &  $18 \;[20]$&  $20 \;[22]$
 & $24 \;[27]$ & $ <2200$\\
$B^0 \to \rho^0 \rho^0$  &  II & $1.3 \;[1.3 ]$ &  $0.59 \;[0.59]$
  &$2.5 \;[2.5 ]$ &$<280$\\
$ B^0 \to \omega \omega$  & II & $0.87 \;[0.96]$  & $0.15\;[0.17]$
 &$0.86 \;[0.96]$ & $ <19$\\
$ B^+ \to \rho^+ \rho^0$  & III & $14 \;[15]$& $11 \;[12]$
 &$6.1 \;[6.8 ]$ & $<1000$\\
$ B^+ \to \rho^+ \omega$  & III & $15 \;[16]$  & $12 \;[13]$
 &$6.6 \;[7.3 ]$ & $<67 $\\

$B^0 \to K^{*+} \rho^-$  & IV & $5.4 \;[6.0 ]$ & $5.9 \;[6.6 ]$ &
$7.0 \;[7.8 ]$ &$ -$\\
$B^0 \to K^{*0} \rho^0$  & IV & $1.1 \;[1.2 ]$ & $1.3 \;[1.4 ]$
 & $1.9 \;[ 1.9]$  &$ <460 $ \\
$ B^+ \to K^{*+ }\rho^0$  & IV & $5.0 \;[5.8 ]$  & $5.5 \;[6.3 ]$  &
  $6.6 \;[7.6 ]$  &$ <900$\\
$ B^+ \to \rho^+ K^{*0}$  & IV & $5.1 \;[ 5.6]$ &$6.3 \;[6.9 ]$ &
 $9.1 \;[10 ]$ & $-$\\
$ B^+ \to K^{*+ }  \bar K^{*0}$  & IV &  $0.29 \;[0.38]$ & $0.37 \;[0.47]$ 
& $0.53 \;[0.68]$ & $-$\\
$ B^0 \to K^{*0} \bar K^{*0}$ & IV & $0.28 \;[0.36]$ & $0.35 \;[0.45]$ 
& $0.51 \;[0.65] $ & $-$\\

$ B^0 \to \rho^0 \omega$  & V & $0.018\;[0.020] $  & $0.005 \;[0.006 ]$
 &$0.23 \;[0.26]$ & $ <11$\\
$ B^0 \to K^{*0}\omega $ & V & $10 \;[12]$  & $3.6 \;[4.0 ]$
  &    $0.63 \;[1.1]$  &$  <23$\\
$ B^+ \to K^{*+} \omega $  & V &  $11 \;[13]$  & $3.7 \;[4.1 ]$  &
    $1.7 \;[2.4 ]$  &$ <87$\\
$ B^+ \to K^{*+} \phi$  & V & $ 16 \;[20]$ &$8.2 \;[10 ]$   
& $0.45 \;[0.57]$ &  $<41$\\
$ B^0 \to K^{*0} \phi$  & V & $ 15 \;[19]$ &$7.9 \;[10 ]$
& $0.43 \;[0.55]$ &  $<21$\\
$ B^+ \to \rho^{+} \phi$  & V & $ 0.039 \;[0.043] $ &$0.004 \;[0.005 ]$
& $0.35 \;[0.38]$ &  $ <16$\\
$ B^0 \to \rho^{0} \phi$  & V & $ 0.019 \;[0.021] $ &$0.002 \;[0.002 ]$
& $0.17 \;[0.18]$ &  $ <13$\\
$ B^0 \to \omega \phi$  & V &$ 0.019\;[0.020] $ &$0.002 \;[0.002]$
& $0.17 \;[0.18]$ &  $ <21$\\
\hline
\end{tabular}
\end{center}
\end{table}

\begin{figure}
    \epsfig{file=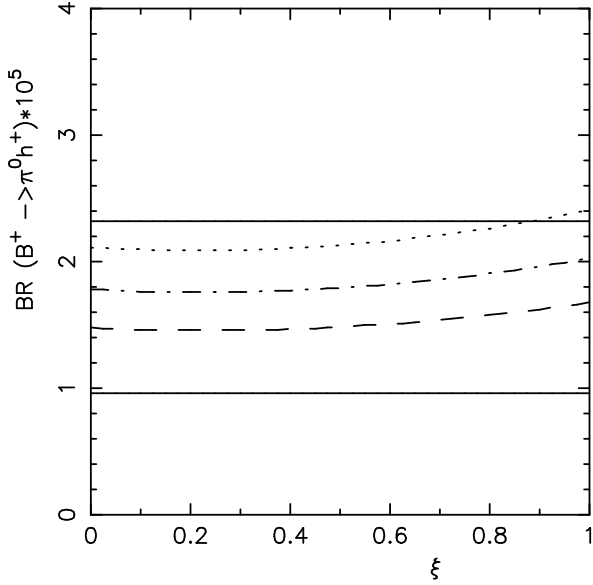,bbllx=2cm,bblly=7.5cm,bburx=21cm,bbury=19cm,%
width=12cm,angle=0}
\caption{Branching ratio for the decays $ B^+ \to \pi^0 h^+$ 
($h^+=\pi^+, K^+$) as a function of $\xi$ for three different sets of 
form factors: BSW Model (dashed-dotted curve), Lattice-QCD/QCD-sum rules 
with central values in Table 4 (dashed curve),
with the values $F_{0,1}^{B \to \pi}=0.36$ and $F_{0,1}^{B \to K}=0.41$
(dotted curve).
 The horizontal solid lines are the $\pm 1\sigma$ measurements from
experiment \protect\cite{cleo}.}
\label{fph}
\end{figure}

\begin{figure}
\epsfig{file=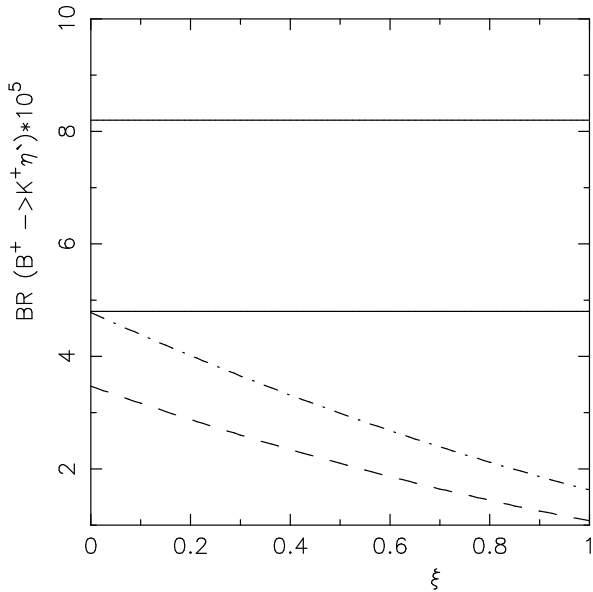,bbllx=2cm,bblly=7.5cm,bburx=21cm,bbury=19cm,
width=12cm,angle=0}
\caption{Branching ratio for $ B^+ \to K^+\eta^\prime$ as a 
function of $\xi=1/N_c$. The dash-dotted and dashed curves correspond to 
the choice $F_1^{B \to \eta^\prime}(0)=F_0^{B \to \eta^\prime} (0)= 0.15$, 
$m_s(\mu=2.5 ~\mbox{GeV}) = 100$ MeV, and 
$F_1^{B \to \eta^\prime} (0)=F_0 ^{B \to \eta^\prime}(0)= 0.135$,
$m_s(\mu=2.5 ~\mbox{GeV}) = 122$ MeV, respectively.
The horizontal solid lines are the $\pm 1\sigma$ measurements from
experiment \protect\cite{cleo}.}
\label{fke}
\end{figure}
\item The two observed $B \to PV$ decays, $B^+ \to \omega K^+$ and
$B^+ \to \omega h^+$, $h^+=\pi^+,K^+$, show 
strong $N_c$-dependence as anticipated. The decay $B^+ \to \omega \pi^+$,
a class-III decay, has not yet been measured and the mode
$B^+ \to K^+ \omega$ (a class-V decay) has a $3.9 \sigma$ experimental 
significance.
The branching ratios of $B^+ \to \omega K^+$ and $B^+ \to \omega \pi^+$
are plotted as 
functions of $\xi$ in Figs.~\ref{fkm2} and \ref{fpm2}, respectively, 
showing the variations
on other parameters (form factors and CKM matrix elements) as well.
Taking the CLEO measurement ${\cal B}(B^+ \to \omega K^+) = 
(1.5^{+0.7}_{-0.6} \pm 0.2) \times 10^{-5}$ on face value, this mode
suggests that $\xi \leq 0.1$ or $\xi \geq 0.6$. The present CLEO upper limit
${\cal B}(B^+ \to \omega \pi^+) < 2.3 \times 10^{-5}$ is not yet restrictive
enough. 
 The branching ratio for the combined decay
$B^+ \to \omega h^+ (h^+=\pi^+,K^+)$ is shown in Fig~\ref{fkm} as a
function of $\xi$ for two values of the form factors $F_1^{B \to K}$ and
$F_1^{B \to \pi}$ and two sets of values for the CKM parameters 
$\rho$ and $\eta$. The values of these form factors correspond to the
BSW model and the upper limit in Table 4 to the Lattice-QCD/QCD sum rule
case. Again, one sees that there is a tendency in the data to 
prefer larger values of the form factors. We note that both
small values $\xi \simeq 0$ and $\xi \geq 0.5$ are compatible with
data in this decay, with the theoretical branching ratio rising above
$1 \times 10^{-5}$.  The 
value corresponding to the naive factorization,
$N_c=3$ (or $\xi=0.33$) is definitely too low compared to the
data on the two measured $B \to PV$ decays. This is in line with  
earlier observations in the literature \cite{ag,deshpande,CT98}.

\begin{figure}
    \epsfig{file=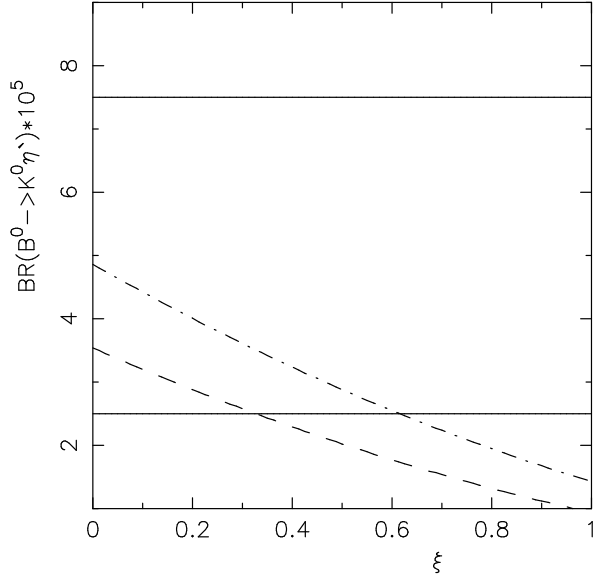,bbllx=2cm,bblly=7.5cm,bburx=21cm,bbury=19cm,%
width=12cm,angle=0}
    \caption{Branching ratio for $ B^0 \to 
K^0 \eta^\prime$ as a function of 
$\xi$. The legends are the same as in Figure \ref{fke}.}
    \label{fk0e}
\end{figure}
\begin{figure}
    \epsfig{file=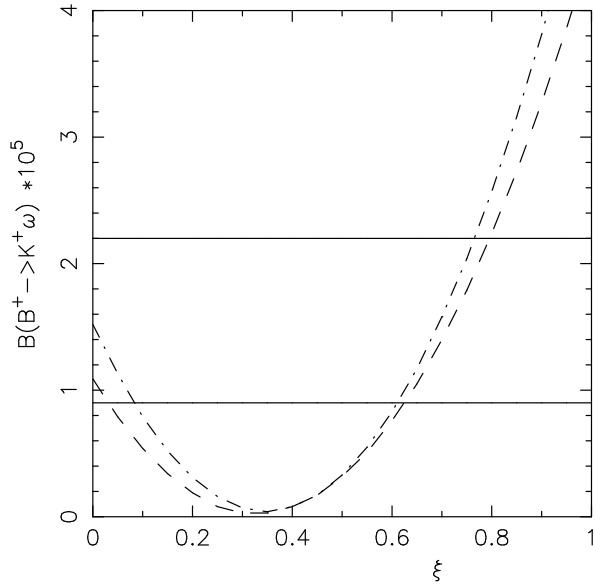,bbllx=2cm,bblly=7.5cm,bburx=21cm,bbury=19cm,%
width=12cm,angle=0}
\caption{Branching ratio for $ B^+ \to K^+ \omega$ and
as a function of $\xi$. The legends are as follows:
$\rho=0.30, \eta=0.42, F_1^{B \to K}=0.44$ (dashed-dotted curve),
$\rho=0.12, \eta=0.34, F_1^{B \to K}=0.38$ (dashed curve).
The horizontal solid lines are the $\pm 1\sigma$ measurements from
experiment \protect\cite{cleobok}. }
\label{fkm2}
\end{figure}

\item No other $B \to PV$ decays have been measured yet. However, an
interesting upper bound 
${\cal B}(B^+ \to K^+ \phi) < 0.5 \times 10^{-5}$ (at $90\%$ C.L.) has been 
put by the CLEO collaboration \cite{cleobok}. This and the related decay
$B^0 \to K^0 \phi$ are both penguin dominated and their decay rates are
expected to be almost equal. The only worthwhile 
CKM-dependence is on the Wolfenstein parameter $A$ (hence weak). However, 
being class-V decays, their branching ratios depend strongly on 
$\xi$, with both having their lowest values at $\xi=0$. The branching 
ratio
${\cal B}(B^+ \to K^+ \phi)$ is shown as a function of $\xi$ in 
Fig.~\ref{fkph} for $A=0.81$ (dashed curve) and $A=0.75$ (dashed-dotted 
curve) and the CLEO $90\%$ C.L. upper bound is also indicated. This 
shows that values $\xi \geq 0.4$ are disfavored by the present data. In 
fact, taken the
data on their face value the measured branching ratios for the decays
$B^+ \to \omega h^+ (h^+=\pi^+,K^+)$ and $B^+ \to \omega K^+$, as well 
as the upper bounds on the branching ratios for $B^+ \to K^+ \phi$ and
$B^+ \to \omega \pi^+$ can be accommodated for a value of $\xi$, close to 
$\xi =0$. All other decay modes in Tables 9 and 10 (for the $B \to PV$ case)
are consistent with their respective upper limits. However, we do expect
that the decay modes $B^+ \to \rho^+ \eta$, $B^+ \to \rho^+ 
\eta^\prime$, $B^0 \to K^{*0} \pi^0$, $B^+ \to K^{*0} \pi^+$ and 
$B^+ \to \rho^+ \omega$ should be observed in the next round of experiments
at CLEO and at B factories.

\begin{figure}
    \epsfig{file=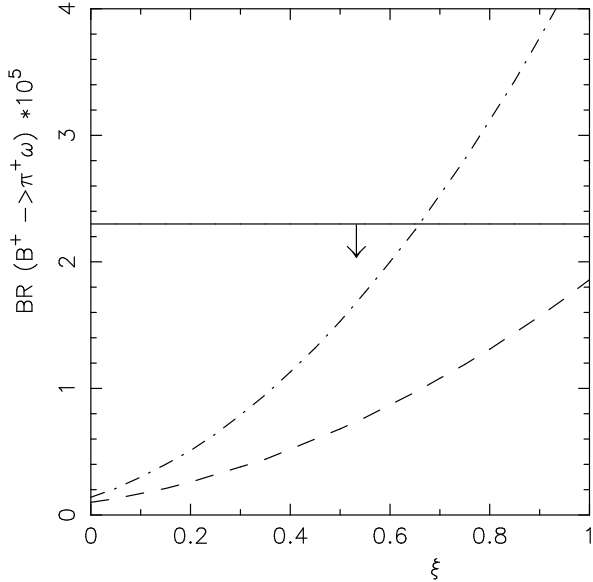,bbllx=2cm,bblly=7.5cm,bburx=21cm,bbury=19cm,%
width=12cm,angle=0}
\caption{Branching ratio for 
$ B^+ \to \pi^+ \omega$ as a function of
$\xi$. The legends are as follows:
 $\rho=0.30, \eta=0.42,
F_1^{B \to \pi}=0.38$ (dashed-dotted curve),
$\rho=0.12, \eta=0.34, F_1^{B \to K}=0.34$ (dashed curve). 
The horizontal solid line is the 90\% C.L. upper limit from
experiment \protect\cite{cleobok}.}
\label{fpm2}
\end{figure}
\item There is one $B \to VV$ decay mode $B \to \phi K^*$, for which 
some experimental evidence exists, and an averaged branching ratio
${\cal B}(B \to \phi K^*)=(1.1 ^{+0.6}_{-0.5} \pm 0.2) \times 10^{-5}$ 
has been posted by the CLEO collaboration \cite{cleobok}.
The decay modes $B^+ \to \phi K^{*+}$ and $B^0 \to \phi K^{*0}$ are
dominated by penguins and are expected to be almost equal (see Table 11).
They also belong to class-V decays, showing very strong $\xi$-dependence 
(almost a factor 35!), with the branching ratios having their smallest
values at $\xi=0$. A comparison of data and factorization-based estimates 
is shown in Fig.~\ref{fkp}. In this case, data favors $0.4 \leq \xi \leq 
0.6$, apparently different from the values of $\xi$ suggested by the $B \to 
PV$ decays discussed earlier. In fact, the branching ratios of the decays
$B^+ \to \phi K^+$, $B^0 \to \phi K^0$, $B^+ \to \phi K^{*+}$ and $B^0 \to 
\phi K^{*0}$ are almost equal in the factorization approach and they all
belong to class-V. Hence, their measurements will be rather crucial
in testing this framework.

\begin{figure}
    \epsfig{file=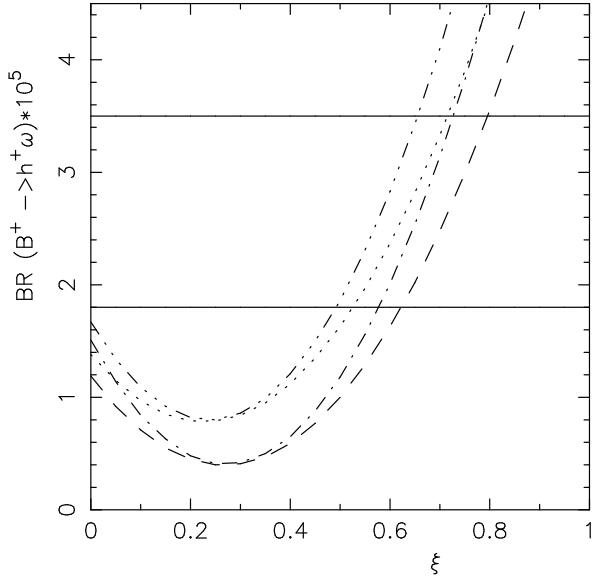,bbllx=2cm,bblly=7.5cm,bburx=21cm,bbury=19cm,%
width=12cm,angle=0}
\caption{Branching ratio for $ B^+ \to h^+ \omega$ as a function of
$\xi$. The legends are as follows:
$\rho=0.30, \eta=0.42, F_1^{B \to \pi}=0.38,
F_1^{B \to K}=0.44$ (dashed-triple dotted curve),
$\rho=0.30, \eta=0.42, F_1^{B \to \pi}=0.33, F_1^{B \to K}=0.38$
(dotted curve),
$\rho=0.12, \eta=0.34, F_1^{B \to \pi}=0.38, F_1^{B
\to K}=0.44$ (dashed-dotted curve),
 $\rho=0.12, \eta=0.34, F_1^{B \to \pi}=0.33, F_1^{B \to K}=0.38$
(dashed curve).
The horizontal solid lines are the $\pm 1\sigma$ measurements from
experiment \protect\cite{cleo}. }
\label{fkm} 
\end{figure}
\item Based on the present measurements of the $B \to PV$ and $B \to VV$
decay modes, we summarize that all of them belong to the class-V (and one
to class-III) decays, for which the factorization-based estimates show 
strong sensitivity to $\xi$. This implies that they are harder to predict.
The classification given above, however, does not imply that the class-V 
decays are necessarily small. In
fact for $N_c=2$, the measured class-IV decays and a number of class-V
$B \to PV$ and $B \to VV$  decays such as the ones mentioned above are 
comparable in rates (within a factor 2). For the class-V decays,
the amplitudes can become very small in some range of $\xi$,
implying large non-perturbative renormalizations which are harder to
quantify in this framework. Also,
many class-V penguin decays may  have significant contributions from 
annihilation and/or FSI, as the factorization-based amplitudes, depending 
on $\xi$, may not dominate the decay rates.
 This is generally not foreseen for the  
class-I (tree-dominated) and class-IV (penguin-dominated) decays
and most of the class-III decays. Hence, these decays can be predicted with
greater certainty. 

\begin{figure}
    \epsfig{file=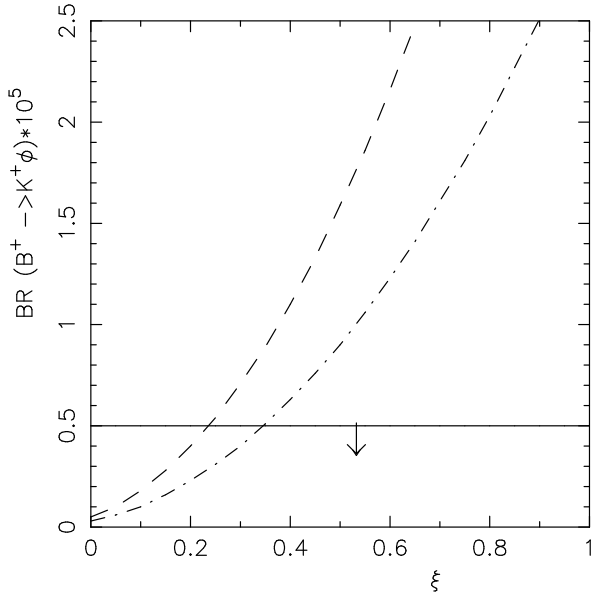,bbllx=2cm,bblly=7.5cm,bburx=21cm,bbury=19cm,%
width=12cm,angle=0}
\caption{
Branching ratio for $ B^+ \to K^+ \phi$ as a function of
$\xi$. The legends are as follows:
Upper curve: Wolfenstein parameter $A=0.81$, $F_1^{B \to K}=0.38$.
Lower curve: Wolfenstein parameter $A=0.75$, $F_1^{B \to K}=0.31$.
The horizontal solid line is the 90\% C.L. upper limit from
experiment \protect\cite{cleo}.}
    \label{fkph}
\end{figure}
\begin{figure}
    \epsfig{file=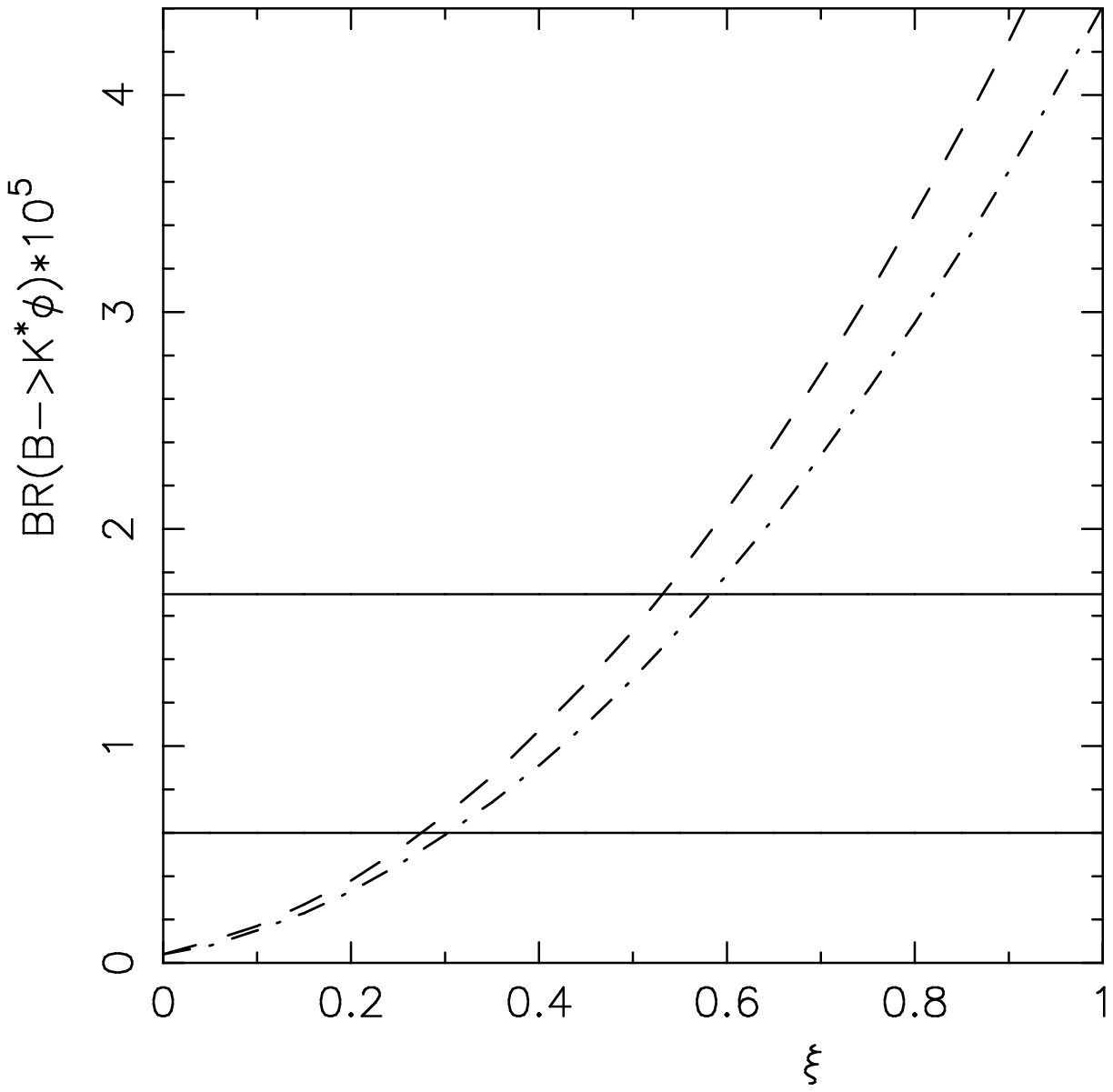,bbllx=2cm,bblly=7.5cm,bburx=21cm,bbury=19cm,%
width=12cm,angle=0}
\caption{
Branching ratio for $ B \to K^* \phi$ as a function of
$\xi$, after averaging over the $B^+$ and $B^0$ decay rates. The legends 
are as follows:
Upper curve: Wolfenstein parameter $A=0.81$.
Lower curve: Wolfenstein parameter $A=0.75$. For the form factors, we use
the BSW model.
The horizontal solid lines represent the CLEO measurement with $\pm 1\sigma$
errors.
 \protect\cite{cleo}.}
    \label{fkp}
\end{figure}    

\item Concerning comparison of our results with the earlier ones in
\cite{ag,acgk}, we note that we
have made use of the theoretical work presented in these papers.
We reproduce all the numerical results for the same values of the input
parameters. Our decay amplitudes agree with the ones presented in 
\cite{CT98}, though our estimates of the matrix elements of
pseudoscalar densities $\langle 0 | \bar{u} \gamma_5 u| 
\eta^{(\prime)}\rangle$ and $\langle 0 | \bar{d} \gamma_5 d|
\eta^{(\prime)}\rangle$ differ from the ones used in \cite{CT98}.
Our expressions are given explicitly in Appendix A. The disagreement in
the decay rates for $B^0 \to \rho^0 \eta$ and $B^0 \to \rho^0 \eta^\prime$
between our results and the ones given in \cite{CT98} has now been resolved
\footnote{We thank Hai-Yang Cheng for a correspondence on this
point.}. However, we do not subscribe to the notion that 
$N_c(V+A)$ induced by the $(V-A)(V+A)$ penguin operators is different
from the $N_c(V-A)$ arising from the  $(V-A)(V-A)$ operators, advocated
in \cite{CT98} and continue to use the same $N_c$ irrespective of the
chiral structure of the four-quark operators. We have discussed at length 
the difficulties in predicting class-V decays some of which, in our opinion, 
may require annihilation and/or FSI effects. 
 
Comparison of our numerical results in the
branching ratios for the $B \to PV$ modes with the ones presented in 
\cite{deshpande} requires a
more detailed comment. First of all, our input parameters are significantly
different from those of \cite{deshpande}. For the same values of
input parameters, our results in charged $B^+ \to (PV)^+$ decays are
in reasonable accord. However, significant differences exist in the 
neutral $B^0 \to (PV)^0$ decay rates, which persist also if we
adopt the input values used in \cite{deshpande}. In particular, in this case
we find for $N_c=\infty$: 
${\cal B}(B^0 \to \rho^0 \eta) = 2.7 \times 10^{-7}$ compared to
$6.7 \times 10^{-6}$ \cite{deshpande}, ${\cal B}(B^0 \to \rho^0 
\eta^\prime) = 1.2 \times 10^{-7}$ compared to
$3.6 \times 10^{-6}$ \cite{deshpande}, ${\cal B}(B^0 \to \omega \eta) = 
6.9 \times 10^{-7}$ compared to
$7.1 \times 10^{-6}$ \cite{deshpande}, and ${\cal B}(B^0 \to \omega 
\eta^\prime) = 1.3 \times 10^{-7}$ compared to
$3.6 \times 10^{-6}$ \cite{deshpande}. For our input values, the differences
in branching ratios are even more drastic, as   
can be seen by comparing our results with the ones in \cite{deshpande} 
for these decays. We have given sufficient details in our paper to 
enable a comparison of the formulae, including 
matrix elements of the pseudoscalar densities, and hence it should not be 
too difficult to figure out the source of the present discrepancy.
Such details are not given  in \cite{deshpande}.   

\item Within the present framework, we have calculated the relative 
importance of electroweak penguins in all the $B \to PP$, $B \to PV$ and 
$B \to VV$ decays studied in this paper. The decay modes where the
electroweak penguins may make a significant contribution 
are shown in Table 12 where we give the ratio
\begin{equation}
R_{\mbox{W}} \equiv \frac{\B ( B \to h_1 h_2) (\mbox{with}~ 
a_7,...,a_{10}=0)} {\B ( B \to h_1 h_2)}~. \\ \nonumber
\label{rewpenguin}
\end{equation}
\end{itemize}
In the $B \to PP$ case, there are five such decays
whose rates show moderate
dependence on the electroweak penguins. The decay
$B^0 \to \pi^0 \pi^0$ receives significant electroweak penguin contribution
for $N_c=3$.
 In the class-IV $B \to PP$ decays, three decays,
namely $B^0 \to K^0 \pi^0$, $ B^0 \to K^0 \eta$ and $B^+ \to K^+ \eta$
(all having branching ratios 
 in excess of $10^{-6}$) have significant electroweak
contributions. The presence of electroweak penguins in these decays
reduces the decay rate by about $\sim 20\%$ to $\sim 40\%$.

%
%
%
%
\begin{table}[htb]
\label{bew}
\begin{center}  
\caption{Ratios of branching ratios $R_{\mbox{W}}$ defined in
 eq.~(\ref{rewpenguin})
for  $N_c=2,3,\infty$ for the form factors in the BSW  model
[Lattice-QCD/QCD sum rule method]. The horizontal lines demarcate the
decays $B \to PP$, $B \to PV$ and $B \to VV$.}
\begin{tabular}{|l|c|c|c|c|}
\hline
Channel & Class & $N_c=2$ & $N_c=3$ &$N_c=\infty$ \\
\hline
$B^0 \to \pi^0 \pi^0$ & II     & $1.2 ~[1.2] $   & $1.5 ~[1.5]$    & 1.1 
[1.1]\\
$B^0 \to \pi^0 \eta ^\prime$ & II     & $1.3 ~[1.3] $   & $1.3 ~[1.3]$    & 
1.4 [1.4]\\
$B^0 \to K^0 \pi^0$ & IV         & $1.5 ~[1.4] $   & $1.4 ~[1.4 ]$   & 1.3 
[1.3]\\
$B^0 \to K^0 \eta $  & IV        & $1.5 ~[1.5 ]$   & $1.5 ~[1.5 ]$   & 1.4 
[1.4]\\
$B^+ \to K^+\eta$  & IV  & $1.6 ~[1.6 ]$   & $1.5 ~[1.5 ]$   & 1.3 
[1.3]\\
\hline
$B^0 \to \rho^0 \pi^0 $ & II    & $1.0 ~[1.0 ]$   & $1.9 ~[1.9 
]$   & 1.1 [1.1]\\
$B^0 \to \rho^0 \eta $ & II    & $1.4 ~[1.4 ]$   & $1.5 ~[1.5 ]$   
& 1.1 [1.1]\\
$B^0 \to \rho^0 \eta ^\prime$ & II    & $1.1 ~[1.2 ]$   & $4.7 ~[4.9 ]$   
& 1.3 [1.2]\\
$B^0 \to K^{*0} \pi^0 $ & IV        & $1.7 ~[1.8 ]$   & $1.6 ~[1.7 ]$   & 1.4
[1.5]\\
$B^0 \to \rho^0 K^0$ & IV         & $0.077~[0.077]$ & $0.008~[0.008]$ & 0.11
[0.11]\\
$B^0 \to K^{*0} \eta$  & IV     & $0.69 ~[0.66 ]$ & $0.70 ~[0.67 ]$ & 0.71
[0.69]\\
$B^+ \to K^{*+}\pi^0$ & IV & $0.63~[0.61]$  & $0.68~[0.66 ]$  & 0.78
[0.75]\\
$B^+\to\rho^0K^+$ & IV   & $0.83~[0.83]$  & $0.59~[0.59 ]$  & 0.13
[0.13]\\
$B^+ \to K^{*+}\eta$ & IV & $0.60 ~[0.58 ]$ & $0.66 ~[0.63 ]$ & 0.78
[0.76]\\
$B^+ \to \rho^+ K^0$& IV  & $0.45 ~[0.45 ]$ &  $0.60 ~[0.60 ]$ & 0.66
[0.66] \\
$B^0 \to K^{*0} \eta ^\prime$ & V     & $0.97 ~[0.54 ]$ & $1.8 ~[1.6 ] $
& 1.1 [1.2]\\
$B^0 \to \omega K^0  $ & V      & $0.83~[0.83 ]$  & $0.42 ~[0.42 ]$ & 1.2
[1.2]\\
$B^0 \to \phi \pi^0$  & V      &  $1.7 ~[1.7 ]$ & $0.002 ~[0.002]$ & 
0.78 [0.78]\\
$B^0 \to \phi \eta $  & V      & $1.7 ~[1.7 ]$  & $0.002 ~[0.002]$ &
0.78 [0.78] \\
$B^0 \to \phi \eta^\prime $ & V      & $1.7 ~[1.7 ]$  & $0.002 
~[0.002]$ & 0.78 [0.78] \\ 
$B^0 \to \phi K ^0$  & V      & $1.2 ~[1.2 ]$  & $1.3   ~[1.3]$ & 2.1  
[2.1] \\
$B^0 \to K^{*0} \bar K^0$ & V   & $0.46 ~[0.46 ]$ & $0.84 ~[0.84 ]$ & 0.73
[0.73]\\
$B^+ \to K^{*+} \bar K^0$ & V   & $0.46 ~[0.46 ]$ & $0.84 ~[0.84 ]$ & 0.73
[0.73]\\
$B^+ \to \phi \pi^+ $& V &  $1.7 ~[1.7 ]$ & $0.002 ~[0.002]$ & 0.78 
[0.78] \\
$B^+ \to \phi K^+ $& V  & $1.2 ~[1.2 ]$ & $1.3   ~[1.3]$   & 2.1  [2.1]
\\
\hline
$B^0 \to \rho^0  \rho^0$ & II   & $0.58 ~[0.58 ]$ & $0.31 ~[ 0.31]$ & 1.0  
[1.0] \\
$B^0 \to \rho^0 K^{*0}$ & IV    & $2.5 ~[2.7 ]$   & $2.4 ~[ 2.6]$   & 2.1
[2.2]\\
$B^+ \to\rho^0K^{*+}$ & IV & $0.54 ~[0.52]$  & $0.61 ~[0.58 ]$ & 0.74
[0.72]\\
$B^0 \to \rho^0 \omega$ & V   & $1.9 ~[1.9 ]$   & $0.08 ~[ 0.08]$ & 0.77
[0.77]\\
$B^0 \to  \rho \phi $ & V     & $1.7 ~[1.7 ]$   & $0.002~[0.002]$ & 0.78
[0.78]\\
$B^0 \to  \omega\phi $ & V     & $1.7 ~[1.7 ]$   & $0.002~[0.002 ]$ & 0.78
[0.78]\\
$B^0 \to K^{*0}  \omega$ & V   & $0.93 ~[0.92 ]$ & $0.84 ~[ 0.82]$ & 1.7
[1.6]\\
$B^0 \to K^{*0}  \phi$ & V     & $1.2 ~[1.2 ]$   & $1.3  ~[ 1.3]$  & 2.1
[2.1]\\
$B^+ \to \rho^+\phi$& V & $1.7 ~[1.7 ]$  & $0.002~[0.002]$ & 0.78 [0.78]
\\
$B^+ \to K^{*+}\phi$ & V & $1.2 ~[1.2 ]$   & $1.3  ~[ 1.3]$  & 2.1
[2.1]\\
\hline
\end{tabular}
\end{center}
\end{table}

In the $B \to PV$ decays, the three class-II decays which may have significant 
electroweak penguin amplitudes are $B^0 \to \rho^0 \pi^0$ and $B^0 \to 
\rho^0 \eta^{(\prime)}$.
 Most striking among the class-IV decays is $B^0 \to \rho^0 K^0$, which is 
completely
dominated by the electroweak penguins for all values of $N_c$.
This decay is estimated to have a
branching ratio of $O(10^{-6})$. Measurement of this decay mode will 
enable us to determine the largest electroweak-penguin coefficient $a_9$.
In the $B \to VV$ decays, the class-II decay $B^0 \to \rho^0 \rho^0$ is
sensitive to the electroweak penguins. Likewise, the two class-IV decays,
$B^0 \to \rho^0 K^{*0}$ and $B^+ \to \rho^0 K^{*+}$ are
sensitive to electroweak penguins. 
All of them are expected to have branching ratios of $O(10^{-6})$ or
larger, and can in principle all be used to determine the coefficients
of the electroweak penguins. Once again, a large number of class-V decays
show extreme sensitivity to the electroweak penguins, as can be seen in
Table 12. 
\section{Stringent tests of the factorization approach and determination of
form factors}

  In the preceding section, we have compared available data with 
estimates based on the factorization approach and have already commented
on the tendency of data to favor somewhat higher values of the form factors
$F_{0,1}^{B \to P}$, than, for example, the central values given in
Table 4. 
However, as the decay rates depend on a number of
parameters and the various parametric dependences are correlated,
it is worthwhile, in our opinion, to measure some ratios of branching ratios
in which many of the parameters endemic to the factorization 
framework cancel. In line with this, 
we propose three different types of ratios which can be helpful
in a quantitative test of the present framework:
\begin{itemize}
\item Ratios which do not depend on the effective coefficients $a_i$, and
which will allow to determine the form factors more precisely in the
factorization framework.
\item Ratios which depend on the parameters $a_i$,
and whose measurements will determine these effective coefficients.
\item Ratios whose measurements will impact on the CKM phenomenology,
i.e., they will help determine the CKM parameters $\rho$ and $\eta$
(equivalently $\sin \alpha, ~\sin \beta$ and $\sin \gamma$).  
\end{itemize}
\subsection{Ratios of branching ratios independent of the coefficients 
$a_i$}
 We start with the ratios of branching 
ratios in which the effective coefficients $a_1, ...,a_{10}$ 
cancel. In the present approach, these ratios depend on the form factors
and hadronic coupling constants. Their measurements will allow us to
discriminate among models, determine some of the hadronic quantities and 
improve the quality of theoretical
predictions for a large number of other decays where these hadronic
quantities enter.

 In what follows, we shall illustrate this by giving 
complete expressions for the relative decay widths of the decay modes in
question. These expressions  
can be derived in a straightforward way from the matrix elements 
given in the Appendices. Then, we shall present simple formulae, which are 
approximate but instructive, and highlight the particular form factors which 
play dominant
roles in these decays. Finally, we shall compare the numerical results
for these ratios obtained from the complete expressions, which have been 
used in calculating the entries in Tables 8 - 11, and the corresponding ones 
obtained from the  simple formulae to judge the quality of the 
approximation in each case.
As practically an almost endless number of ratios can be formed from the
seventy six branching ratios given in Tables 8 - 11, some thought has 
gone into selecting the eleven ratios which we discuss below. Our criterion
is based on the theoretical simplicity and experimental 
feasibility of these ratios. To be specific, these
ratios involve those decays whose branching ratios are expected to be 
$ O(10^{-6})$ or higher, with the ratios of branching ratios of order one
so that a reasonable experimental accuracy could be achieved, 
and whose decay widths are dominated by a single form factor.

We start with the discussion of decay modes involving the final states
$\pi \pi$, $\rho\pi $ and $\rho \rho$.
These ratios are listed below:

\begin{eqnarray}
P_1 &\equiv &\frac{\B(B^0 \to \rho^+ \pi^-) }{\B (B^0 \to 
\rho^+ \rho^-)}\label{p1}\\
&=&\frac{x^2 f(\pi,\rho)^3 | F_1^ {B\to \pi}(m_\rho^2)|^2
}{f(\rho,\rho)^3\left[\frac{1}{4} (\frac{3x^4}{f(\rho,\rho)^2}+1)
(1+x)^2 A_1^2+ 
\frac{f(\rho,\rho)^2A_2^2}{(1+x)^2} + \frac{2  x^4 V^2}
{(1+x)^2} - (\frac{1}{2}-x^2)A_1A_2 \right]},\nonumber
\end{eqnarray}
where $x=m_\rho /m_B$. The form factors  $A_1$, $A_2$ and $V$  
involve the $B\to \rho$ transition. The function $f(a,b)$ is the momentum 
fraction carried by the final particles, $f (a,b)<1/2$.
$$f(a,b)=\frac
{\sqrt{(m_B^2- m_a^2 -m_b^2)^2 - 4  m_a^2 m_b^2}}{2m_B^2}.$$
Since $f(\pi,\rho)\simeq f(\rho,\rho)\simeq 1/2- x^2$,
and in almost all models one expects $A_1\simeq A_2$, the expression
given in eq.~(\ref{p1}) gets considerably simplified. Neglecting  
the terms proportional to $x^4$ in the denominator, one has:
\begin{equation}
\label{P1approx}
P_1\simeq\frac{ | F_1^ {B\to \pi}(m_\rho^2)|^2
}{(1+x) |A_1^{B\to \rho}(m_\rho^2)|^2}\label{p1p}~,
\end{equation}
which is essentially determined by the ratios of the form 
factors $F_1^{B\to \pi}$ and $A_1^{B\to \rho}$.
We show the values of the ratio $P_1$ in Table 13 for the BSW model and the
lattice-QCD/QCD sum rules method for both the full widths and following
from the approximate relation given in eq.~(\ref{P1approx}). 
 There are various other relations
of a similar kind. For example, neglecting the small QCD penguin
 contribution and the very small difference in phase space, we get the
relations:
\begin{equation}
\label{P2approx}
P_2 \equiv \frac{\B(B^0
\to \pi^{-} \pi^+)}{\B (B^0\to
\rho^{+} \pi^-) }\simeq \left( \frac{f_\pi F_0^{B\to \pi}(m_\pi^2)}
{f_\rho F_1^{B\to \pi}(m_\rho^2)}\right)^2~,
\end{equation}
\begin{equation}
\label{P3approx}
P_3 \equiv \frac{\B(B^0
\to \pi^{+} \rho^-)}{\B (B^0\to  
\rho^{+} \pi^-)}\simeq \left(\frac{f_\pi A_0^{B\to \rho}(m_\pi^2)}{
f_\rho F_1^{B\to \pi}(m_\rho^2)}
\right)^2~.
\end{equation}
 
 As can be seen in Table 13, both eqs.~(\ref{P2approx}) and
(\ref{P3approx}) are excellent 
approximations and, for the two models in question, we get
an almost form-factor independent prediction, namely $P_2 \simeq 0.4$
and $P_3 \simeq 0.28$. It must be remarked here that one must disentangle
$B^0$ decays from the $\overline{B^0}$ decays as both $P_2$ and $P_3$ are 
defined for the decays of $B^0$. 

%
%
\begin{table}
\begin{center}\label{pi}
\caption{Values of $P_i$'s calculated with the form factors from the BSW
model and the hybrid lattice-QCD/QCD-sum rule method.
The numbers in square brackets are calculated using the approximate
formulae derived in the text.}
\begin{tabular}{|c|c|c|}
\hline
Ratio & BSW model & Lattice-QCD/QCD-Sum rules \\
\hline
 $P_1$ &  1.19 [1.21] & 1.27 [1.55] \\
 $P_2$ &  0.43 [0.39] & 0.43 [0.39] \\
 $P_3$ &  0.28 [0.28] & 0.27 [0.27] \\
 $P_4$ &  0.49 [0.47] & 0.53 [0.61] \\
 $P_5$ &  0.52 [0.47] & 0.55 [0.61] \\
 $P_6$ &  1.11 [1.21] & 1.19 [1.55] \\
 $P_7$ &  1.11 [1.21] & 1.19 [1.55] \\
 $P_8$ &  1.08 [1.14] & 0.99 [1.18] \\
 $P_9$ &  1.09 [1.14] & 0.99 [1.18]\\
 $P_{10}$ &  1.01 [1.15] & 0.92 [1.19] \\
 $P_{11}$ &  1.01 [1.15] & 0.92 [1.19] \\
\hline
\end{tabular}\end{center}
\end{table}
In the same vein, we define the ratios $P_4$ and $P_5$ involving the $\pi 
\pi$ and $\rho \rho$ modes:
 \begin{equation}
P_4 \equiv  \frac{\B({B^+}\to  \pi^+
 \pi^0)}{\B (B^+\to \rho^+
 \rho^0)} ,
\end{equation}
\begin{equation}
P_{5} \equiv  \frac{\B(B^0
\to \pi^{-} \pi^+)}{\B (B^0\to
\rho^{-} \rho^+)}.
\end{equation}
  Neglecting
the QCD penguin contribution  in $P_{4}$ and the EW penguin
in $P_{5}$, which are excellent approximations (see Table 13), we can
obtain these ratios as:
\begin{equation}
P_4\simeq P_{5}\simeq \left(\frac{f_\pi}{f_\rho}\right)^2
\frac{x^2 (1-m_\pi^2/m_B^2)
f(\pi,\pi) | F_0^ {B\to \pi}(m_{\pi}^2)|^2
}{f(\rho,\rho)^3\left[\frac{1}{4} (\frac{3x^4 }{f(\rho,\rho)^2}+1)
(1+x)^2 A_1^2+
\frac{f(\rho,\rho)^2 A_2^2}{(1+x)^2} + \frac{ 2 x^4 V^2}
{(1+x)^2} -\frac{1}{2} (1-2x^2)A_1A_2 \right]}.
\nonumber
\end{equation}
Neglecting higher order terms in $x$, we  get:
\begin{equation}
P_4 \simeq P_{5} \simeq  \left(\frac{f_\pi}{f_\rho}\right)^2\frac{ | F_1^ 
{B\to
\pi}(m_{\pi}^2)|^2 }{(1+x) |A_1^{B\to \rho}(m_{\rho}^2)|^2}\label{p9s},
\end{equation}
very similar to the relation for $P_1$, except for the ratio of the
decay constants.

The next ratios are defined for the final states involving $K^* \pi$ and
$K^* \rho$.  
\begin{eqnarray}
P_6 &\equiv & \frac{\B(B^0
\to K^{*+} \pi^-)}{\B (B^0\to 
K^{*+} \rho^-)}~, \nonumber \\
P_7 & \equiv & \frac{\B({B^+} \to\pi^+
K^{*0})}{\B (B^+\to \rho^+
 K^{*0} )}~.
\end{eqnarray}
One can express these ratios as:
\begin{eqnarray}
P_6 &=& P_7 \\
&=&\frac{x^2 f(\pi,{K^*})^3 | F_1^ {B\to \pi}(m_{K^*}^2)|^2
}{f(\rho,{K^*})^3\left[\frac{1}{4} (\frac{3x^2y^2}{f(\rho,{K^*})^2}+1)(1+x)^2 
A_1^2+ \frac{f(\rho,{K^*})^2 A_2^2}{(1+x)^2} + 
\frac{ 2 x^2 y^2 V^2}
{(1+x)^2} -\frac{1}{2} (1-x^2-y^2)A_1A_2 \right]},\nonumber
\end{eqnarray}
where $y=m_{K^*} /m_B$, and we have neglected the small phase space 
difference. Similar to the expression for 
$P_1$, we can derive a simple formula by dropping higher powers in $x$
 \begin{equation}
P_6=P_7\simeq \frac{ | F_1^ {B\to \pi}(m_{K^*}^2)|^2
}{(1+x) |A_1^{B\to \rho}(m_{K^*}^2)|^2}\label{p2s}.
\end{equation}

Again, neglecting the small phase space factor and the extrapolations of
the form factors between $q^2=m_{\rho}^2$ and $q^2 = m_{K^*}^2$, the
near equality 
$P_1\simeq P_6 \simeq P_7$ holds in the factorization assumption.

The next ratios, called $P_8$ and $P_9$, involve the final states $K 
\bar{K}^*$ and $K^{*}\bar{K}^*$, respectively. 
Defining 
\begin{eqnarray}
P_8& \equiv& \frac{\B({B^+}\to  K^+ \bar K^{*0})}
{\B (B^+\to K^{*+} \bar K^{*0} )}~,\nonumber\\
P_9 & \equiv &
 \frac{\B(B^0 \to   K^0
  \bar K^{*0})}{\B ( B^0\to  K^{*0}
 \bar K^{*0} )} ~,
\end{eqnarray}
we now have
\begin{equation}
P_8 \simeq P_9
= \frac{y^2 | F_1^ {B\to K}(m_{K^*}^2)|^2 |f(K,{K^*})/f({K^*},{K^*})|^3
}{\frac{1}{4} (\frac{3y^4}{f({K^*},{K^*})^2}+1)
(1+y)^2 |A_1^{K^*}|^2+ \frac{f({K^*},{K^*})^2 |A_2^{K^*}|^2}{(1+y)^2} + 
\frac{ 2 y^4 |V^{K^*}|^2}
{(1+y)^2} -\frac{1}{2} (1-2y^2)A_1^{K^*}A_2^{K^*}
 } ~.
\end{equation}
The form factors $A_1^{K^*}$,  $A_2^{K^*}$, $V^{K^*}$ 
are abbreviations for $A_1^{B\to K^*}$ etc., and 
again small phase space differences have been neglected.  
Expanding in $y$ and dropping higher order terms, we  get:
\begin{equation}
P_8 \simeq P_9 \simeq \frac{ | F_1^ {B\to K}(m_{K^*}^2)|^2
}{(1+y) |A_1^{B\to K^*}(m_{K^*}^2)|^2}\label{p4s},
\end{equation}
which involves ratios of the form 
factors $F_{1}^{B\to K}$ and $A_1^{B\to K^*}$.

Finally, in this series we define the ratio $P_{10}$ and $P_{11}$ involving 
the states $K\phi$ and $K^*\phi$, respectively:
 \begin{eqnarray}
P_{10} &\equiv& \frac{\B({B^+}\to  K^+
 \phi)}{\B (B^+\to K^{*+}
 \phi )} ~,\nonumber\\
P_{11} & \equiv &
 \frac{\B(B^0 \to  K^0
 \phi)}{\B ( B^0\to  K^{*0}
 \phi )} ~.
\end{eqnarray}
Ignoring the small phase space difference, we get
\begin{equation}
P_{10} \simeq P_{11}=
\frac{y^2  | F_1^ {B\to K}(m_{\phi}^2)|^2 |f(K,{\phi})/f({K^*},\phi)|^3
}{\frac{1}{4} (\frac{3y^2 z^2}{f({K^*},\phi)^2}+1)
(1+y)^2 |A_1^{K^*}|^2+ \frac{f({K^*},\phi)^2 |A_2^{K^*}|^2}{(1+y)^2} + 
\frac{ 2 y^2 z^2 |V^{K^*}|^2}
{(1+y)^2} -\frac{1}{2} (1-y^2-z^2)A_1^{K^*}A_2^{K^*} 
},\nonumber
\end{equation}
where  $z=m_{\phi} /m_B$.
Expanding in $y$ and $z$ and again neglecting higher order terms in $y$ and 
$z$, we  get:
\begin{equation}
P_{10} \simeq P_{11} \simeq \frac{ | F_1^ {B\to K}(m_{\phi}^2)|^2
}{(1+y) |A_1^{B\to K^*}(m_{\phi}^2)|^2}\label{p5s}~.
\end{equation}
So, in the factorization approximation and ignoring the small
extrapolation between $q^2=m_{K^*}^2$ and $q^2=m_{\phi}^2$, in
the form factors, we  have the near equality $P_8\simeq P_9\simeq P_{10} 
\simeq P_{11}$.
 These ratios are all proportional to the ratios of the form 
factors $F_{1}^{B\to K}$ and $A_1^{B\to K^*}$.

The ratios $P_1,...,P_{11}$ involve decays in which at least one of the $0^-$
mesons is replaced by the corresponding vector $1^-$ particle. If these
particles in the decay $B \to h_1 h_2$ were heavy, such as $D,D^*, 
D_s,D_s^*$, one could use the large energy ($1/E)$ expansion to
derive the ratios $P_i$. We have not investigated this point and hence can 
not claim that these ratios
are at the same theoretical footing as the corresponding relations involving
the decays $B \to D(D^{*}) \pi(\rho)$, studied, for example, in \cite{NS97}.
However, as the energy released in $B \to h_1 h_2$ decays is large, and no
fine tuning among the various amplitudes is involved, which is the case
in class-V decays, we think that the above relations are likely to hold.
 The ratios of branching ratios are 
 also independent of the CKM matrix elements, therefore they constitute 
good test of the factorization hypotheses. In Table~13, 
we have presented the numerical values of the ratios $P_i$, $i=1,...,11$. 
 This table  shows that 
almost all the ratios are remarkably close for the two models
used for the form factors. This, however, reflects our choice of the 
specific values of the form factors, which is influenced by the present CLEO 
data. In general, the ratios $P_i$ are measures of the ratios of the form 
factors, which could vary quite significantly from model to model, and hence 
they can be used to distinguish between them.
It can also be seen that in most cases, the
simple formulae are good approximations and would enable us to draw 
quantitative conclusions about the ratios of dominant form factors in these 
decays.

\subsection{Determination of the effective coefficients $a_i$}

In this section, we aim at measuring the effective coefficients $a_i$ of
the factorization framework. To that end, we shall study some ratios
of branching ratios which are largely free of hadronic form factors and 
decay constants.
In general, these ratios depend on the effective coefficients $a_i$
and the CKM parameters in a rather entangled fashion. To disentangle this
and gain some insight, we will have to make approximations, whose accuracy, 
however, we specify quantitatively within the present framework.

\subsubsection{Determination of the tree coefficients $a_1$ and $a_2$} 
\begin{figure}
    \epsfig{file=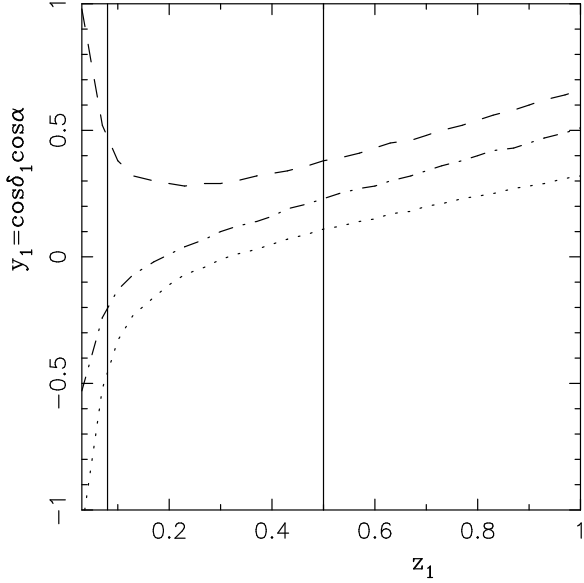,bbllx=2cm,bblly=7.5cm,bburx=21cm,bbury=19cm,%
width=12cm,angle=0}
    \caption{$y_1=\cos\delta_1 \cos \alpha$ as a function of
$z_1$ in the factorization approach.
The dotted, dashed-dotted and dashed curves correspond to $N_c=\infty$
and $\vert V_{ub}/V_{cb} \vert =0.11$, $N_c=3$ and $\vert V_{ub}/V_{cb} 
\vert =0.08$ and  $N_c=2$ and $\vert V_{ub}/V_{cb} \vert =0.06$, yielding in 
the BSW model the values  $S_1=2.07$, $S_1=0.94$ and $S_1=0.59$, respectively.
The two vertical lines indicate the bounds on $z_1$ from our model 
and the CKM unitarity fits $0.08<z_1<0.50$.}
    \label{zz2}
\end{figure}

We start with a discussion of the decays $B^0 
\to \pi^+ \pi^-$ and $B^+ \to \pi^+ \pi^0$, which are on the verge of
measurements \cite{cleo}.
Neglecting the electroweak contributions, which we have checked is a very
good approximation in these decays, we can derive from 
eqs.~(\ref{p+p-}) and (\ref{p-p0}) the following relation:
\begin{equation}
S_1 \equiv \frac{{\cal B}(B^0 
\to \pi^+ \pi^-)}{{2}{\cal B}(B^+ \to 
\pi^+ \pi^0)}\simeq \frac{\tau_{B^0}}{\tau_{B^+}}\left[
\left(\frac{a_1}{a_1+a_2}\right)^2 -2\frac{a_1}{a_1+a_2}
z_1\cos \alpha \cos \delta_1 + z_1^2\right],
\end{equation}
where 
$$z_1=\left|\frac{V_{tb}V_{td}^*}{V_{ub}V_{ud}^*}\right|
\left|\frac{a_4+a_6R_1}{a_1+a_2}\right|~.$$ Here, the quantities
$\tau_{B^0}$ and $\tau_{B^+}$ are the lifetimes of the $B^0$ and $B^+$
mesons, which, within present experimental accuracy, are equal to each
other. The implicit dependence on the CKM matrix elements in the quantity 
$a_4 + a_6R_1$ is not very marked (see section 2). The explicit CKM 
 factor is bounded from the unitarity fits in the
range (at $95\%$ C.L.):
$1.4<|{V_{tb}V_{td}^*}|/|{V_{ub}V_{ud}^*}|<4.6$. Varying then $N_c$ from  
$N_c=2$ to $N_c= \infty$, we get $0.08<z_1 <0.50$. This would suggest
that one might be able to determine the quantity $\cos \alpha$ from this
ratio.  However, the value of
$z_1$ is strongly correlated with that of the product
$y_1\equiv\cos \delta_1 \cos \alpha$, as shown in Fig.~\ref{zz2}
where the dependence of this product
is shown as a function of $z_1$, indicating the allowed range of 
$z_1$ for assumed values of the ratio $S_1$.
 As a result of this correlation, which is specific
to the factorization approach, 
the ratio $z_1 \cos \delta_1 \cos \alpha$ remains small in the entire
allowed parameter space. The quantity $z_1\cos \delta_1 \cos \alpha$ is
bounded from above to lie below $0.16$, which corresponds to using $N_c=2$
and $\vert V_{ub}/V_{cb} \vert =0.06$. 
 This is then bad news for determining the quantity
$\cos \alpha$ from the ratio $S_1$ but good news as far as the 
determination of 
the effective coefficients $a_1/(a_1 + a_2)$ from $S_1$ is concerned.
 Taking this as a generic case for other decays of interest,
 our best bet in the determination of the effective coefficients is to find 
ratios of branching ratios in which 
the quantity $z_i \cos \delta_i \cos \phi_i$ (here $\phi_i=
\alpha, \beta$ or $\gamma$) as well as $z_i^2$ are both small. Within the 
factorization framework, and using the present constraints on the CKM 
parameters, this can be systematically studied.
 With this in mind, we shall present
a number of approximate formulae for the ratios $S_i$, which are expected to 
hold in the limit: $z_i \cos \delta_i \cos \phi_i \ll 1$ and $z_i^2 \ll 1$.
To quantify the quality of our approximation, we shall make detailed
numerical comparisons between the numerical results for $S_i$,
obtained with the complete expressions for the respective decay widths, 
and the ones following from our approximate formulae.

   There are some ratios of branching ratios in which, within our 
theoretical 
framework, the factors $z_i \cos \delta_i \cos \phi_i$ are large, or else 
the CKM dependence of the ratios factorizes in a simple way. We shall use 
these ratios
to determine the CKM parameters in non-leptonic two-body decays $B \to
h_1 h_2$. This kind of analysis has already been suggested in the literature
\cite{fm,ag,gr}. We add a number of interesting decay modes to the
cases already studied in the literature and make quantitative predictions
for them in the present model.  

Returning to the determination of the coefficients $a_i$, we note that a
ratio similar to  $S_1$ can be defined with the $\rho \rho$ 
final states: \begin{equation}
S_2 \equiv \frac{{\cal B}(B^0 
\to \rho^+ \rho^-)}{2{\cal B}(B^+ \to 
\rho^+ \rho^0)}\simeq \frac{\tau_{B^0}}{\tau_{B^+}}\left[
\left(\frac{a_1}{a_1+a_2}\right)^2 -2\frac{a_1}{a_1+a_2}
z_2\cos \alpha \cos \delta_2 + z_2^2\right],
\end{equation}
where 
$$z_2=\left|\frac{V_{tb}V_{td}^*}{V_{ub}V_{ud}^*}\right|
\left|\frac{a_4}{a_1+a_2}\right|~.$$
This is not expected to exceed its maximum value $z_2^{max}=0.26$, the
central value being around $z_2\simeq 0.08$. Hence, one could 
use an approximate formulae for $S_1$ and $S_2$ by keeping the 
dominant term arising from the tree contributions (setting $\tau_{B^0}=
\tau_{B^+}$):
\begin{equation}
S_1 \equiv \frac{{\cal B}(B^0 
\to \pi^+ \pi^-)}{{2}{\cal B}(B^+ \to
\pi^+ \pi^0)}\simeq 
\left(\frac{a_1}{a_1+a_2}\right)^2 ,
\label{eqs1}
\end{equation}

 \begin{equation}
S_2 \equiv \frac{{\cal B}(B^0 
\to \rho^+ \rho^-)}{2{\cal B}(B^+ \to 
\rho^+ \rho^0)}\simeq 
\left(\frac{a_1}{a_1+a_2}\right)^2.
\label{eqs2}
\end{equation}

%
%
%
\begin{table}[htb]
\caption{The ratios $S_i$ calculated using  
the indicated values of $N_c$ and different values of $\rho$ and $\eta$. 
The values  are
calculated using the approximate formula (Approx.) derived in the text also.}
\begin{center}
\begin{tabular}{||l|c||ccc|ccc|ccc||}
\hline\hline
 & $N_c$ & \multicolumn{3}{|c|}{$N_c=2$} &  \multicolumn{3}{|c|}{$N_c=3$}
 &  \multicolumn{3}{|c||}{$N_c=\infty$}\\
\hline
&$|V_{ub}/V_{cb}|$ & $0.06$ & $0.08$ & $0.11$  & $0.06$ & $0.08$ & 
$0.11$ 
 & $0.06$ & $0.08$ & $0.11$ \\
\hline\hline
$ S_1 $&Exact & 0.59 & 0.66 & 0.68 & 0.83 & 0.94 & 0.95 & 1.81 & 2.03 & 
2.07\\
$     $&Approx. & 0.64 & 0.64 & 0.64 & 0.91 & 0.91 & 0.91 & 1.97 & 1.97 & 
1.97\\ \hline
$ S_2 $&Exact & 0.60 & 0.64 & 0.65  & 0.85 & 0.90 & 0.91 & 1.84 & 1.95 & 
1.98\\
$     $&Approx. & 0.64 & 0.64 & 0.64  & 0.91 & 0.91 & 0.91 & 1.97 & 1.97 & 
1.97\\ \hline
$S_{3}$&Exact & 1.33 & 1.32 & 1.32  & 1.09 & 1.09 & 1.09 & 0.74 & 0.75 & 
0.75\\
$     $&Approx. & 1.29 & 1.29 & 1.29  & 1.06 & 1.06 & 1.06 & 0.71 & 0.71 & 
0.71\\ \hline
$S_{4}$&Exact & 2.41 & 2.20 & 2.13  & 1.40 & 1.23 & 1.17 & 0.32 & 0.23 & 
0.19\\
$     $&Approx. & 2.13 & 2.13 & 2.13  & 1.20 & 1.20 & 1.20 & 0.22 & 0.22 & 
0.22\\ \hline
$S_{5}$&Exact & 0.55 & 0.97 & 1.96  & 0.37 & 0.66 & 1.33 & 0.16 & 0.28 & 
0.56\\ 
$     $&Approx. & 0.55 & 0.95 & 1.89  & 0.38 & 0.66 & 1.31 & 0.17 & 0.29 & 
0.58\\ \hline
$S_{6}$&Exact & 3.07 & 5.46 & 11.01& 1.95 & 3.47 & 7.00 & 0.75 & 1.34 & 
2.71\\
$      $&Approx. & 3.10 & 5.30 & 10.56& 2.06 & 3.53 & 7.03 & 0.86 & 1.46 & 
2.92\\ \hline
$S_{7}$&Exact & 1.97 & 3.73 & 7.62 & 1.77 & 3.35 & 6.84 & 1.49 & 2.82 & 
5.76\\
$      $&Approx. & 1.99 & 3.40 & 6.78 & 1.87 & 3.19 & 6.36 & 1.68 & 2.88 & 
5.74\\ \hline
$S_{8}$&Exact & 1.84 & 3.47 & 7.10 & 1.65 & 3.12 & 6.37 & 1.39 & 2.62 & 
5.36\\
$      $&Approx. & 1.99 & 3.40 & 6.78 & 1.87 & 3.19 & 6.36 & 1.68 & 2.88 & 
5.74\\ \hline
$S_{9}$&Exact & 0.32 & 0.65 & 1.33 & 0.31 & 0.62 & 1.27 & 0.28 & 0.57 & 
1.17\\
$      $&Approx. & 0.36 & 0.61 & 1.22 & 0.35 & 0.59 & 1.18 & 0.33 & 0.57 & 
1.13\\ \hline
$S_{10}$&Exact & 0.22 & 0.18 & 0.13 & 0.15 & 0.14 & 0.13 & 0.09 & 0.12 &
0.18\\
$      $&Approx. & 0.14 & 0.14 & 0.14 & 0.13 & 0.13 & 0.13 & 0.11 & 0.11 &
0.11\\ \hline
$S_{11}$&Exact & 0.37 & 0.17 & 0.06 & 0.28 & 0.15 & 0.07 & 0.20 & 0.16 &  
0.12\\
$      $&Approx. & 0.26 & 0.15 & 0.07 & 0.25 & 0.14 & 0.07 & 0.23 & 0.13 &
0.07\\
 \hline\hline
\end{tabular}\label{tr}
\end{center}
\end{table}
Likewise, neglecting the penguin contributions, 
which  give only several percent uncertainties, 
the value $a_2/a_1$ can also be measured from the following ratios, 
\begin{equation}
S_{3} \equiv \frac{2{\cal B}(B^+ 
\to \rho^+ \pi^0)}{{\cal B}(B^0 \to 
\rho^+ \pi^-)}\simeq  \left(1+
\frac{1}{x}\frac{a_2}{a_1}\right)^2~,
\label{eqs3}
\end{equation}
\begin{equation}
S_{4} \equiv \frac{2{\cal B}(B^+ 
\to \pi^+ \rho^0)}{{\cal B}(B^0 \to 
\pi^+ \rho^-)}\simeq  \left(1+
{x}\frac{a_2}{a_1}\right)^2~,
\label{eqs4}
\end{equation}
where the quantity $x=(f_\rho F_1^{B\to \pi})/(f_\pi A_0^{B\to \rho})$  
can be measured by measuring the ratio $P_3$.

\subsubsection{Determining the penguin coefficients}

 Concerning the coefficients $a_3,...,a_6$, we recall that the dominant
QCD penguin amplitudes are proportional to $a_4$ and $a_6$. The others 
($a_3$ and $a_5$) enter 
either as small corrections in class-IV decays, or else enter in class-V
decays, which in most cases are highly unstable due to large cancellations 
in the respective amplitudes, hence rendering this exercise not very 
trustworthy for determining the smaller coefficients. In view of
this we concentrate on relations involving the QCD-penguin coefficients 
$a_4$ and $a_6$.
For this purpose, quite a few class-IV decays listed in Tables 8 - 11
suggest themselves. Here, we take the ratios between some of the
representative decays from this class and from class-I or class-III decays.
These ratios and their approximate dependence on the coefficients of
interest are as follows:
\begin{equation}
S_{5} \equiv \frac{2{\cal B}(B^+
\to \pi^+ \pi^0)}{{\cal B}(B^+ \to 
\pi^+ K^0)} \simeq \left(\frac{f_\pi}{f_K} \right)^2
\left|\frac{V_{ub}V_{ud}^*}{V_{tb}V_{ts}^*}\right|^2 
\left|\frac{a_1+a_2}{a_4+a_6R_5}\right|^2~,
\end{equation}
\begin{equation}
S_{6} \equiv \frac{2{\cal B}(B^+ 
\to \rho^+ \rho^{0} )}{{\cal B}(B^+ \to 
\rho^+ K^{*0})} \simeq \left(\frac{f_\rho }{f_{K^*}} \right)^2
\left|\frac{V_{ub}V_{ud}^*}{V_{tb}V_{ts}^*}\right|^2 
\left|\frac{a_1+a_2}{a_4}\right|^2~,
\end{equation}

\begin{equation}
S_{7} \equiv \frac{{\cal B}(B^0
\to \pi^- \rho^+  )}{{\cal B}( B^+ \to 
\pi^+  K^{*0} )} \simeq \left(\frac{f_\rho }
{f_{K^*} } \right)^2 \left|\frac{V_{ub}V_{ud}^*}{V_{tb}V_{ts}^*}\right|^2
 \left|\frac{a_1}{a_4} \right|^2~,
\end{equation}
\begin{equation}
S_{8} \equiv \frac{{\cal B}(B^0
\to \rho^- \rho^+  )}{{\cal B}( B^+ \to 
\rho^+  K^{*0} )} \simeq \left(\frac{f_\rho}{f_{K^*}}
 \right)^2 \left|\frac{V_{ub}V_{ud}^*}{V_{tb}V_{ts}^*}\right|^2
 \left|\frac{a_1}{a_4} \right|^2~,
\end{equation}
\begin{equation}
S_{9} \equiv \frac{{\cal B}(B^0
\to \pi^+ \pi^-  )}{{\cal B}( B^+ \to 
\pi^+  K^0 )} \simeq \left(\frac{f_\pi }
{f_{K} } \right)^2 \left|\frac{V_{ub}V_{ud}^*}{V_{tb}V_{ts}^*}\right|^2
 \left|\frac{a_1}{a_4+a_6R_5} \right|^2~.
\end{equation}
Here, the quantity $R_5$ is defined as $R_5\equiv 
2m_{K^0}^2/(m_b-m_d)(m_d+m_s)$. %
 As is obvious from the formulae given above, the determination of the 
effective coefficients through these ratios is
correlated with the values  of the CKM factors, which in all
cases are given essentially by the ratio
 $|V_{ub}/V_{ts}| \simeq |V_{ub}/V_{cb}| \simeq 0.08 \pm 0.02$. We expect
that the CKM matrix element $|V_{ub}/V_{cb}|$ will be very precisely
measured in forthcoming experiments. Hence, a better use of these
ratios is to determine the effective coefficients.  To give a quantitative 
content 
to the approximations made in reaching the simple expressions for $S_i$,
$i=1,...,9$, we display in Table 14 the numerical values of these
ratios, together with the ratios $S_{10}$ and $S_{11}$ discussed below, as a 
function of $|V_{ub}/V_{cb}|$, taking a rather generous
error on this quantity in the range $0.06 \leq |V_{ub}/V_{cb}| \leq 0.11$,
for three values of $N_c$. The rows labeled as ``Exact" are the results
obtained by using the complete amplitudes and the rows labeled as
``Approx." are the results following from the simple relations given above
for these ratios. As one can see, these formulae are quite accurate
over a large parameter space,
with the deviations
mostly remaining well within $10\%$. One can also check that the ratios 
$S_5$ - $S_9$ for the complete result scale almost quadratically with 
$V_{ub}/V_{cb}$, as follows from the simple formulae, which 
shows that the CKM dependence displayed in the approximate formulae is
actually quite accurate.   

Concerning the measurements of the electroweak coefficients, 
$a_7,...,a_{10}$, we recall that the dominant contribution of the
electroweak penguin amplitudes is proportional to $a_9$. The rest of
the electroweak coefficients are
either small or they enter in combinations which render them very sensitive
to the variation in $N_c$. 
 It is instructive to consult Table~12, where the
decays in which electroweak penguins may make a significant contribution
to the branching ratios are listed. In line with our argument, we will
concentrate only on class-IV penguin-decays, and pick up the decay mode 
$B^0 \to \rho^0 K^0$ as an illustrative example. To that
end, we define the following two ratios involving a class-I and a class-IV
processes, dominated by the tree and QCD-penguins, respectively.:
\begin{eqnarray}
S_{10} & \equiv & \frac{2{\cal B}(B^0 \to \rho^0 K^0)}
{{\cal B}(B^+ \to \pi^+ K^{*0} )} \simeq \frac{9}{4} \left|
\frac{f_\rho F_1^{B \to 
K}(m_\rho^2)}{f_{K^*} F_1^{B \to \pi}(m_K^2)}\right|^2 
\left|\frac{a_9}{a_4}\right|^2~,
\nonumber\\
S_{11} & \equiv & \frac{2{\cal B}(B^0 \to \rho^0 K^0)}
{{\cal B}(B^0 \to \rho^-\pi^+ )} \simeq \frac{9}{4}
\left|\frac{V_{tb}V_{ts}^*}{V_{ub}V_{ud}^*}\right|^2 \left|\frac{f_\rho 
F_1^{B \to
K}(m_\rho^2)}{f_\pi A_0^{B \to \rho}(m_\pi^2)}\right|^2 
\left|\frac{a_9}{a_1}\right|^2~.
\label{S1011}
\end{eqnarray}
We show the numerical values of these ratios in Table 14 for the three
indicated values of the ratio $|V_{ub}/V_{cb}|$, both for the exact and
approximate cases. The approximate relations are reliable over most part of
the parameter space. Other similar ratios can be written down in a 
straightforward way.
Measurements of the ratios $S_1$ - $S_{11}$ will overconstrain the
coefficients $a_4$, $a_6$ and $a_9$, testing both the factorization
hypothesis and determining these crucial penguin coefficients.
Note that $S_{10}$ depends only slightly on the CKM factors, compared to
the others discussed above, and $S_1$ to $S_4$ do not depend on $\vert 
V_{ub}/V_{cb}\vert$ when we use the approximations in eqs.~(\ref{eqs1}) -
(\ref{eqs4}).

\subsection{Potential impact of $B \to h_1 h_2$ decays on CKM
 phenomenology}

{\bf (i) $B\to \pi K$ channels:}

 In this subsection, we consider the ratios of branching ratios which can be
gainfully used to get information on the CKM parameters. The
most celebrated one in this class is the ratio discussed by
Fleischer and Mannel recently \cite{fm}, involving the decays
$B^0  \to K^+ \pi^-$ and $B^+ \to
K^0 \pi^+$. Ignoring the electroweak penguin contribution, which is
estimated to be small in our model, one can write this ratio as: 
\begin{equation}
\label{S12}
S_{12} \equiv \frac{{\cal B}(B^0 
\to  K^+ \pi^- )}{{\cal B}(B^+ \to
K^0 \pi^+ )} \simeq 1-2z_{12}\cos \delta_{12} \cos \gamma +z_{12}^2,
\end{equation}
with
$$z_{12}=\frac{|T|}{|P|}=\left|\frac{V_{ub}V_{us}^*}{V_{tb}V_{ts}^*}\right|
\left|\frac{a_1}{a_4+a_6R_5} \right|.$$

\begin{figure}
  \begin{center}
    \epsfig{file=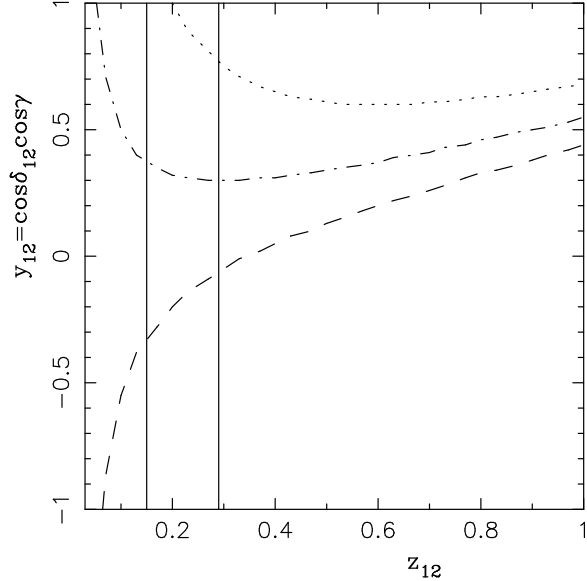,bbllx=2cm,bblly=7.5cm,bburx=21cm,bbury=19cm,%
width=12cm,angle=0}
    \caption{$y_{12}=\cos\delta_{12} \cos \gamma$ as a function of 
$z_{12}$ in the factorization approach.
The dotted, dashed-dotted and dashed curves correspond to $N_c=\infty$
and $\vert V_{ub}/V_{cb} \vert =0.11$, 
$N_c=3$ and $\vert V_{ub}/V_{cb} \vert =0.08$,
and $N_c=2$ and $\vert V_{ub}/V_{cb} \vert =0.06$, yielding in the BSW model 
the values $S_{12}=0.46$, $S_{12}=0.91$ and $S_{12}=1.12$, respectively.
The two vertical lines indicate the bounds on $z_{12}$ from our model
and the CKM factors discussed in the text, yielding $0.15<z_{12}<0.29$.}
    \label{zz1}
  \end{center}
\end{figure}

The branching ratios involved in $S_{12}$ have  been measured by the CLEO 
collaboration and their values can be seen in Table 8. The ratio
$S_{12}$ itself has the following value:
\begin{equation}
S_{12}= 0.65 \pm 0.39 \quad .
\end{equation}
For the central values of the CKM parameter ($\rho=0.12, \eta=0.34)$, the
value of $S_{12}$ is found to be $0.80 \leq S_{12} \leq 1.0$ varying
$N_c$ and using the two form factor models displayed in Table 8.
However, varying the CKM parameters in their presently allowed range, we
find $0.46 \leq S_{12} \leq 1.12$, where the lower  and upper values 
correspond to $\vert V_{ub}/V_{cb}\vert =0.11$ and $\vert 
V_{ub}/V_{cb}\vert =0.06$, respectively.
The ratio $S_{12}$ is, formally speaking, very similar to the one defined 
for the ratio $S_1$.
However, the difference between $S_1$ and $S_{12}$ is that the product 
$z_{12} \cos \delta_{12} \cos \gamma$, as opposed to the corresponding 
quantity $z_1 \cos \delta_1 \cos \alpha$ in $S_1$,  is 
not small in the allowed region of $z_{12}$. The range $0.15 \leq z_{12} 
\leq 0.29$ is estimated in the factorization approach varying the CKM
matrix element ratio in the range 
$0.013<|{V_{ub}V_{us}^*}|/|{V_{tb}V_{ts}^*}|<0.023$ and $N_c$. 
This is shown in Fig.~\ref{zz1}.
Hence, the ratio $S_{12}$ and its kinds, discussed below, do provide, in
principle, a constraint on $\cos \gamma$. This figure also shows that
the ratio $S_{12}$ is in quite good agreement with the measured ratio
by CLEO.

\begin{figure}
    \epsfig{file=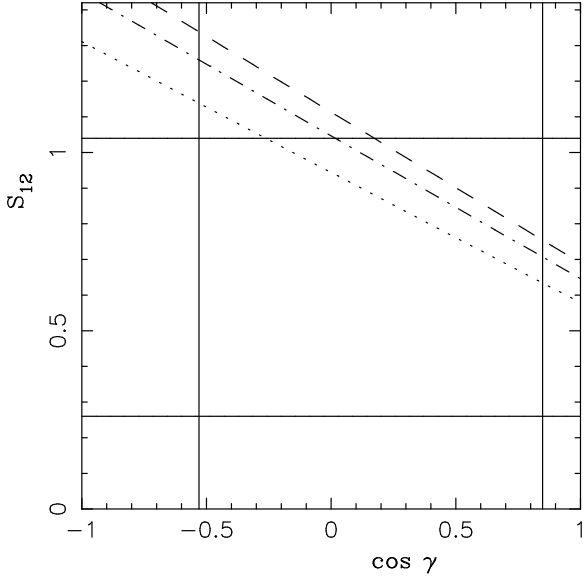,bbllx=2cm,bblly=7.5cm,bburx=21cm,bbury=19cm,%
width=12cm,angle=0}
    \caption{$S_{12}$ as a function of $\cos \gamma$ in the factorization
approach.
The dotted, dashed-dotted and dashed curves correspond to $N_c=\infty$,   
$N_c=3$ and $N_c=2$, respectively. The horizontal lines are the CLEO
$(\pm 1\sigma)$ measurements of $S_{12}$.
The two vertical lines correspond to $32^\circ < \gamma < 122^\circ$. } 
\label{ss1}
\end{figure}

 In the context of the
factorization models, the CLEO data was analyzed in \cite{ag} and it was
shown that theoretical estimates in this framework are in agreement
with data. The ratio $S_{12}$ (called $R_1$ in \cite{ag}) provides a
constraint on the CKM parameter $\rho$ (equivalently  $\cos \gamma$). 
Taking data at $\pm 1 \sigma$
value, the CLEO data disfavored the negative-$\rho$ region. The allowed
values of this parameter resulting from the measurement of $S_{12}$ were
found to be in comfortable agreement with the ones
allowed by the CKM unitarity fits. In addition, the dependence of
$S_{12}$ on the CKM parameter $\eta$ 
was found to be weak. This overlap in the value of $\rho$ following from
the analysis of the ratio $S_{12}$ in the factorization approach and from
the CKM unitarity fits has 
also been confirmed
recently in \cite{Parodi98}.
We show here the ratio $S_{12}$ plotted as a function of $\cos \gamma$
for $N_c=2, 3$ and $\infty$ and fixed value of the ratio
$|V_{ub}/V_{cb}|=0.08$ in Fig.~\ref{ss1}. 
The form factor dependence of this ratio is
rather weak (as can be seen in Table 8) and for the sake of definiteness we 
display the result for the
BSW form factors. It is seen that for all values of $N_c$, the CLEO data
provides a constraint on $\cos \gamma$, which is compatible with the one
allowed by the CKM fits, yielding  $32^\circ \leq \gamma \leq 122^\circ$ 
\cite{aliapctp97}. This is in line with what has already been reported in 
\cite{ag}.

 The ratio $S_{12}$ given in eq.~(\ref{S12}) is a generic example of the
kind of relations that one can get from the ratios of branching ratios
in which the quantity $z_i \cos \delta_i \cos \gamma$ is not small.
We have argued, in line with \cite{ag}, that the factorization model
gives an adequate account of $S_{12}$. We discuss below some
related ratios, which, once measured, could be used to determine $\cos 
\gamma$ as well as further test the consistency of the factorization 
approach.

{\bf (ii) Ratios for $B \to \pi K^*$ modes:}

\begin{figure}
    \epsfig{file=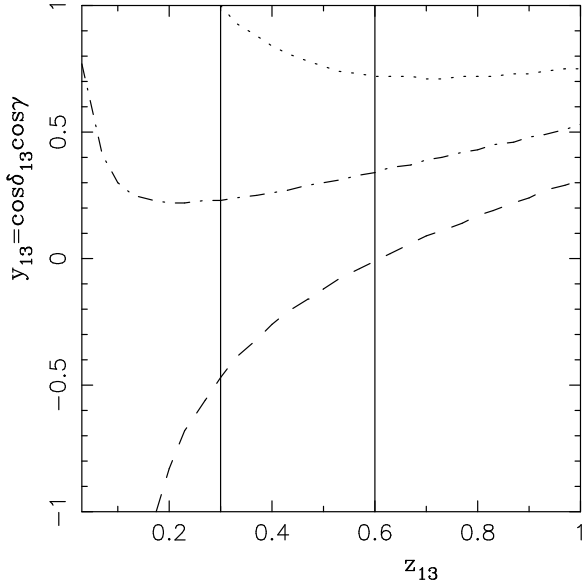,bbllx=2cm,bblly=7.5cm,bburx=21cm,bbury=19cm,%
width=12cm,angle=0}
    \caption{$y_{13}=\cos\delta_{13} \cos \gamma$ as a function of 
$z_{13}$ in the factorization approach.
The dotted, dashed-dotted and dashed curves correspond to $N_c=\infty$
and $\vert V_{ub}/V_{cb} \vert =0.11$,
$N_c=3$ and $\vert V_{ub}/V_{cb} \vert =0.08$,
and $N_c=2$ and $\vert V_{ub}/V_{cb} \vert =0.06$, yielding in the BSW 
model 
the values $S_{13}=0.49$, $S_{13}=0.95$ and $S_{13}=1.37$, respectively.
The two vertical lines indicate the bounds on $z_{13}$ from our model
and the CKM factors $0.30<z_{13}<0.60$.}  
\label{zz7}
\end{figure}

\begin{figure}
    \epsfig{file=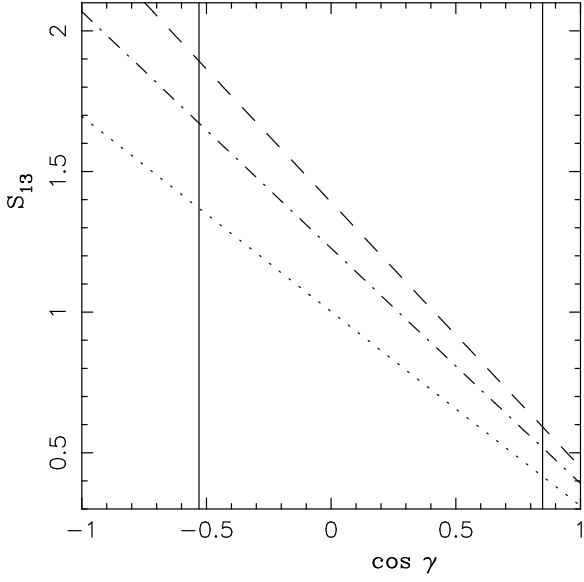,bbllx=2cm,bblly=7.5cm,bburx=21cm,bbury=19cm,%
width=12cm,angle=0}
    \caption{$S_{13}=S_{15}$ as a function of
$\cos \gamma$.  The dotted, dashed-dotted and dashed lines correspond to
results with $N_c=\infty$, $N_c=3$ and $N_c=2$, respectively.
The two vertical lines correspond to $32^\circ < \gamma < 122^\circ$.}
\label{ss4}
\end{figure}

One can define analogous to eq.~(\ref{S12}), the ratio $S_{13}$, involving
the decays $B^0 
\to \pi^- K^{*+}$ and $B^+ \to K^{*0}\pi^+ $:
\begin{equation}
S_{13} \equiv \frac{{\cal B}(B^0 
\to \pi^- K^{*+} )}{{\cal B}(B^+ \to 
\pi^+ K^{*0})} \simeq 1-2z_{13} \cos \delta_{13} \cos \gamma +z_{13}^2,
\end{equation}
with 
$$z_{13}=\frac{|T|}{|P|}=\left|\frac{V_{ub}V_{us}^*}{V_{tb}V_{ts}^*}\right|
\left|\frac{a_1}{a_4} \right|$$
Using $0.013<|{V_{ub}V_{us}^*}|/|{V_{tb}V_{ts}^*}|<0.023$, and from $N_c=2$ to 
$N_c= \infty$, we get $0.30<z_{13} <0.60$, indicated in Fig.\ref{zz7}.
The ratio $S_{13}$ is plotted in Fig.~\ref{ss4} as a function of $\cos 
\gamma$ for three different values of $N_c$ and $\vert V_{ub}/V_{cb} \vert$.
When measured, this ratio will provide a constraint on the phase $\cos 
\gamma$. Varying the CKM parameters and $N_c$ in the indicated range, we
find the ratio $S_{13}$ to lie in the range $0.49 \leq S_{13} \leq 1.37$.
The upper bound is larger than the one for $S_{12}$ given earlier, reflecting
that the QCD-penguin contributions in the two ratios are similar but not 
identical.

{\bf (iii) Ratios for $B \to \rho K$ modes}

\begin{equation}
S_{14} \equiv \frac{{\cal B}(B^0 
\to \rho^- K^{+ } )}{{\cal B}(B^+ \to 
\rho^+ K^0)} \simeq 1-2z_{14} \cos \delta_{14} \cos \gamma +z_{14}^2,
\end{equation}
with 
$$z_{14}=\frac{|T|}{|P|}=\left|\frac{V_{ub}V_{us}^*}{V_{tb}V_{ts}^*}\right|
\left|\frac{a_1}{a_4+a_6Q_4} \right|~.$$
The central value of the quantity $z_{14}$ is $z_{14} \simeq 5.07$.
However, being very large, the ratio $S_{14}$ implies that the branching
ratio in the denominator is appreciably smaller and
perhaps lot more difficult to measure. In view of this, we are less sure of
its utility of the ratio $S_{14}$ in the foreseeable future.

 {\bf (iv) Ratios for $B \to \rho K^*$ modes}

Finally, we note that the ratio $S_{15}$ defined below provides, within
our model, a very similar constraint on $\cos \gamma$ as the one following 
from the ratio $S_{13}$: 
\begin{equation}
S_{15} \equiv \frac{{\cal B}(B^0 
\to \rho^- K^{*+} )}{{\cal B}(B^+ \to 
\rho^+ K^{*0})} \simeq 1-2z_{15} \cos \delta_{15} \cos \gamma +z_{15}^2~,
\end{equation}
where $z_{15}=z_{13}$ and $\delta_{15}=\delta_{13}$.
This will be a further test of the factorization Ansatz. 

Finally, in conclusion of this section, we mention that a method of
measuring the CKM matrix element ratio $|V_{td}/V_{ts}|$ using exclusive
non-leptonic $B$ decays has been proposed in ref.~\cite{gr}. Some of these
ratios have modest theoretical uncertainties due to SU(3) breaking effects.
These relations hold in the factorization framework as well, and we list a
few of them below:
\begin{equation}
 \frac{{\cal B}(B^+
\to K^+ \bar K^0 )}{{\cal B}(B^+ \to 
\pi^+  K^0)} \simeq  \frac{{\cal B}(B^+
\to K^+ \bar K^{*0} )}{{\cal B}(B^+ \to 
\pi^+  K^{*0})} \simeq \frac{{\cal B}(B^+
\to K^{*+} \bar{K}^0  )}{{\cal B}(B^+ \to 
\rho^+  K^0)}
\end{equation}
$$ \simeq \frac{{\cal B}(B^+
\to K^{*+} \bar K^{*0} )}{{\cal B}(B^+ \to 
\rho^+ \ K^{*0})} 
\simeq\left|\frac{V_{td}}{V_{ts}}\right|^2.$$

\section{Summary and Conclusions}

  We have presented estimates of the decay rates  
in two-body non-leptonic decays $B \to h_1 
h_2$ involving pseudoscalar and vector light hadrons in which QCD and 
electroweak penguins play a significant role.  
This work partly overlaps with studies done earlier along these lines
on branching ratios, in particular in \citer{ag,CT98}. We make use of the 
theoretical framework detailed in \cite{ag,acgk} but we think that this 
is the most comprehensive study of its 
kind in the factorization framework.

 Using the sensitivity 
on $N_c$ as a criterion of theoretical stability, we have classified all
the decays $B \to h_1 h_2$ into five different classes involving penguin
and tree amplitudes. This extends the classification of tree amplitudes
{\it en vogue} in the literature \cite{BSW87,NS97}.
We hope that the detailed anatomy of the decays $B \to h_1 h_2$
presented here, in particular concerning the QCD and
electroweak penguins, will serve to have a more critical view of what can
be reasonably calculated in the factorization framework and what
involves a good deal of theoretical fine tuning.
Following the classification discussed here,  we think that class-I and 
class-IV decays, 
and probably also class-III decays, can be calculated with a reasonable
theoretical uncertainty, typically a factor 2. However, most class-II and
class-V decays deserve a
careful theoretical reappraisal to establish the extent of 
non-factorizing contributions. 
In particular, we have outlined the pattern of power suppression in
annihilation contributions to two-body non-leptonic $B$ decays. Being 
suppressed by $m_{h}^4/m_B^4$, the
annihilation contributions are small in the decays $B \to PP$ but since
this suppression is only $m_{h}^2/m_B^2$ in $B \to PV$ and $B \to VV$ 
decays, in specific cases this can be easily overcome by the favorable
effective coefficients. Hence, annihilation contributions can be significant
in some $B \to h_1 h_2$ decays involving vector mesons. 

 Our results can be summarized as follows.
\begin{itemize}
\item The recently measured decay modes $B^0 \to K^+\pi^-, B^+ \to K^+ 
\eta^\prime, B^0 \to K^0 \eta^\prime, B^+ \to \pi^+ K^0$, and $B^+ \to 
\omega K^+$ can be explained in the factorization framework. The first 
four of these belong
to the QCD-penguin dominated class-IV decays, which we argue can be reliably 
calculated. The last belongs to the $N_c$-unstable
class-V decays, which may receive significant FSI and/or annihilation 
contributions.
Taken the present theory and data on face value, all measured decay modes
are consistently accommodated, with some preference for $\xi=1/N_c \leq 
0.2$. Data on the combined decay modes $B \to \phi K^*$ prefers somewhat 
higher
value for $\xi$. However, we caution against drawing too quantitative
conclusions at this stage.
\item A number of decays is tantalizingly close to the present experimental
upper limits. We think that with $O(10^{8})$ $B/\bar{B}$ hadrons,
available in the next three to five years, a good
fraction of the seventy six decay modes worked out here will be measured
providing a detailed test of the factorization approach. 
\item To further quantify these tests, we have put forward numerous proposals
which involve measurements of the ratios of branching ratios. Carefully 
selecting the decay modes, one
could determine the effective coefficients $a_1,a_2,a_4,
a_6$ and $a_9$ from data on $B \to h_1 h_2$ decays in the future.
A consistent determination of these 
coefficients will greatly help in developing a completely
quantitative theory of non-leptonic $B$ decays.
Leaving out $a_2$ from this list, which depends significantly on $N_c$,
we do not expect that the rest will be greatly modified by
non-perturbative effects.
It will be difficult to quantitatively determine the smaller penguin
coefficients not listed explicitly.
\item We have proposed a number of ratios involving the decays $B \to h_1 
h_2$, relating the final states in which a pseudoscalar meson is replaced
by a vector meson. They will help in determining the form factors for the 
various decays considered here. While these relations are derived in the
factorization approach, perhaps their validity is more general. 
\item The current and impending interest in
two-body non-leptonic decays for the CKM 
phenomenology is illustrated, arguing that they provide potentially
non-trivial constraints on the CKM parameters. While ultimately not 
competitive to more precise determinations of the CKM parameters from 
$B^0$-$\overline{B^0}$ mixings and radiative and semileptonic $B$ decays,
they are of current phenomenological interest as the constraints following
from them are already complementary to the ones from the
CKM unitarity fits.
\item Finally, within the factorization framework which gives an 
adequate account of the present data on decay rates, it will be 
instructive to study direct and indirect CP violation in all two-body
non-leptonic $B$ decays discussed here. We hope to return to this in
a forthcoming publication \cite{AKL98-2}.
\end{itemize}

{\bf Acknowledgments}\\
We thank Christoph Greub and Jim Smith for helpful 
discussions on various aspects of theory and data discussed in this paper.
We also thank Jim Smith for critically reading the manuscript.

\begin{appendix}
\section{Matrix elements for B decays to two pseudo-scalar mesons}

(1) $b\to d$ processes:
\begin{eqnarray}
M(\bar B^0 \to \pi^-  \pi^+) & =& -i\frac{G_F}{\sqrt{2}}
f_\pi F_0^{B\to \pi } (m_\pi^2) (m_B^2-m_\pi^2) \nonumber \\
&\times &\left\{V_{ub}V_{ud}^*  a_1 -V_{tb}V_{td}^* [a_4+a_{10} 
+(a_6+a_8)R_1]\right\}, \label{p+p-}
\end{eqnarray}
with $R_1=\frac{2m_\pi^2}{(m_b-m_u)(m_u+m_d)}$.

\begin{eqnarray}
M(\bar B^0 \to \pi^0 \pi^0 ) &=&  i\frac{G_F}{\sqrt{2}}
f_\pi F_0^{B\to \pi } (m_\pi^2) (m_B^2-m_\pi^2) 
\left\{V_{ub}V_{ud}^* a_2\right . \nonumber \\
& &\left .+V_{tb}V_{td}^*[a_4-\frac{1}{2} 
a_{10}+\frac{3}{2}a_7-\frac{3}{2}a_9
 +(a_6-\frac{1}{2} a_8)R_2]\right \},
\end{eqnarray}
with $R_2=\frac{2m_{\pi^0}^2}{(m_b-m_d)(m_d+m_d)}$.
After squaring of the matrix element, the decay rate should be divided by 2,
for the symmetric factor of identical particles in the final states.

\begin{eqnarray}
M(B^- \to \pi^- \pi^0 ) & =& -i\frac{G_F}{{2}}
f_\pi F_0^{B\to \pi } (m_\pi^2) (m_B^2-m_\pi^2) \nonumber \\
&\times &\left\{V_{ub}V_{ud}^*  (a_1+a_2) -V_{tb}V_{td}^* \times 
\frac{3}{2}[a_9+a_{10}-a_7+a_8R_2] \right \}. \label{p-p0}
\end{eqnarray}

\begin{eqnarray}
M( B^- \to \pi^- \eta ^{(\prime)} )  & =& -i\frac{G_F}{\sqrt{2}}
f_\pi F_0^{B\to \eta ^{(\prime)} } (m_\pi^2) (m_B^2-m_{\eta ^{(\prime)}}^2)
\left \{ V_{ub}V_{ud}^* a_1 \right .\nonumber \\
&&~~~~\left . -V_{tb}V_{td}^*
[a_4+a_{10} +(a_6+a_8)R_1]\right \} \\
&-& i\frac{G_F}{\sqrt{2}}
f_{\eta ^{(\prime)}}^u F_0^{B\to \pi } (m_{\eta^{(\prime)}}^2) (m_B^2-m_\pi^2) 
 \left\{V_{ub}V_{ud}^*  a_2 
+V_{cb}V_{cd}^*  a_2 \frac{ f_{\eta ^{(\prime)}}^c} {f_{\eta ^{(\prime)}}^u}
 \right .\nonumber \\
&&-V_{tb}V_{td}^*\left[ a_4-\frac{1}{2}a_{10} 
+2a_3-2a_5 +\frac{1}{2}(a_9-a_{7})+(a_6-\frac{1}{2}a_8)R_3^{(\prime)}
\left( 1-\frac{f_{\eta^{(\prime)}}^u}{f_{\eta^{(\prime)}}^s}\right)\right. 
\nonumber \\
&&~~~~\left.\left.+(a_3-a_5+a_9-a_7)\frac{f_{\eta^{(\prime)}}^c}
{f_{\eta^{(\prime)}}^u}+
\left( a_3-a_5 -\frac{1}{2}(a_9-a_{7}) \right )\frac{ f_{\eta ^{(\prime)}}^s} 
{f_{\eta ^{(\prime)}}^u}\right] \right \}, \nonumber
\end{eqnarray}
with $R_3 ^{(\prime)}=\frac{m_{\eta ^{(\prime)} }^2}{(m_b-m_d)m_s}$.
The definitions of the decay constants involving
$\eta$ and $\eta^\prime$ are as follows:
\begin{equation}
\langle 0 | \bar{u} \gamma_\mu \gamma_5 u | \eta^{(\prime)}(p) \rangle =
i f_{\eta^{(\prime)}}^u p_\mu~, ~~~
\langle 0 | \bar{s} \gamma_\mu \gamma_5 s | \eta^{(\prime)}(p) \rangle =
i f_{\eta^{(\prime)}}^s p_\mu~, ~~~
\langle 0 | \bar{c} \gamma_\mu \gamma_5 c | \eta^{(\prime)}(p) \rangle =
i f_{\eta^{(\prime)}}^c p_\mu~.
\end{equation}
The quantities $f_{\eta^{(\prime)}}^u$ and $f_{\eta^{(\prime)}}^s$ 
in the two-angle mixing formalism are:
\begin{equation}
  f_{\eta^\prime} ^u =\frac{f_8}{\sqrt{6}} \sin \theta_8 + 
\frac{f_0}{\sqrt{3}} \cos \theta_0 , ~~~~~   
f_{\eta^\prime} ^s =-2 \frac{f_8}{\sqrt{6}} \sin \theta_8 + 
\frac{f_0}{\sqrt{3}} \cos \theta_0 ;
\end{equation}
\begin{equation}
  f_{\eta } ^u =\frac{f_8}{\sqrt{6}} \cos \theta_8 - 
\frac{f_0}{\sqrt{3}} \sin \theta_0 , ~~~~~   
f_{\eta } ^s =-2 \frac{f_8}{\sqrt{6}} \cos \theta_8 - 
\frac{f_0}{\sqrt{3}} \sin \theta_0 .
\end{equation}
We shall also need the matrix elements of the
pseudoscalar densities for which we use the following  equations:
\begin{eqnarray}
\frac{\langle 0| \bar{u} \gamma_5 u | \eta \rangle}
{\langle 0| \bar{s} \gamma_5 s | \eta \rangle} &=& 
\frac{f_{\eta}^u}{f_{\eta}^s} ~, \nonumber\\
\frac{\langle 0| \bar{u} \gamma_5 u | \eta^\prime \rangle}
{\langle 0| \bar{s} \gamma_5 s | \eta^\prime \rangle} &=&
\frac{f_{\eta'}^u}{f_{\eta'}^s} ~,
\end{eqnarray}
which differ from the corresponding equations in \cite{BFT95}, which have
been sometimes used in the literature. In the approximation of setting 
$f_8=f_0$,
and $\theta_8=\theta_0$, the relations given above, however, agree with the 
results derived in
\cite{AF89}. The results for the densities $\langle 0| \bar{s} \gamma_5 s 
| \eta^\prime \rangle$ and $\langle 0| \bar{s} \gamma_5 s | \eta 
\rangle$ have been derived in \cite{ag} which we use here:
\begin{eqnarray}
\langle 0| \bar{s} \gamma_5 s | \eta^\prime \rangle &=& 
-i\frac{(f_{\eta^\prime}^s-f_{\eta^\prime}^u) m_{\eta^\prime}^2}{2 m_s} ~,
\nonumber\\
\langle 0| \bar{s} \gamma_5 s | \eta \rangle &=&
-i\frac{(f_{\eta}^s-f_{\eta}^u) m_{\eta}^2}{2 m_s} ~.   
\end{eqnarray}
We point out that the anomaly 
contributions have been taken into account in deriving these expressions.
They are numerically important. The relevant form factors for the $B \to 
\eta^\prime$ and $B \to \eta$ transitions are:
\begin{equation}
F_{0,1}^{B\to \eta^\prime}  =F_{0,1}^\pi \left( \frac{ \sin \theta_8 
}{\sqrt{6}} + \frac{ \cos \theta_0}{\sqrt{3}} \right) , ~~~~~   
F_{0,1}^{B\to \eta }  =F_{0,1}^\pi \left(\frac{ \cos \theta_8 }{\sqrt{6}} - 
\frac{ \sin \theta_0}{\sqrt{3}}\right).
\end{equation}
The mixing angles that we have used in the numerical calculations are 
$\theta_8 =-22.2^\circ $, $\theta_0 =-9.1^\circ $ \cite{FK97}.

\begin{eqnarray}
M( \bar B^0 \to \pi^0 \eta ^{(\prime)} )  & =& -i\frac{G_F}{2}
f_\pi F_0^{B\to \eta ^{(\prime)} } (m_\pi^2) (m_B^2-m_{\eta ^{(\prime)}}^2)
\left \{ V_{ub}V_{ud}^* a_2 \right .\nonumber \\
&&~~~~\left . -V_{tb}V_{td}^*
\left[-a_4+\frac{1}{2}a_{10} +(-a_6+\frac{1}{2}a_8)R_2 +\frac{3}{2}
(a_9-a_7)\right]\right \} \\
&+& i\frac{G_F}{2}
f_{\eta ^{(\prime)}}^u F_0^{B\to \pi } (m_{\eta^{(\prime)}}^2) (m_B^2-m_\pi^2) 
 \left\{V_{ub}V_{ud}^*  a_2 
+V_{cb}V_{cd}^*  a_2 \frac{ f_{\eta ^{(\prime)}}^c} {f_{\eta ^{(\prime)}}^u}
 \right .\nonumber \\
&&-V_{tb}V_{td}^*\left[ a_4
+2a_3-2a_5 +\frac{1}{2}(a_9-a_{7}-a_{10} )+(a_6-\frac{1}{2}a_8)R_3^{(\prime)}
\left( 1-\frac{f_{\eta^{(\prime)}}^u}{f_{\eta^{(\prime)}}^s}\right)
\right. \nonumber \\
&&~~~~\left.\left.+(a_3-a_5+a_9-a_7)\frac{f_{\eta^{(\prime)}}^c}
{f_{\eta^{(\prime)}}^u}+
\left( a_3-a_5 -\frac{1}{2}(a_9-a_{7}) \right )\frac{ f_{\eta ^{(\prime)}}^s} 
{f_{\eta ^{(\prime)}}^u}\right] \right \}. \nonumber
\end{eqnarray}

\begin{eqnarray}
M( \bar B^0 \to \eta \eta ^\prime )  & =& -i\frac{G_F}{\sqrt{2}}
f^u_\eta F_0^{B\to \eta ^\prime } (m_\eta^2) (m_B^2-m_{\eta ^\prime}^2)
\left \{ V_{ub}V_{ud}^* a_2 
+V_{cb}V_{cd}^*  a_2 \frac{ f_{\eta}^c} {f_{\eta }^u}\right . \\\nonumber
&& -V_{tb}V_{td}^*
\left[a_4 +2a_3-2a_5 +\frac{1}{2}(a_9-a_7-a_{10} )+
(a_6-\frac{1}{2}a_8)R_3
\left( 1-\frac{f_{\eta}^u}{f_{\eta}^s}\right)
\right.\\\nonumber
&&~~~~\left.\left.+(a_3-a_5+a_9-a_7)\frac{f_{\eta}^c}
{f_{\eta}^u}+
\left( a_3-a_5 -\frac{1}{2}(a_9-a_{7}) \right )\frac{ f_{\eta }^s} 
{f_{\eta }^u}\right]\right \} \\
&-& i\frac{G_F}{\sqrt{2}}
f_{\eta ^\prime}^u F_0^{B\to \eta } (m_{\eta^\prime}^2) (m_B^2-m_\eta^2) 
 \left\{V_{ub}V_{ud}^*  a_2 
+V_{cb}V_{cd}^*  a_2 \frac{ f_{\eta ^\prime}^c} {f_{\eta ^\prime}^u}
 \right .\nonumber \\
&&-V_{tb}V_{td}^*\left[ a_4 
+2a_3-2a_5 +\frac{1}{2}(a_9-a_{7}-a_{10} )+(a_6-\frac{1}{2}a_8)R_3^{\prime}
\left( 1-\frac{f_{\eta^{\prime}}^u}{f_{\eta^{\prime}}^s}\right)
\right. \nonumber \\
&&~~~~\left.\left.+(a_3-a_5+a_9-a_7)\frac{f_{\eta^{\prime}}^c}
{f_{\eta^{\prime}}^u}+
\left( a_3-a_5 -\frac{1}{2}(a_9-a_{7}) \right )\frac{ f_{\eta ^\prime}^s} 
{f_{\eta ^\prime}^u}\right] \right \}. \nonumber
\end{eqnarray}

\begin{eqnarray}
M( \bar B^0 \to \eta^\prime \eta ^\prime )  & =& -i\frac{2G_F}{\sqrt{2}}
f^u_{\eta^\prime} F_0^{B\to \eta ^\prime } (m_{\eta^\prime}^2) 
(m_B^2-m_{\eta ^\prime}^2)
\left \{ V_{ub}V_{ud}^* a_2 
+V_{cb}V_{cd}^*  a_2 \frac{ f_{\eta^\prime}^c} {f_{\eta^\prime }^u}
\right .\\\nonumber 
&&~~~~ -V_{tb}V_{td}^*
\left[a_4 +2a_3-2a_5 +\frac{1}{2}(a_9-a_7-a_{10})+(a_6-\frac{1}{2}a_8)R_3^{\prime}
\left( 1-\frac{f_{\eta^{\prime}}^u}{f_{\eta^{\prime}}^s}\right)
\right.\\\nonumber
&&~~~~\left.\left.+(a_3-a_5+a_9-a_7)\frac{f_{\eta^{\prime}}^c}
{f_{\eta^{\prime}}^u}+
\left( a_3-a_5 -\frac{1}{2}(a_9-a_{7}) \right )\frac{ f_{\eta^\prime }^s} 
{f_{\eta^\prime }^u}\right]\right \} .
\end{eqnarray}
The matrix elements for $M( \bar B^0 \to \eta \eta  )$ 
are the same with the above equation with $\eta' \to \eta$.

(2) $b\to s$ processes:
\begin{eqnarray}
M(\bar B^0 \to K^- \pi^+  ) & =& -i\frac{G_F}{\sqrt{2}}
f_K F_0^{B\to \pi } (m_K^2) (m_B^2-m_\pi^2) \nonumber \\
&\times &\left\{V_{ub}V_{us}^* a_1 -V_{tb}V_{ts}^*
[a_4+a_{10} +(a_6+a_8)R_4]\right \}, 
\end{eqnarray}
with $R_4=\frac{2m_{K}^2}{(m_b-m_u)(m_u+m_s)}$.

\begin{eqnarray}
M(\bar B^0 \to \bar K^0 \pi^0 ) &=&- i\frac{G_F}{{2}}
f_K F_0^{B\to \pi } (m_K^2) (m_B^2-m_\pi^2)
 V_{tb}V_{ts}^*  \left
[a_4-\frac{1}{2}a_{10} +(a_6-\frac{1}{2}a_8)R_5\right]\\
& -& i\frac{G_F}{{2}}
f_\pi F_0^{B\to K } (m_\pi^2) (m_B^2-m_K^2) 
 \left\{V_{ub}V_{us}^*   a_2  -V_{tb}V_{ts}^*\times\frac{3}{2}(a_9-a_{7})
\right \} \nonumber , 
\end{eqnarray}
with $R_5=\frac{2m_{K^0}^2}{(m_b-m_d)(m_d+m_s)}$.

\begin{eqnarray}
M( B^- \to K^- \pi^0 )  & =& -i\frac{G_F}{{2}}
f_K F_0^{B\to \pi } (m_K^2) (m_B^2-m_\pi^2)\left \{ V_{ub}V_{us}^*
a_1 \right .\nonumber \\
&&~~~~\left . -V_{tb}V_{ts}^*
[a_4+a_{10} +(a_6+a_8)R_4]\right \} \\
&-& i\frac{G_F}{{2}}
f_\pi F_0^{B\to K } (m_\pi^2) (m_B^2-m_K^2) 
 \left\{V_{ub}V_{us}^*  a_2  -V_{tb}V_{ts}^*\times\frac{3}{2}(a_9-a_{7})
\right \}. \nonumber
\end{eqnarray}

\begin{eqnarray}
M( B^- \to K^- \eta ^{(\prime)} )  & =& -i\frac{G_F}{\sqrt{2}}
f_K F_0^{B\to \eta ^{(\prime)} } (m_K^2) (m_B^2-m_{\eta ^{(\prime)}}^2)
\left \{ V_{ub}V_{us}^* a_1 \right .\nonumber \\
&&~~~~\left . -V_{tb}V_{ts}^*
[a_4+a_{10} +(a_6+a_8)R_4]\right \} \\
&-& i\frac{G_F}{\sqrt{2}}
f_{\eta ^{(\prime)}}^u F_0^{B\to K } (m_{\eta^{(\prime)}}^2) (m_B^2-m_K^2) 
 \left\{V_{ub}V_{us}^*  a_2 
+V_{cb}V_{cs}^*  a_2 \frac{ f_{\eta ^{(\prime)}}^c} {f_{\eta ^{(\prime)}}^u}
 \right .\nonumber \\
&&-V_{tb}V_{ts}^*\left[ 2a_3-2a_5 +\frac{1}{2}(a_9-a_{7})-
(a_6-\frac{1}{2}a_{8})R_6 ^{(\prime)}\right. \nonumber \\
&&~~~+(a_3-a_5+a_9-a_7)\frac{f_{\eta^{(\prime)}}^c}
{f_{\eta^{(\prime)}}^u}+\nonumber\\
&&~~\left.\left.
\left( a_3-a_5+a_4 +\frac{1}{2}(a_7-a_{9}-a_{10})
+(a_6-\frac{1}{2}a_{8})R_6 ^{(\prime)} \right )\frac{ f_{\eta ^{(\prime)}}^s} 
{f_{\eta ^{(\prime)}}^u}\right] \right \}, \nonumber
\end{eqnarray}
with $R_6 ^{(\prime)}=\frac{2m_{\eta ^{(\prime)} }^2}{(m_b-m_s)(m_s+m_s)}$.

\begin{eqnarray}
M( \bar B^0 \to \bar K^0 \eta ^{(\prime)} )  &=&i\frac{G_F}{\sqrt{2}}
f_K F_0^{B\to \eta ^{(\prime)} } (m_K^2) (m_B^2-m_{\eta ^{(\prime)} }^2)
V_{tb}V_{ts}^*\left
[a_4-\frac{1}{2}a_{10} +(a_6-\frac{1}{2}a_8)R_5\right] \nonumber\\
& -&  i\frac{G_F}{\sqrt{2}}
f_{\eta  ^{(\prime)}}^u F_0^{B\to K } (m_{\eta ^{(\prime)}}^2) (m_B^2-m_K^2) 
 \left\{V_{ub}V_{us}^*  a_2 
+V_{cb}V_{cs}^*  a_2 \frac{ f_{\eta ^{(\prime)}}^c} {f_{\eta^{(\prime)} }^u}
 \right .\nonumber \\
&& -V_{tb}V_{ts}^* \left[ 2a_3-2a_5 +\frac{1}{2}(a_9-a_{7})-
(a_6-\frac{1}{2}a_{8})R_6 ^{(\prime)}\right.  \\
&&~~~+(a_3-a_5+a_9-a_7)\frac{f_{\eta^{(\prime)}}^c}
{f_{\eta^{(\prime)}}^u}+\nonumber\\
&&~~~\left.\left.
\left( a_3-a_5 +a_4+\frac{1}{2}(a_7-a_{9}-a_{10})
+(a_6-\frac{1}{2}a_{8})R_6 ^{(\prime)} \right )
\frac{ f_{\eta ^{(\prime)} }^s} {f_{\eta ^{(\prime)}
 }^u}\right]
\right \} . \nonumber
\end{eqnarray}

(3) Pure penguin processes:
\begin{equation}
M( B^- \to \pi^- \bar K^0)  =i\frac{G_F}{\sqrt{2}}
f_K F_0^{B\to \pi } (m_K^2) (m_B^2-m_\pi^2)V_{tb}V_{ts}^*
\left \{a_4-\frac{1}{2}a_{10}+(a_6-\frac{1}{2}a_8)R_5 \right \}.
\end{equation}

\begin{equation}
M( B^- \to K^- K^0) =i\frac{G_F}{\sqrt{2}}
f_K F_0^{B\to K } (m_{K^0}^2) (m_B^2-m_K^2)V_{tb}V_{td}^*
\left \{a_4-\frac{1}{2}a_{10}+(a_6-\frac{1}{2}a_8)R_7\right \},
\end{equation}
with $R_7=\frac{2m_{K^0}^2}{(m_b-m_s)(m_d+m_s)}$.

\begin{equation}
M( \bar  B^0 \to K^0 \bar K^0) =  i\frac{G_F}{\sqrt{2}}
f_K F_0^{B\to K } (m_{K^0}^2) (m_B^2-m_{K^0}^2)V_{tb}V_{td}^*
\left \{  a_4-\frac{1}{2}a_{10}+(a_6-\frac{1}{2}a_8)R_7\right \}.
\end{equation}

\section{Matrix elements for B decays to a vector and a pseudo-scalar meson}

(1) $b\to d$  processes:
\begin{equation}
M(\bar B^0 \to \rho^-  \pi^+)  = \sqrt{2}G_F f_\rho F_1^{B\to \pi } (m_\rho^2)
m_\rho (\epsilon \cdot p_\pi )\left \{ V_{ub}V_{ud}^* a_1 
 -V_{tb}V_{td}^*[a_4+a_{10}]\right \}.
\end{equation}

\begin{eqnarray}
M(\bar B^0 \to \rho^+  \pi^-)  &=& \sqrt{2}G_F f_\pi A_0^{B\to \rho }(m_\pi^2)
m_\rho (\epsilon \cdot p_\pi )\left \{ V_{ub}V_{ud}^* a_1\right .\nonumber \\
&& ~~~~\left . -V_{tb}V_{td}^*
 [a_4+a_{10}+(a_6+a_8)Q_1]\right \},
\end{eqnarray}
with $Q_1= \frac{-2m_\pi^2}{(m_b+m_u)(m_u+m_d)}$.
 
\begin{eqnarray}
M(\bar B^0 \to \pi^0 \rho^0 )&= &-\frac{G_F}{\sqrt{2}} m_\rho 
(\epsilon \cdot p_\pi ) \left (
f_\pi A_0^{B\to \rho }(m_\pi^2)
\left \{ V_{ub}V_{ud}^* a_2 \right .\right . \nonumber\\
&&~~~~~~\left .+V_{tb}V_{td}^* \left[a_4-\frac{1}{2}a_{10}+(a_6
-\frac{1}{2}a_8)Q_2+ \frac{3}{2}(a_7-a_9)\right] 
\right \} \\
 & &+ \left . f_\rho F_1^{B\to \pi } (m_\rho^2)
\left \{ V_{ub}V_{ud}^*a_2
+V_{tb}V_{td}^*[a_4-\frac{1}{2}a_{10}-\frac{3}{2}(a_7+a_9)]\right \}\right), 
\nonumber\end{eqnarray}
with $Q_2= \frac{-2m_{\pi^0}^2}{(m_b+m_d)(m_d+m_d)}$.

\begin{eqnarray}
M( B^- \to \pi^- \rho^0 )&= &G_F m_\rho (\epsilon \cdot p_\pi ) \left (
f_\pi A_0^{B\to \rho }(m_\pi^2)
\left \{ V_{ub}V_{ud}^* a_1 
-V_{tb}V_{td}^* [a_4+a_{10}+(a_6+a_8)Q_1] \right \}\right . \nonumber\\
 & &+ \left . f_\rho F_1^{B\to \pi } (m_\rho^2)
\left \{ V_{ub}V_{ud}^*a_2
-V_{tb}V_{td}^*[-a_4+\frac{1}{2}a_{10}+\frac{3}{2}(a_7+a_9)]\right \}\right). 
\end{eqnarray}

\begin{eqnarray}
M( B^- \to \rho^- \pi^0 )&=&G_F m_\rho (\epsilon \cdot p_\pi )\left(
f_\pi A_0^{B\to \rho }(m_\pi^2) \left \{ V_{ub}V_{ud}^* a_2 
-V_{tb}V_{td}^* \left[-a_4+\frac{1}{2}a_{10}\right .\right .\right .
 \nonumber\\
&&~~~~~~~~~~~\left .\left .
+(-a_6+\frac{1}{2}a_8)Q_2+\frac{3}{2}(a_9-a_7)\right]\right \} \nonumber\\
 &+&  \left . f_\rho F_1^{B\to \pi } (m_\rho^2)
\left \{ V_{ub}V_{ud}^*  a_1 
-V_{tb}V_{td}^* [a_4+a_{10}]\right \} \right).
\end{eqnarray}

\begin{eqnarray}
M( \bar B^0 \to \pi^0 \omega)  &= & \frac{G_F}{\sqrt{2}}
m_\omega (\epsilon \cdot p_\pi ) \left(
 f_\pi A_0^{B\to \omega }(m_\pi^2)  
 \left \{ V_{ub}V_{ud}^* a_2\right .\right .\nonumber\\
&&~~~~~~\left .-V_{tb}V_{td}^*\left[
-a_4+\frac{1}{2}a_{10}+(\frac{1}{2}a_8-a_6)Q_2
+\frac{3}{2} (a_9-a_7)\right]\right 
\} \nonumber\\
 & -&  f_\omega F_1^{B\to \pi } (m_\omega^2)
\left \{ V_{ub}V_{ud}^*a_2 \right .\nonumber\\
&&~~~~~~\left .\left .
-V_{tb}V_{td}^*\left[a_4+2(a_3+a_5)+\frac{1}{2}(a_7+a_9-a_{10})
\right] \right \} \right).
\end{eqnarray}

\begin{eqnarray}
M( B^- \to \pi^- \omega)  &= &G_F m_\omega (\epsilon \cdot p_\pi )\left(
f_\pi A_0^{B\to \omega }(m_\pi^2)   \left \{ V_{ub}V_{ud}^* a_1
-V_{tb}V_{td}^*[a_4+a_{10}+(a_6+a_8)Q_1]\right \}\right. \nonumber\\
 & +&   f_\omega F_1^{B\to \pi } (m_\omega^2)
\left \{ V_{ub}V_{ud}^*a_2 \right .\nonumber\\
&&~~~~~~\left .\left .
-V_{tb}V_{td}^*\left[a_4+2(a_3+a_5)+\frac{1}{2}(a_7+a_9-a_{10})
\right] \right \} \right) .
\end{eqnarray}

\begin{eqnarray}
M( B^- \to \rho^- \eta ^{(\prime)} )  & =& \sqrt{2}G_F m_\rho (\epsilon \cdot
p_{\eta^{(\prime)}})\left(
f_\rho F_1^{B\to \eta ^{(\prime)} } (m_\rho^2) 
\left \{ V_{ub}V_{ud}^* a_1  -V_{tb}V_{td}^*
[a_4+a_{10} ]\right \}\right. \nonumber\\
&+& 
f_{\eta ^{(\prime)}}^u A_0^{B\to \rho } (m_{\eta^{(\prime)}}^2) 
 \left\{V_{ub}V_{ud}^*  a_2 
+V_{cb}V_{cd}^*  a_2 \frac{ f_{\eta ^{(\prime)}}^c} {f_{\eta ^{(\prime)}}^u}
 \right .\nonumber \\
&&-V_{tb}V_{td}^*\left[ a_4 
+2a_3-2a_5 +\frac{1}{2}(a_9-a_{7}-a_{10})+(a_6-\frac{1}{2}a_8)Q_3^{(\prime)}
\left( 1-\frac{f_{\eta^{(\prime)}}^u}{f_{\eta^{(\prime)}}^s}\right)
\right. \nonumber \\
&&~~~~\left.\left.\left.+(a_3-a_5+a_9-a_7)\frac{f_{\eta^{(\prime)}}^c}
{f_{\eta^{(\prime)}}^u}+
\left( a_3-a_5 -\frac{1}{2}(a_9-a_{7}) \right )\frac{ f_{\eta ^{(\prime)}}^s} 
{f_{\eta ^{(\prime)}}^u}\right] \right \}\right) , 
\end{eqnarray}
where $Q_3^{(\prime)}= -\frac{m^2_{\eta ^{(\prime)}}}{m_s (m_b+m_d)}$.

\begin{eqnarray}
M( \bar B^0 \to \rho^0 \eta ^{(\prime)} )  & =& G_Fm_\rho (\epsilon \cdot
p_{\eta^{(\prime)}})\left(
f_\rho F_1^{B\to \eta ^{(\prime)} } (m_\rho^2) 
\left \{ V_{ub}V_{ud}^* a_2  \right.\right.\nonumber\\
&&~~~~\left.-V_{tb}V_{td}^*
\left[-a_4+\frac{1}{2}a_{10} +\frac{3}{2}(a_9+a_{7})\right]\right \} 
\nonumber\\
&-&
f_{\eta ^{(\prime)}}^u A_0^{B\to \rho } (m_{\eta^{(\prime)}}^2) 
 \left\{V_{ub}V_{ud}^*  a_2 
+V_{cb}V_{cd}^*  a_2 \frac{ f_{\eta ^{(\prime)}}^c} {f_{\eta ^{(\prime)}}^u}
 \right .\nonumber \\
&&-V_{tb}V_{td}^*\left[ a_4 
+2a_3-2a_5 +\frac{1}{2}(a_9-a_{7}-a_{10})+(a_6-\frac{1}{2}a_8)Q_3^{(\prime)}
\left( 1-\frac{f_{\eta^{(\prime)}}^u}{f_{\eta^{(\prime)}}^s}\right)
\right. \nonumber \\
&&~~~~\left.\left.\left.+(a_3-a_5+a_9-a_7)\frac{f_{\eta^{(\prime)}}^c}
{f_{\eta^{(\prime)}}^u}+
\left( a_3-a_5 -\frac{1}{2}(a_9-a_{7}) \right )\frac{ f_{\eta ^{(\prime)}}^s} 
{f_{\eta ^{(\prime)}}^u}\right] \right \}\right ). 
\end{eqnarray}

\begin{eqnarray}
M( \bar B^0 \to \omega \eta ^{(\prime)} )  & =& G_Fm_\omega (\epsilon \cdot
p_{\eta^{(\prime)}})\left(
f_\omega F_1^{B\to \eta ^{(\prime)} } (m_\omega^2) 
\left \{ V_{ub}V_{ud}^* a_2  \right. \right.\nonumber\\
&&~~~~\left.-V_{tb}V_{td}^*
\left[a_4+2(a_3+a_5) +\frac{1}{2}(a_7+a_9-a_{10})\right]\right \} 
\nonumber\\
&+& 
f_{\eta ^{(\prime)}}^u A_0^{B\to \omega } (m_{\eta^{(\prime)}}^2) 
 \left\{V_{ub}V_{ud}^*  a_2 
+V_{cb}V_{cd}^*  a_2 \frac{ f_{\eta ^{(\prime)}}^c} {f_{\eta ^{(\prime)}}^u}
 \right .\nonumber \\
&&-V_{tb}V_{td}^*\left[ a_4
+2a_3-2a_5 +\frac{1}{2}(a_9-a_{7}-a_{10})+(a_6-\frac{1}{2}a_8)Q_3^{(\prime)}
\left( 1-\frac{f_{\eta^{(\prime)}}^u}{f_{\eta^{(\prime)}}^s}\right)
\right. \nonumber \\
&&~~~~\left.\left.\left.+(a_3-a_5+a_9-a_7)\frac{f_{\eta^{(\prime)}}^c}
{f_{\eta^{(\prime)}}^u}+
\left( a_3-a_5 -\frac{1}{2}(a_9-a_{7}) \right )\frac{ f_{\eta ^{(\prime)}}^s} 
{f_{\eta ^{(\prime)}}^u}\right] \right \}\right). 
\end{eqnarray}

(2) $b\to s$  processes:
\begin{equation}
  M(\bar B^0 \to K^{*-} \pi^+) =\sqrt{2}G_F f_{K^*} F_1^{B\to \pi } 
(m_{K^*}^2) m_{K^*}  (\epsilon \cdot p_\pi )  \left \{ V_{ub}V_{us}^*
 a_1- V_{tb}V_{ts}^*[a_4+a_{10}] \right \} .
\end{equation}

\begin{eqnarray}
M(\bar B^0 \to K^{-} \rho^+ ) &=& \sqrt{2}G_F f_K A_0^{B\to \rho }(m_K^2)
m_\rho (\epsilon \cdot p_K )\left \{ V_{ub}V_{us}^* a_1\right . \nonumber\\
 & & ~~~~\left .
-V_{tb}V_{ts}^*  [a_4+a_{10}+(a_6+a_8)Q_4]\right \},
\end{eqnarray}
with $Q_4= \frac{-2m_K^2}{(m_b+m_u)(m_u+m_s)}$.

\begin{eqnarray}
M(\bar B^0 \to \bar K^{*0} \pi^0)  &=&  G_F m_{K^{*0}} (\epsilon \cdot p_\pi )
\left \{f_\pi A_0^{B\to K^* }(m_\pi^2) \left [ V_{ub}V_{us}^*  a_2 - 
V_{tb}V_{ts}^*\frac{3}{2}(a_9-a_{7}) \right ]\right. \nonumber\\
 &+&\left .  f_{K^*} F_1^{B\to \pi } (m_{K^{*0}}^2)
V_{tb}V_{ts}^*  \left[a_4-\frac{1}{2}a_{10}\right ] \right \}. 
\end{eqnarray}

\begin{eqnarray}
M(\bar B^0 \to \bar K^0 \rho^0)  & =&   G_F m_\rho (\epsilon \cdot p_K )
\left \{f_{K} A_0^{B\to \rho } (m_{K^{0}}^2)
V_{tb}V_{ts}^*
\left[a_4-\frac{1}{2}a_{10} 
+(a_6-\frac{1}{2}a_8)Q_5\right]\right.\nonumber\\
 &&+ \left.  f_\rho F_1^{B\to K }(m_\rho^2)\left [ V_{ub}V_{us}^* 
 a_2 - V_{tb}V_{ts}^*\times\frac{3}{2}(a_9+a_{7}) \right ]\right\}, 
\end{eqnarray}
with $Q_5= \frac{-2m_{K^0}^2}{(m_b+m_d)(m_d+m_s)}$.

\begin{eqnarray}
M( B^- \to K^{*- }\pi^0)  &= &  G_F m_{K^{*}} (\epsilon \cdot p_\pi )\left [
f_\pi A_0^{B\to K^* }(m_\pi^2)
\left \{ V_{ub}V_{us}^*  a_2-
V_{tb}V_{ts}^*\times\frac{3}{2}
(a_9-a_{7})\right \} \right .\nonumber\\
&+&  \left . f_{K^*} F_1^{B\to \pi } (m_{K^{*}}^2) \left \{
V_{ub}V_{us}^* a_1 -V_{tb}V_{ts}^* (a_4+a_{10}) \right \}\right ] .
\end{eqnarray}
\begin{eqnarray}
M( B^- \to K^- \rho^0)  &=&  G_F m_\rho (\epsilon \cdot p_K )\left[
f_{K} A_0^{B\to \rho } (m_{K}^2)
\left \{ V_{ub}V_{us}^*  a_1 \right .\right .\nonumber\\
&&~~~~\left .-V_{tb}V_{ts}^*[a_4+a_{10} +(a_6+a_8)Q_4]\right \}\\
&+&  \left .  f_\rho F_1^{B\to K }(m_\rho^2)
\left \{ V_{ub}V_{us}^* 
 a_2 - V_{tb}V_{ts}^*\times\frac{3}{2}(a_9+a_{7}) \right \}\right]. \nonumber
\end{eqnarray}
\begin{eqnarray}
M(\bar B^0 \to \bar K^0 \omega)   &=& G_F m_\omega (\epsilon \cdot p_K )\left(
-  f_{K} A_0^{B\to \omega } (m_{K^{0}}^2)
V_{tb}V_{ts}^*\left[a_4-\frac{1}{2}a_{10} 
+(a_6-\frac{1}{2}a_8)Q_5\right] \right. \nonumber\\
& +&   f_\omega F_1^{B\to K }(m_\omega^2)
\left \{ V_{ub}V_{us}^* 
 a_2 \right. \nonumber\\
&&~~~~\left.\left.
 - V_{tb}V_{ts}^* \left[2(a_3+a_5)+\frac{1}{2}(a_9+a_{7}) \right]
\right \} \right).
\end{eqnarray}
\begin{eqnarray}
M( B^- \to K^- \omega)  &=&  G_F m_\omega (\epsilon \cdot p_K )\left [
 f_{K} A_0^{B\to \omega } (m_{K}^2)\left \{ V_{ub}V_{us}^*  a_1 
\right .\right .\nonumber\\
&&~~~~\left .-V_{tb}V_{ts}^*(a_4+a_{10} +(a_6+a_8)Q_4)\right \}\\
&&+ \left .   f_\omega F_1^{B\to K }(m_\omega^2)
\left \{ V_{ub}V_{us}^* 
 a_2 - V_{tb}V_{ts}^*\left(2(a_3+a_5)+\frac{1}{2}(a_9+a_{7})\right)
 \right \}\right ]. \nonumber
\end{eqnarray}

\begin{eqnarray}
M( B^- \to K^{*-} \eta ^{(\prime)})  &=& \sqrt{2} G_F m_{K^*}
 (\epsilon \cdot p_{B} )\left(
 f_{K^*} F_1^{B\to \eta  ^{(\prime)}} (m_{K}^2)\left \{ V_{ub}V_{us}^*  a_1 
-V_{tb}V_{ts}^*(a_4+a_{10} )\right \}\right.  \nonumber\\
&+&     f_{\eta ^{(\prime)}}^u A_0^{B\to K^* }(m_{\eta ^{(\prime)}}^2)
\left \{ V_{ub}V_{us}^* 
 a_2 
+V_{cb}V_{cs}^*  a_2 \frac{ f_{\eta ^{(\prime)}}^c} {f_{\eta^{(\prime)} }^u}
\right.\nonumber\\
&&- V_{tb}V_{ts}^*\left[2(a_3-a_5)+\frac{1}{2}(a_9-a_{7})-(
a_6-\frac{1}{2}a_{8})Q_6^{(\prime)}\right .\label{kpep}\\
&&~~~~ +(a_3-a_5+a_9-a_7)\frac{f_{\eta^{(\prime)}}^c}
{f_{\eta^{(\prime)}}^u}+\nonumber\\
&&\left.\left.\left.
 \left(a_3-a_5-\frac{1}{2}(a_9-a_{7})+a_4-\frac{1}{2}a_{10}+(
a_6-\frac{1}{2}a_{8})Q_6^{(\prime)}\right)\frac{f_{\eta ^{(\prime)}}^s}
{f_{\eta ^{(\prime)}}^u}\right]
 \right \}\right ), \nonumber
\end{eqnarray}
with $Q_6 ^{(\prime)}=-\frac{2m_{\eta ^{(\prime)} }^2}{(m_b+m_s)(m_s+m_s)}$.

\begin{eqnarray}
M(\bar B^0 \to \bar K^{*0} \eta ^{(\prime)})  &=&  \sqrt{2} G_F m_{K^*}
 (\epsilon \cdot p_{B} ) \left( -
 f_{K^*} F_1^{B\to \eta  ^{(\prime)}} (m_{K}^2)
V_{tb}V_{ts}^*\left[a_4-\frac{1}{2} a_{10} \right] \right .\nonumber\\
&+&    f_{\eta ^{(\prime)}}^u A_0^{B\to K^* }(m_{\eta ^{(\prime)}}^2)
\left \{ V_{ub}V_{us}^* 
 a_2 
+V_{cb}V_{cs}^*  a_2 \frac{ f_{\eta ^{(\prime)}}^c} {f_{\eta^{(\prime)} }^u}
\right.\nonumber\\
&&- V_{tb}V_{ts}^*\left[2(a_3-a_5)+\frac{1}{2}(a_9-a_{7})-(
a_6-\frac{1}{2}a_{8})Q_6^{(\prime)}+\right .\\
&&~~~~ +(a_3-a_5+a_9-a_7)\frac{f_{\eta^{(\prime)}}^c}
{f_{\eta^{(\prime)}}^u}+\nonumber\\
&&\left.\left.\left.
 \left(a_3-a_5-\frac{1}{2}(a_9-a_{7})+a_4-\frac{1}{2}a_{10}+(
a_6-\frac{1}{2}a_{8})Q_6^{(\prime)}\right)\frac{f_{\eta ^{(\prime)}}^s}
{f_{\eta ^{(\prime)}}^u}\right]
 \right \} \right) . \nonumber
\end{eqnarray}

(3) Pure penguin processes:
\begin{equation}
  M( B^- \to \pi^- \bar K^{*0}) =-\sqrt{2} G_F f_{K^*} F_1^{B\to \pi } 
(m_{K^{*}}^2)
m_{K^{*}} (\epsilon \cdot p_\pi ) V_{tb}V_{ts}^*  \left [
a_4-\frac{1}{2}a_{10} \right ].
\end{equation}

\begin{equation}
  M( B^- \to \rho^- \bar K^0) =-\sqrt{2} G_F f_{K} A_0^{B\to \rho } 
(m_{K^{0}}^2) m_{\rho} (\epsilon \cdot p_K ) V_{tb}V_{ts}^* 
 \left [ a_4-\frac{1}{2}a_{10}+(a_6-\frac{1}{2}a_8)Q_5 \right ].
\end{equation}

 \begin{eqnarray}
  M(B^- \to K^-  K^{*0}) &=& \nonumber\\
M(\bar B^0 \to \bar K^0  K^{*0}) &=&-\sqrt{2} G_F f_{K^*} F_1^{B\to K } 
(m_{K^{*}}^2)
m_{K^{*}} (\epsilon \cdot p_K ) V_{tb}V_{td}^*  \left [a_4-\frac{1}{2}a_{10}
 \right ].
\end{eqnarray}

\begin{eqnarray}
  M( B^- \to K^{*-}  K^0)&=&  \nonumber\\
M( \bar B^0 \to \bar K^{*0}  K^0)&=&
- \sqrt{2} G_F f_{K} A_0^{B\to K^* } 
(m_{K^{0}}^2) m_{K^*} (\epsilon \cdot p_K )\nonumber\\
&\times & V_{tb}V_{td}^*
\left [ a_4-\frac{1}{2}a_{10}+(a_6-\frac{1}{2}a_8)Q_7 \right ],
\end{eqnarray}
with $Q_7= \frac{-2m_{K^0}^2}{(m_b+m_s)(m_d+m_s)}$.

\begin{equation}
  M( \bar B^0 \to \pi^0 \phi) = G_F f_\phi F_1^{B\to \pi }(m_\phi^2)
m_{\phi} (\epsilon \cdot p_\pi ) V_{tb}V_{td}^*  \left \{
 a_3+a_5-\frac{1}{2}(a_7+a_9)\right \}. 
\end{equation}
\begin{equation}
  M( B^- \to \pi^- \phi) = -\sqrt{2} M( \bar B^0 \to \pi^0 \phi). 
\end{equation}

\begin{equation}
  M( \bar B^0 \to \eta^{(\prime)} \phi) = -\sqrt{2}G_F f_\phi 
  F_1^{B\to \eta^{(\prime)} }(m_\phi^2)
m_{\phi} (\epsilon \cdot p_\eta^{(\prime)} ) V_{tb}V_{td}^*  \left \{
 a_3+a_5-\frac{1}{2}(a_7+a_9)\right \}. 
\end{equation}

\begin{eqnarray}
M( B^- \to K^- \phi)  & =&  \nonumber\\
M( \bar B^0 \to \bar K^0 \phi)  & =& -\sqrt{2}G_F f_\phi F_1^{B\to K }(m_\phi^2)
m_{\phi} (\epsilon \cdot p_K ) V_{tb}V_{ts}^*  \nonumber\\
&\times &\left \{
a_3+a_4+a_5-\frac{1}{2}(a_7+a_9+a_{10})\right \}. 
\end{eqnarray}

\section{Matrix elements for B decays to two vector mesons}

(1) $b\to d$ processes:
\begin{eqnarray}
M(\bar B^0 \to \rho^-  \rho^+) & =& -i\frac{G_F}{\sqrt{2}}
f_\rho m_\rho \left \{ (\epsilon _+ \cdot \epsilon _- ) (m_B+m_\rho)
A_1^{B\to \rho} (m_\rho^2)\right .\nonumber \\
&&\left .
 -(\epsilon _+ \cdot p_B) (\epsilon _- \cdot p_B)
\frac{2A_2^{B\to \rho}(m_\rho^2)}{(m_B+m_\rho)}
-i\epsilon_{\mu\nu\alpha\beta} \epsilon_-^\mu \epsilon_+^\nu p_B^\alpha
p_+^\beta \frac{ 2V^{B\to \rho}(m_\rho^2)}{(m_B+m_\rho)}\right \}\nonumber \\
&\times &\left\{V_{ub}V_{ud}^*  a_1 -V_{tb}V_{td}^* [a_4+a_{10} ]\right\}.
\end{eqnarray}

\begin{eqnarray}
M(\bar B^0 \to \rho^0 \rho^0 ) &=&  i\frac{G_F}{\sqrt{2}}
f_\rho m_\rho \left \{ (\epsilon _1 \cdot \epsilon _2 ) (m_B+m_\rho)
A_1^{B\to \rho} (m_\rho^2)\right .\nonumber \\
&&\left .
 -(\epsilon _1 \cdot p_B) (\epsilon _2 \cdot p_B)
\frac{2A_2^{B\to \rho}(m_\rho^2)}{(m_B+m_\rho)}
-i\epsilon_{\mu\nu\alpha\beta} \epsilon_1^\mu \epsilon_2^\nu p_B^\alpha
p_2^\beta \frac{ 2V^{B\to \rho}(m_\rho^2)}{(m_B+m_\rho)}\right \}\nonumber \\
&\times & 
\left\{V_{ub}V_{ud}^* a_2 +V_{tb}V_{td}^*[a_4-\frac{1}{2} 
a_{10}-\frac{3}{2}a_7-\frac{3}{2}a_9 ]\right \}.
\end{eqnarray}

\begin{eqnarray}
M(B^- \to \rho^- \rho^0 ) & =& -i\frac{G_F}{{2}}
f_\rho m_\rho \left [ (\epsilon _0 \cdot \epsilon _- ) (m_B+m_\rho)
A_1^{B\to \rho} (m_\rho^2)\right .\nonumber \\
&&\left .
 -(\epsilon _0 \cdot p_B) (\epsilon _- \cdot p_B)
\frac{2A_2^{B\to \rho}(m_\rho^2)}{(m_B+m_\rho)}
-i\epsilon_{\mu\nu\alpha\beta} \epsilon_-^\mu \epsilon_0^\nu p_B^\alpha
p_0^\beta \frac{ 2V^{B\to \rho}(m_\rho^2)}{(m_B+m_\rho)}\right ]\nonumber \\
&\times &
\left\{V_{ub}V_{ud}^*  (a_1+a_2) -V_{tb}V_{td}^* \times 
\frac{3}{2}[a_7+a_9+a_{10}] \right \}.
\end{eqnarray}

\begin{eqnarray}
M(\bar B^0 \to \omega \omega ) & =& -i\frac{G_F}{\sqrt{2}}
f_\omega m_\omega \left \{ (\epsilon _1 \cdot \epsilon _2 ) (m_B+m_\omega)
A_1^{B\to \omega} (m_\omega^2)\right .\nonumber \\
&&\left .
 -(\epsilon _1 \cdot p_B) (\epsilon _2 \cdot p_B)
\frac{2A_2^{B\to \omega}(m_\omega^2)}{(m_B+m_\omega)}
-i\epsilon_{\mu\nu\alpha\beta} \epsilon_1^\mu \epsilon_2^\nu p_B^\alpha
p_2^\beta \frac{ 2V^{B\to \omega}(m_\omega^2)}{(m_B+m_\omega)}
\right \}\nonumber \\
&\times &
\left\{V_{ub}V_{ud}^*  a_2 -V_{tb}V_{td}^* [a_4
+2(a_3+a_5)+\frac{1}{2} (a_7+a_9-a_{10})] \right \}.
\end{eqnarray}

\begin{eqnarray}
M(\bar B^0 \to \rho^0 \omega ) & =& -i\frac{G_F}{2\sqrt{2}}
f_\rho m_\rho \left \{ (\epsilon _0 \cdot \epsilon _\omega ) (m_B+m_\omega)
A_1^{B\to \omega} (m_\rho^2)\right .\nonumber \\
&&\left .
 -(\epsilon _0 \cdot p_B) (\epsilon _\omega \cdot p_B)
\frac{2A_2^{B\to \omega}(m_\rho^2)}{(m_B+m_\omega)}
-i\epsilon_{\mu\nu\alpha\beta} \epsilon_0^\mu \epsilon_\omega^\nu p_B^\alpha
p_\omega^\beta \frac{ 2V^{B\to \omega}(m_\rho^2)}{(m_B+m_\omega)}
\right \}\nonumber \\
&\times &
\left\{V_{ub}V_{ud}^*  a_2 -V_{tb}V_{td}^* [-a_4+\frac{1}{2}a_{10}
+\frac{3}{2} (a_7+a_9)] \right \}\nonumber\\ 
 & +& i\frac{G_F}{2\sqrt{2}}
f_\omega m_\omega \left \{ (\epsilon _0 \cdot \epsilon _\omega ) (m_B+m_\rho)
A_1^{B\to \rho} (m_\omega^2)\right .\nonumber \\
&&\left .
 -(\epsilon _0 \cdot p_B) (\epsilon _\omega \cdot p_B)
\frac{2A_2^{B\to \rho}(m_\omega^2)}{(m_B+m_\rho)}
-i\epsilon_{\mu\nu\alpha\beta} \epsilon_\omega^\mu \epsilon_0^\nu p_B^\alpha
p_-^\beta \frac{ 2V^{B\to \rho}(m_\omega^2)}{(m_B+m_\rho)}\right \}\nonumber \\
&\times &
\left\{V_{ub}V_{ud}^*  a_2 -V_{tb}V_{td}^* [a_4+2(a_3+a_5)+
\frac{1}{2}[a_7+a_9-a_{10}] \right \}.
\end{eqnarray}

\begin{eqnarray}
M(B^- \to \rho^- \omega ) & =& -i\frac{G_F}{{2}}
f_\rho m_\rho \left \{ (\epsilon _0 \cdot \epsilon _- ) (m_B+m_\omega)
A_1^{B\to \omega} (m_\rho^2)\right .\nonumber \\
&&\left .
 -(\epsilon _0 \cdot p_B) (\epsilon _- \cdot p_B)
\frac{2A_2^{B\to \omega}(m_\rho^2)}{(m_B+m_\omega)}
-i\epsilon_{\mu\nu\alpha\beta} \epsilon_-^\mu \epsilon_0^\nu p_B^\alpha
p_\omega^\beta \frac{ 2V^{B\to \omega}(m_\rho^2)}{(m_B+m_\omega)}
\right \}\nonumber \\
&\times &
\left\{V_{ub}V_{ud}^*  a_1 -V_{tb}V_{td}^* [a_4+a_{10}] \right \}\nonumber\\ 
 & -& i\frac{G_F}{{2}}
f_\omega m_\omega \left \{ (\epsilon _0 \cdot \epsilon _- ) (m_B+m_\rho)
A_1^{B\to \rho} (m_\omega^2)\right .\nonumber \\
&&\left .
 -(\epsilon _0 \cdot p_B) (\epsilon _- \cdot p_B)
\frac{2A_2^{B\to \rho}(m_\omega^2)}{(m_B+m_\rho)}
-i\epsilon_{\mu\nu\alpha\beta} \epsilon_0^\mu \epsilon_-^\nu p_B^\alpha
p_-^\beta \frac{ 2V^{B\to \rho}(m_\omega^2)}{(m_B+m_\rho)}\right \}\nonumber \\
&\times &
\left\{V_{ub}V_{ud}^*  a_2 -V_{tb}V_{td}^* \left[a_4+2(a_3+a_5)+
\frac{1}{2}(a_7+a_9-a_{10})\right] \right \}.
\end{eqnarray}

(2) $b\to s$  processes:
\begin{eqnarray}
M(\bar B^0 \to K^{*-} \rho^+  ) & =& -i\frac{G_F}{\sqrt{2}}
f_{K^*} m_{K^*} \left \{ (\epsilon _+ \cdot \epsilon _- ) (m_B+m_\rho)
A_1^{B\to \rho} (m_{K^*}^2)\right .\nonumber \\
&&
 -(\epsilon _+ \cdot p_B) (\epsilon _- \cdot p_B)
\frac{2A_2^{B\to \rho}(m_{K^*}^2)}{(m_B+m_\rho)}\nonumber \\
&&\left .
-i\epsilon_{\mu\nu\alpha\beta} \epsilon_-^\mu \epsilon_+^\nu p_B^\alpha
p_+^\beta \frac{ 2V^{B\to \rho}(m_{K^*}^2)}{(m_B+m_\rho)}\right \}\nonumber \\
&\times &\left\{V_{ub}V_{us}^* a_1 -V_{tb}V_{ts}^*
[a_4+a_{10} ]\right \}.
\end{eqnarray}

\begin{eqnarray}
M(\bar B^0 \to \bar K^{*0} \rho^0 ) & =& -i\frac{G_F}{{2}}
f_\rho m_\rho \left \{ (\epsilon _\rho \cdot \epsilon _K ) (m_B+m_{K^*})
A_1^{B\to K^* } (m_\rho^2)\right .\nonumber \\
&&\left .
 -(\epsilon _\rho \cdot p_B) (\epsilon _K \cdot p_B)
\frac{2A_2^{B\to  K^*}(m_\rho^2)}{(m_B+m_{K^*})}
-i\epsilon_{\mu\nu\alpha\beta} \epsilon_\rho^\mu \epsilon_K^\nu p_B^\alpha
p_K^\beta \frac{ 2V^{B\to  K^*}(m_\rho^2)}{(m_B+m_ {K^*})}\right \}\nonumber \\
&\times &
 \left\{V_{ub}V_{us}^*   a_2  -V_{tb}V_{ts}^*\times\frac{3}{2}(a_9+a_{7})
\right \} \nonumber \\
&-&
 i\frac{G_F}{{2}}
f_{K^*} m_{K^*} \left \{ (\epsilon _\rho \cdot \epsilon _K ) (m_B+m_\rho)
A_1^{B\to \rho} (m_{K^*}^2)\right .\nonumber \\
&&\left .
 -(\epsilon _\rho \cdot p_B) (\epsilon _K \cdot p_B)
\frac{2A_2^{B\to \rho}(m_{K^*}^2)}{(m_B+m_\rho)}
-i\epsilon_{\mu\nu\alpha\beta} \epsilon_K^\mu \epsilon_\rho^\nu p_B^\alpha
p_\rho^\beta \frac{ 2V^{B\to \rho}(m_{K^*}^2)}{(m_B+m_\rho)}\right \}
\nonumber \\
&\times &
 V_{tb}V_{ts}^*  \left
[a_4-\frac{1}{2}a_{10}\right]. 
\end{eqnarray}

\begin{eqnarray}
M( B^- \to K^{*-} \rho^0 )  &=&- i\frac{G_F}{{2}}
f_\rho m_\rho \left \{ (\epsilon _0 \cdot \epsilon _- ) (m_B+m_{K^*})
A_1^{B\to K^* } (m_\rho^2)\right .\nonumber \\
&&\left .
 -(\epsilon _0 \cdot p_B) (\epsilon _- \cdot p_B)
\frac{2A_2^{B\to  K^*}(m_\rho^2)}{(m_B+m_{K^*})}
-i\epsilon_{\mu\nu\alpha\beta} \epsilon_0^\mu \epsilon_-^\nu p_B^\alpha
p_-^\beta \frac{ 2V^{B\to  K^*}(m_\rho^2)}{(m_B+m_ {K^*})}\right \}\nonumber \\
&\times &
 \left\{V_{ub}V_{us}^*  a_2  -V_{tb}V_{ts}^*\times\frac{3}{2}(a_9+a_{7})
\right \}\\
& -& i\frac{G_F}{{2}}
f_{K^*} m_{K^*} \left \{ (\epsilon _0 \cdot \epsilon _- ) (m_B+m_\rho)
A_1^{B\to \rho} (m_{K^*}^2)\right .\nonumber \\
&&\left .
 -(\epsilon _0 \cdot p_B) (\epsilon _- \cdot p_B)
\frac{2A_2^{B\to \rho}(m_{K^*}^2)}{(m_B+m_\rho)}
-i\epsilon_{\mu\nu\alpha\beta} \epsilon_-^\mu \epsilon_0^\nu p_B^\alpha
p_0^\beta \frac{ 2V^{B\to \rho}(m_{K^*}^2)}{(m_B+m_\rho)}\right \}\nonumber \\
&\times &
\left \{ V_{ub}V_{us}^* a_1 -V_{tb}V_{ts}^*
[a_4+a_{10}]\right \}\nonumber  .
\end{eqnarray}

\begin{eqnarray}
M( \bar B^0 \to \bar K^{*0} \omega )  & =& -i\frac{G_F}{{2}}
f_{K^*} m_{K^{*0}} \left \{ (\epsilon _0 \cdot \epsilon _\omega) (m_B+m_\omega)
A_1^{B\to \omega} (m_{K^{*0}}^2)\right .\nonumber \\
&&\left .
 -(\epsilon _0 \cdot p_B) (\epsilon _\omega \cdot p_B)
\frac{2A_2^{B\to \omega}(m_{K^*}^2)}{(m_B+m_\omega)}
-i\epsilon_{\mu\nu\alpha\beta} \epsilon_0^\mu \epsilon_\omega^\nu p_B^\alpha
p_0^\beta \frac{ 2V^{B\to \omega}(m_{K^*}^2)}{(m_B+m_\omega)}
\right \}\nonumber \\
&\times & V_{tb}V_{ts}^*
[-a_4+\frac{1}{2}a_{10}]\nonumber \\
&-& i\frac{G_F}{{2}}
f_\omega m_\omega \left \{ (\epsilon _0 \cdot \epsilon _\omega ) (m_B+m_{K^*})
A_1^{B\to K^* } (m_\omega^2)\right .\nonumber \\
&&\left .
 -(\epsilon _0 \cdot p_B) (\epsilon _\omega \cdot p_B)
\frac{2A_2^{B\to  K^*}(m_\omega^2)}{(m_B+m_{K^*})}
-i\epsilon_{\mu\nu\alpha\beta} \epsilon_\omega^\mu \epsilon_0^\nu p_B^\alpha
p_-^\beta \frac{ 2V^{B\to  K^*}(m_\omega^2)}{(m_B+m_ {K^*})}\right \}
\nonumber \\
&\times &
 \left\{V_{ub}V_{us}^*  a_2  -V_{tb}V_{ts}^*[2(a_3+a_5)+\frac{1}{2}(a_9+a_{7})
\right \}.
\end{eqnarray}

\begin{eqnarray}
M( B^- \to K^{*-} \omega )  & =& -i\frac{G_F}{{2}}
f_{K^*} m_{K^*} \left \{ (\epsilon _0 \cdot \epsilon _- ) (m_B+m_\omega)
A_1^{B\to \omega} (m_{K^*}^2)\right .\nonumber \\
&&\left .
 -(\epsilon _0 \cdot p_B) (\epsilon _- \cdot p_B)
\frac{2A_2^{B\to \omega}(m_{K^*}^2)}{(m_B+m_\omega)}
-i\epsilon_{\mu\nu\alpha\beta} \epsilon_-^\mu \epsilon_0^\nu p_B^\alpha
p_0^\beta \frac{ 2V^{B\to \omega}(m_{K^*}^2)}{(m_B+m_\omega)}
\right \}\nonumber \\
&\times &
\left \{ V_{ub}V_{us}^* a_1 -V_{tb}V_{ts}^*
[a_4+a_{10}]\right \}\nonumber \\
&-& i\frac{G_F}{{2}}
f_\omega m_\omega \left \{ (\epsilon _0 \cdot \epsilon _- ) (m_B+m_{K^*})
A_1^{B\to K^* } (m_\omega^2)\right .\nonumber \\
&&\left .
 -(\epsilon _0 \cdot p_B) (\epsilon _- \cdot p_B)
\frac{2A_2^{B\to  K^*}(m_\omega^2)}{(m_B+m_{K^*})}
-i\epsilon_{\mu\nu\alpha\beta} \epsilon_0^\mu \epsilon_-^\nu p_B^\alpha
p_-^\beta \frac{ 2V^{B\to  K^*}(m_\omega^2)}{(m_B+m_ {K^*})}\right \}
\nonumber \\
&\times &
 \left\{V_{ub}V_{us}^*  a_2  -V_{tb}V_{ts}^*[2(a_3+a_5)+\frac{1}{2}(a_9+a_{7})
\right \}.
\end{eqnarray}

(3) Pure penguin processes:
\begin{eqnarray}
M( B^- \to \rho^- \bar K^{*0})  &=&i\frac{G_F}{\sqrt{2}}
f_{K^*} m_{K^*} \left \{ (\epsilon _\rho \cdot \epsilon _K ) (m_B+m_\rho)
A_1^{B\to \rho} (m_{K^*}^2)\right .\nonumber \\
&&
 -(\epsilon _\rho \cdot p_B) (\epsilon _K \cdot p_B)
\frac{2A_2^{B\to \rho}(m_{K^*}^2)}{(m_B+m_\rho)}\nonumber \\
&&
\left .
-i\epsilon_{\mu\nu\alpha\beta} \epsilon_K^\mu \epsilon_\rho^\nu p_B^\alpha
p_\rho^\beta \frac{ 2V^{B\to \rho}(m_{K^*}^2)}{(m_B+m_\rho)}\right \}
V_{tb}V_{ts}^*
\left \{a_4-\frac{1}{2}a_{10} \right \}.
\end{eqnarray}

\begin{eqnarray}
M( \bar B^0 \to \omega \phi)& =&i\frac{G_F}{2}
f_\phi m_\phi \left \{ (\epsilon _\phi \cdot \epsilon _\omega ) (m_B+m_{\omega})
A_1^{B\to \omega} (m_\phi^2)\right .\nonumber \\
&&
 -(\epsilon _\phi \cdot p_B) (\epsilon _\omega \cdot p_B)
\frac{2A_2^{B\to \omega}(m_\phi^2)}{(m_B+m_{\omega})}\nonumber \\
&&
\left .
-i\epsilon_{\mu\nu\alpha\beta} \epsilon_\phi^\mu \epsilon_\omega^\nu p_B^\alpha
p_\omega^\beta \frac{ 2V^{B\to \omega}(m_\phi^2)}{(m_B+m_{\omega})}\right \}\nonumber \\
&\times & V_{tb}V_{td}^*
\left \{a_3+a_5-\frac{1}{2}(a_7+a_9)\right \}.
\end{eqnarray}

\begin{eqnarray}
M( \bar B^0 \to \rho^{0} \phi)& =&-i\frac{G_F}{2}
f_\phi m_\phi \left \{ (\epsilon _\phi \cdot \epsilon _\rho ) (m_B+m_{\rho})
A_1^{B\to \rho} (m_\phi^2)\right .\nonumber \\
&&
 -(\epsilon _\phi \cdot p_B) (\epsilon _\rho \cdot p_B)
\frac{2A_2^{B\to \rho}(m_\phi^2)}{(m_B+m_{\rho})}\nonumber \\
&&
\left .
-i\epsilon_{\mu\nu\alpha\beta} \epsilon_\phi^\mu \epsilon_\rho^\nu p_B^\alpha
p_\rho^\beta \frac{ 2V^{B\to \rho}(m_\phi^2)}{(m_B+m_{\rho})}\right \}\nonumber \\
&\times & V_{tb}V_{td}^*
\left \{a_3+a_5-\frac{1}{2}(a_7+a_9)\right \}.
\end{eqnarray}
\begin{eqnarray}
M( B^- \to \rho^{-} \phi)& =&-\sqrt{2} M( \bar B^0 \to \rho^{0} \phi).
\end{eqnarray}

\begin{eqnarray}
M( B^- \to K^{*-} \phi)& =&\nonumber \\
M( \bar B^0 \to \bar K^{*0} \phi)& =&i\frac{G_F}{\sqrt{2}}
f_\phi m_\phi \left \{ (\epsilon _\phi \cdot \epsilon _K ) (m_B+m_{K^*})
A_1^{B\to K^*} (m_\phi^2)\right .\nonumber \\
&&
 -(\epsilon _\phi \cdot p_B) (\epsilon _K \cdot p_B)
\frac{2A_2^{B\to K^*}(m_\phi^2)}{(m_B+m_{K^*})}\nonumber \\
&&
\left .
-i\epsilon_{\mu\nu\alpha\beta} \epsilon_\phi^\mu \epsilon_K^\nu p_B^\alpha
p_K^\beta \frac{ 2V^{B\to K^*}(m_\phi^2)}{(m_B+m_{K^*})}\right \}\nonumber \\
&\times & V_{tb}V_{ts}^*
\left \{a_3+a_4+a_5-\frac{1}{2}(a_7+a_9+a_{10})\right \}.
\end{eqnarray}

\begin{eqnarray}
M( B^- \to K^{*-} K^{*0})& =&\nonumber \\
M( \bar  B^0 \to K^{*0} \bar K^{*0}) &=&  i\frac{G_F}{\sqrt{2}}
f_{K^*} m_{K^*} \left \{ (\epsilon _1 \cdot \epsilon _2 ) (m_B+m_{K^*})
A_1^{B\to K^*} (m_{K^*}^2)\right .\nonumber \\
&&
 -(\epsilon _1 \cdot p_B) (\epsilon _2 \cdot p_B)
\frac{2A_2^{B\to K^*}(m_{K^*}^2)}{(m_B+m_{K^*})}\nonumber \\
&&
\left .
-i\epsilon_{\mu\nu\alpha\beta} \epsilon_1^\mu \epsilon_2^\nu p_B^\alpha
p_2^\beta \frac{ 2V^{B\to K^*}(m_{K^*}^2)}{(m_B+m_{K^*})}\right \}
V_{tb}V_{td}^* \left \{  a_4-\frac{1}{2}a_{10}\right \}.
\end{eqnarray}
\end{appendix}

\end{document}